\documentclass[]{pasj01}
\usepackage{natbib,bm,url,comment}
\usepackage{ulem}
\begin{document} 
\Received{}
\Accepted{}

\title{SIRIUS Project. III. Star-by-star simulations of star cluster formation using a direct $N$-body integrator with stellar feedback}

\author{Michiko S. \textsc{Fujii}\altaffilmark{1}}
\altaffiltext{1}{Department of Astronomy, Graduate School of Science, The University of Tokyo, 7-3-1 Hongo, Bunkyo-ku, Tokyo 113-0033, Japan}
\email{fujii@astron.s.u-tokyo.ac.jp}

\author{Takayuki R. \textsc{Saitoh}\altaffilmark{2,3}}
\altaffiltext{2}{Department of Planetology, Graduate School of Science, Kobe University, 1-1 Rokkodai-cho, Nada-ku, Kobe, Hyogo 657-8501, Japan}
\altaffiltext{3}{Earth-Life Science Institute, Tokyo Institute of Technology, 2-12-1 Ookayama, Meguro-ku, Tokyo 152-8551, Japan}
\email{saitoh@people.kobe-u.ac.jp}

\author{Yutaka \textsc{Hirai}\altaffilmark{4,5}\thanks{JSPS Research Fellow}}
\altaffiltext{4}{Astronomical Institute, Tohoku University, 6-3, Aramaki, Aoba-ku, Sendai, Miyagi 980-8578, Japan}
\altaffiltext{5}{RIKEN Center for Computational Science, 7-1-26 Minatojima-minami-machi, Chuo-ku,
Kobe, Hyogo 650-0047, Japan}
\email{yutaka.hirai@astr.tohoku.ac.jp}

\author{Long \textsc{Wang}\altaffilmark{1,5*}}

\newcommand{\HII}{H\emissiontype{II} }

\maketitle

\begin{abstract}
One of the computational challenges of cluster formation simulations is resolving individual stars and simulating massive clusters with masses of more than $10^4 M_{\odot}$ without gravitational softening. Combining direct $N$-body code with smoothed-particle hydrodynamics (SPH) code, we have developed a new code, \textsc{ASURA+BRIDGE}, in which we can integrate stellar particles without softening. 
We add a feedback model for \HII regions into this code, in which thermal and momentum feedback is given within the Str{\"o}mgren radius. We perform $N$-body/SPH simulations of star cluster formation. Without softening, a portion of massive stars are ejected from the forming clusters. As a result, the stellar feedback works outside the clusters. This enhances/suppresses the star formation in initially sub-virial/super-virial clouds. We find that the formed star clusters are denser than currently observed open clusters, but the mass--density relation is consistent with or even higher than that is estimated as an initial cluster density. We also find that some clusters have multiple peaks in their stellar age distribution as a consequence of their hierarchical formation. Irrespective of the virial ratio of molecular clouds, approximately one-third of stars remain in the star clusters after gas expulsion.
\end{abstract}

\section{Introduction}

Most of stars form in star clusters \citep{2003ARA&A..41...57L}. The formation process of star clusters is closely related to the formation of massive stars. Once massive stars have formed in star clusters, they ionize the surrounding gas, their stellar wind blows out the gas, and finally, supernovae eject all the remaining gas \citep[][ and references therein]{2007A&A...463..981R,2014prpl.conf..291L,2019ARA&A..57..227K}. Such gas expulsion processes are also important for understanding the initial states of observed star clusters and their later dynamical evolution \citep{2014ApJ...794..147P,2015MNRAS.450.2451F,2015MNRAS.454.3872H}. 

To obtain the dynamical evolution of star clusters as collisional systems, we need to integrate the stellar systems without gravitational softening, i.e., we need a direct $N$-body integration of stars. In previous studies based on direct $N$-body simulations, fixed potential for the gas distribution was assumed as background gas distribution \citep{2014ApJ...794..147P,2015MNRAS.450.2451F,2015MNRAS.454.3872H}. Spherical potential is usually assumed in such simulations. In contrast, observations of actual star-cluster forming regions imply hierarchical formation of star clusters in turbulent molecular clouds \citep{2005ApJ...632..397G,2012ApJ...754L..37S,2014prpl.conf..291L,2015AJ....150...78Z}.

Simulations of star-cluster formation originating from turbulent molecular clouds have been dramatically improved in the last 20 years. 
In pioneering work by \citet{2003MNRAS.343..413B} using a smoothed-particle hydrodynamics (SPH) method \citep{1977AJ.....82.1013L,1977MNRAS.181..375G} with a sink method \citep{1995MNRAS.277..362B}, hierarchical formation of star clusters was demonstrated. 
These types of simulations have improved resolution and include radiation feedback from stars. The maximum total stellar mass in the models done by \citet{2007MNRAS.382.1759D,2012MNRAS.424..377D} was $1000 M_{\odot}$.
Although this is close to the typical mass of an open cluster, $500$--$1000 M_{\odot}$ \citep{2009A&A...498L..37P}, the formation process of more massive star clusters, such as young massive clusters ($\sim 10^4 M_{\odot}$) \citep[][ and references therein]{2010ARA&A..48..431P}, have not been fully investigated by using star-by-star simulations with realistic gas distribution.

Simulations of star clusters more massive than $10^4 M_{\odot}$ have been performed using super-particle methods, in which each star particle is assumed to include several stars following a given mass function. Such super-particles are treated as `cluster particles.' The cluster particles are also formed using a sink method and the radiation feedback from the cluster particles is also included using radiation hydrodynamics codes \citep{2018ApJ...859...68K, 2018NatAs...2..725H,2019MNRAS.489.1880H,2020MNRAS.497.3830F}.

There are some simulations of star-cluster formation in which stars are integrated without softening, and the gas dynamics and the feedback from massive stars are included.
\citet{2019ApJ...887...62W} coupled a fourth-order Hermite integrator, \textsc{ph4} \citep{1992PASJ...44..141M}, with an AMR code, \textsc{FLASH}, using the Astrophysical Multi-purpose Software Environment \citep[\textsc{AMUSE},][]{2013CoPhC.183..456P, 2013A&A...557A..84P, AMUSE}. \textsc{AMUSE} is a Python framework that provides us an interface to couple existing codes (community codes) written in different programming languages. The Bridge scheme \citep{2007PASJ...59.1095F} is one of the implemented coupling methods. 
In this scheme, two different integrators are combined by using an extension of a mixed-variable symplectic integrator \citep{2012MNRAS.420.1503P}. When a gas-dynamics code is coupled with a gravity $N$-body code using the Bridge scheme, the gravitational forces from gas are given as velocity kicks to stars in a fixed timestep. Using \textsc{AMUSE}, \cite{2020arXiv200309011W} performed simulations of star cluster formation up to $\sim 1000$ stars.  
\citet{2012MNRAS.420.1503P} also used \textsc{AMUSE} to perform simulations of star clusters embedded in a molecular cloud with a similar number of stars. They did not include star formation during the simulations but investigated the gas expulsion due to the stellar wind from massive stars and supernovae.

Star-by-star simulations of star clusters with gas are important for obtaining the proper timescale of gas expulsion due to stellar feedback. \citet{2020MNRAS.499..748D} performed simulations of star clusters embedded in a molecular cloud with feedback from massive stars using a direct $N$-body code combined with \textsc{FLASH}. They used the Ahmad--Cohen method \citep{1973JCoPh..12..389A} to integrate the motions of stars (sink particles). In their scheme, the force from gas to stars is updated with large (regular) timesteps, whereas the motions of stars are integrated with small (irregular) timesteps. Although their total number of stars was a maximum of $\sim 400$, they demonstrated that off-centered massive stars due to the dynamical scattering of massive stars make the gas expulsion time-scale longer compared with the case in which the feedback source is only in the cluster center. 

\citet{ONC2019MNRAS.484.1843W} argued that ejections of massive stars in cluster-forming regions due to strong binary--single encounters quit the feedback in the cluster center once and result in the formation of the second and third generations of stars \citep{Kroupa2018}. Such a multiple-star-formation event is actually observed in the Orion Nebula Cluster \citep{2017A&A...604A..22B}.   
Thus, star-by-star treatment without gravitational softening is important for understanding the formation process of star clusters.

For this purpose, we have developed a new scheme for star-by-star formation \citep[][ Paper I]{2020arXiv200512906H} and a hybrid integrator \citep[][ Paper II]{2021arXiv210105934F}. These schemes are implemented in our new code, \textsc{ASURA+BRIDGE}. In this work, we implement a \HII region feedback model into this code and perform  simulations of star-by-star cluster formation with the total masses of more than $10^4 M_{\odot}$.

This paper is organized as follows. 
In Section 2, we describe our code, \textsc{ASURA+BRIDGE}, and our model for \HII feedback. The models of star-cluster formation are also described. The results of the simulations are summarized in Section 3.  Section 4 presents the summary. In the Appendix, we present the results of test simulations for the \HII region model.

\section{Methods}
\subsection{\textsc{ASURA+BRIDGE}}
We briefly describe our direct $N$-body/SPH code, \textsc{ASURA+BRIDGE}. This code is based on an $N$-body/SPH code first developed for galaxy simulations, \textsc{ASURA} \citep{2008PASJ...60..667S, 2009PASJ...61..481S}. In Paper II, we implemented the Bridge scheme of \citet{2007PASJ...59.1095F} into \textsc{ASURA}. In the Bridge scheme in \textsc{ASURA+BRIDGE}, all stars are integrated using a higher-order scheme without gravitational softening, and gas particles are integrated using a leapfrog scheme with hierarchical timesteps. At every Bridge timestep, star particles receive a velocity kick due to the gas distribution \citep[see also][]{2012MNRAS.420.1503P}. 

For the integration of stars, we have two choices, a sixth-order Hermite \citep{2006NewA...12..169N} code or \textsc{PeTar} \citep{PeTar2020MNRAS.497..536W}. \textsc{PeTar} integrates stars using a hybrid integrator, the Particle--Particle Particle--Tree (P$^3$T) scheme \citep{2011PASJ...63..881O,2015ComAC...2....6I,2017PASJ...69...81I} with slow-down algorithmic regularization \citep[SDAR;][]{SDAR2020MNRAS.493.3398W}. P$^3$T combines the efficient particle--tree method \citep{1986Natur.324..446B} to calculate the long-range interaction of distant particles with the accurate direct $N$-body method \citep[fourth-order Hermite with block time steps;][]{1992PASJ...44..141M} to evolve the orbits of neighboring particles. The particle--tree method has a cost of $O(N \log{N})$ for force calculation and performs velocity kicks to stars with a shared timestep using the second-order leapfrog scheme. The SDAR method can efficiently and accurately evolve multiple systems (e.g., close encounters, binaries, triples, and quadruples).  \textsc{petar} is parallelized by using the Framework for Developing Particle Simulators \citep[FDPS;][]{Iwasawa2016FDPS}.
With \textsc{petar}, we can integrate star clusters with more than $10^5$ particles including multiple systems in a sufficiently short computational time.

In \textsc{ASURA+BRIDGE}, we use a star formation model similar to the one often used in galaxy simulations, but we apply the formation of individual stars instead of cluster particles.
Star particles are stochastically created from gas particles in a converging flow within a dense ($>5~\times~10^5$ cm$^{-3}$) and cold ($<$ 20\,K) region. We adopt a star formation efficiency per free-fall time of 0.02 following the observations of giant molecular clouds \citep[and references therein]{2019ARA&A..57..227K}. Masses of star particles from 0.1 to 150 $M_{\odot}$ are randomly assigned following the initial mass function (IMF) of \citet{2001MNRAS.322..231K} by using the chemical evolution library \citep[CELib,][]{2017AJ....153...85S}.
More details on our star formation scheme are described in \citet{2020arXiv200512906H}. Similar probabilistic star formation schemes are adopted in some other works \citep{2013MNRAS.435.1701C,2020MNRAS.499..668G} using adaptive mesh refinement (AMR) codes.

\subsection{\HII region model}
Instead of solving radiation transfer, we model the formation and evolution of \HII regions around massive stars.
When ionizing photons are emitted from the central source (i.e., massive star) at a rate $Q$, the radius to the ionization front is given by the Str{\"o}mgren radius \citep{1939ApJ....89..526S}:
\begin{eqnarray}
R_{\mathrm{St}}=\left(\frac{3 Q}{4 \pi \alpha_{\mathrm{B}} n_{\rm H}^{2}}\right)^{1 / 3},\label{eq:Stromgran_radius}
\end{eqnarray}
where $n_{\rm H}$ is the hydrogen number density of the surrounding medium
and $\alpha_{\rm B}$ is the recombination coefficient: $\alpha_{\rm B}=2.6\times 10^{-13}\,{\rm cm}^{3}\,{\rm s}^{-1}$, following \citet{2006agna.book.....O}. We ignore the temperature dependence of $\alpha_{\rm B}$. This radius is the boundary of the \HII region.

In \textsc{ASURA}, the \HII region feedback is modeled as follows \citep[see also ][]{2017MNRAS.464..246B, 2012MNRAS.421.3488H,2013MNRAS.436.1836R}. 
We estimate the Str{\"o}mgren radius using a binary search so that the gas in the radius can absorb all emitted photons. We take into account the inhomogeneity of the gas density in the radius.
Next, we assign thermal energy to gas particles within this radius by setting their temperature to $10^4$\,K.
This type of scheme has also been used in recent star-cluster formation simulations using an AMR code \citep{2013MNRAS.435.1701C,2020MNRAS.499..668G}. In SPH simulations of galaxies, \citet{2017MNRAS.464..246B} used the same method implemented to \textsc{ASURA}, but the ionizing photon counts per unit time ($Q$) was calculated for a bunch of stars assuming the simple stellar population (SSP) approximation. We instead use photon counts from individual massive stars, which are calculated as follows.

To obtain $Q$ for each star, we use OSTAR2002 data \citep{2003ApJS..146..417L}, which gives a grid data of surface gravity, effective temperature, chemical compositions, and ionizing photon flux of massive stars. From the table in \citet{2003ApJS..146..417L}, we take the values of the ionizing photon flux \citep[$q_0$ in ][]{2003ApJS..146..417L} at each effective temperature ($T_{\rm eff}$). We calculate the $T_{\rm eff}$ and radius of each stellar mass using the single stellar evolution code, \textsc{SSE} \citep{2000MNRAS.315..543H} in \textsc{amuse}. Here, we assume the solar metallicity and zero-age main sequence for the stellar radii and ignore the stellar evolution. Using these, we obtain the relation between stellar mass and $Q$.

\begin{figure}
 \begin{center}
  \includegraphics[width=7.8cm]{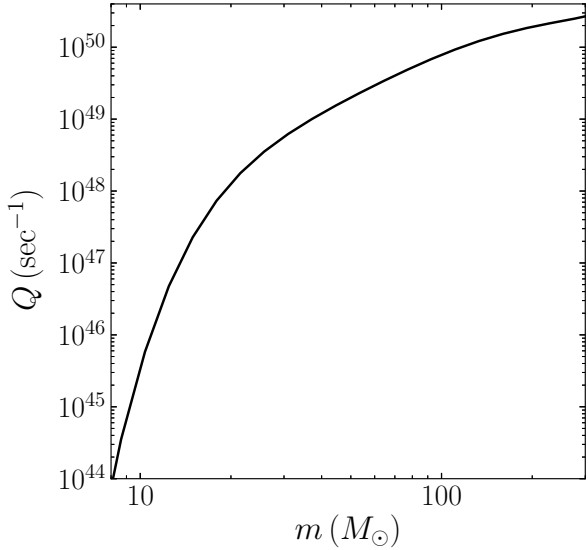}
 \end{center}
\caption{Ionizing photons per unit time as a function of stellar mass (OSTAR2002 model).}\label{fig:Q_mass}
\end{figure}

To use this relation in the simulations, we fit the data points with a polynomial function:
\begin{eqnarray}
\log (Q) = a + bx + cx^2 + dx^3 + ex^4 + fx^5,
\label{eq:Q}
\end{eqnarray}
where $x\equiv \log(m)$; $a$, $b$, $c$, $d$, $e$, and $f$ are coefficients.
The fitted coefficients are listed in table \ref{tb:Coefficients}.
The function we use is shown in Fig.~\ref{fig:Q_mass}, where masses from $5\,M_{\odot}$ to $300\,M_{\odot}$ are covered. If the stellar mass is over $300\,M_{\odot}$, we use the value at $300\,M_{\odot}$.

\begin{table}
\begin{center}
  \tbl{Coefficients of the polynomial fitting of  $\dot{\mathcal{N}}_{\rm LyC}$}{%
  \begin{tabular}{cccccc}
      \hline
      $a$  & $b$ & $c$ & $d$ & $e$ & $f$ \\ 
      \hline
-39.3178 & 221.997 & -227.456 & 117.410 & -30.1511 & 3.06810 \\
      \hline
    \end{tabular}}\label{tb:Coefficients}
\end{center}
\end{table}

\subsection{Radiation pressure}
In \HII regions, radiation pressure pushes the gas outward. This effect can be modeled as a velocity kick toward the radial direction from massive stars. 
\citet{2013MNRAS.436.1836R} treated this momentum feedback as radial velocity kicks, $\Delta v$, over the time interval, $\Delta t$: 
\begin{eqnarray}
\Delta v=s \frac{Q h\nu}{M_{\mathrm{H}_{\mathrm{II}}} c} \Delta t,
\label{eq:kick}
\end{eqnarray}
where $Q$ is the ionizing photon emission from the star, $h$ is the Planck constant, $c$ is the speed of light, $s$ is a constant parameter, $M_{\rm H_{II}}$ is the gas mass in the \HII region, and $\nu$ is the frequency of the flux that represents the most energetic part in the spectrum of the source. 
\citet{2013MNRAS.436.1836R} adopted the luminosity of Lyman $\alpha$ photons and set $\nu = 2.45\times10^{15}$\,s$^{-1}$ ($h\nu \sim 10$\,eV); $s=2.5$. In their model, the luminosity was calculated as a total luminosity from a population of stars. 
The factor $s$ can be changed depending on the feedback other than the radiation. We will discuss this parameter in Section \ref{sec:other}.

Following \citet{2016ApJ...819..137K}, we use the mean photon energy above the Lyman limit, $\langle h\nu\rangle_{\rm i}$, with a value of $18.6$\,eV. 
We therefore assume the velocity kick due to the radiation pressure to be  
\begin{eqnarray}
\Delta v_{\rm i}=\frac{Q \langle h\nu\rangle_{\rm i}}{M_{\mathrm{H}_{\mathrm{II}}} c} \Delta t.
\label{eq:kick_ion}
\end{eqnarray}

\begin{figure}
 \begin{center}
  \includegraphics[width=5cm]{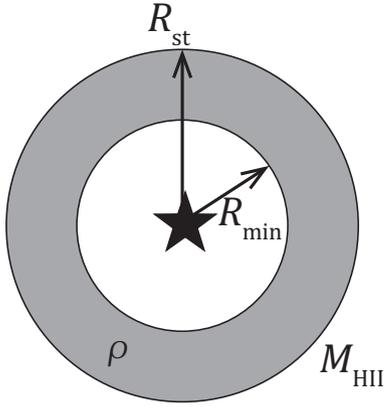}
 \end{center}
\caption{Schematic of our \HII region model.}\label{fig:HII}
\end{figure}

Following the analytic model of \citet{2011ApJ...732..100D} and \citet{2016ApJ...819..137K}, we also model a dusty \HII region, in which dust absorbs photons with $h\nu>13.3$\,eV, and the dust-reprocessed radiation pressure also contributes to the expansion of the \HII region. Through this effect, a central hole can be created as shown in the schematic of the \HII region in Fig.~\ref{fig:HII}. We includes this process in the form of a velocity kick depending on the distance from the central star ($r$) as follows.

In \citet{2011ApJ...732..100D} and \citet{2016ApJ...819..137K}, the balance between the radiation and thermal pressures was considered. They calculated the luminosity as a function of $r$:
\begin{eqnarray}
L=L_{\rm i}\phi(r)+L_{\rm n}e^{-\tau_{\rm d}(r)},\label{eq:luminosity}
\end{eqnarray}
where $L_{\rm i}$ and $L_{\rm n}$ are the luminosity from the hydrogen ionizing and nonionizing photons, respectively, and $\phi(r)$ is a dimensionless quantity for the attenuation of ionizing photons with a range from $0$ to $1$. The optical depth of the nonionizing photons is described as
\begin{eqnarray}
\tau_{\mathrm{d}}(r)=\int_{0}^{r} n \sigma_{\mathrm{d}} d r,    
\label{eq:tau}
\end{eqnarray}
where $\sigma_{\rm d}$ is the absorption cross section per hydrogen nucleus. We adopt a constant value of $\sigma_{\rm d}=1.0\times 10^{21}$\,cm$^{-2}$\,H$^{-1}$, following \cite{2016ApJ...819..137K}. 
Using the relation given in equation (\ref{eq:luminosity}) and $Q=L_{\rm i}\langle h\nu \rangle_{\rm i}$, we replace $Q$ in equation (\ref{eq:kick_ion}) to $Q[\phi + \beta e^{\tau_{\rm d}(r)}]$, where we assume a fixed ratio of luminosity $\beta\equiv L_{\rm n}/L_{\rm i}=1.5$, following \citet{2016ApJ...819..137K}. We also set $\phi$ to be a constant value of $1$ to reduce the calculation cost. Next, it is necessary to estimate $\tau_{\rm d}(r)$ for each massive star.

In our simulations, we identify the innermost gas particle of the massive star (the source of the feedback) and define its distance to the star as $R_{\rm min}$ (see Fig.~\ref{fig:HII}). Here, we assume that the density in the \HII region is uniform. Using the Str{\"o}mgren radius (equation~\ref{eq:Stromgran_radius}), we can estimate the density in the \HII region as
\begin{eqnarray}
\rho = \frac{3M_{\rm \HII}}{4\pi (R_{\rm St}^3-R_{\rm min}^3)},
\end{eqnarray}
where $M_{\rm \HII}$ is the total gas mass within the \HII region.
Using $\rho = nm_{\rm H}$ ($m_{\rm H}$ is hydrogen mass) and equation~(\ref{eq:tau}), we can calculate $\tau_{\mathrm{d}}(r)$. Similar to \citet{2016ApJ...819..137K}, we assume that $\tau$ saturates at a finite value of $\tau_{\rm d, max}=1.97$.

In summary, we apply velocity kicks to gas particles within the Str{\"o}mgren radius depending on the distance from the central star.
The kick velocity is given as:  
\begin{eqnarray}
\Delta v_{\rm dusty}(r)=[\phi+\beta e^{-\tau_{\rm d}(r)}]\frac{Q\langle h\nu\rangle_{\rm i}}{M_{\mathrm{H}_{\mathrm{II}}} c} \Delta t.\label{eq:kick_dusty}
\end{eqnarray}
The timestep for integration, $\Delta t$, is typically less than 1000\,yr in our simulations.

\begin{figure*}
 \begin{center}
  \includegraphics[width=14cm]{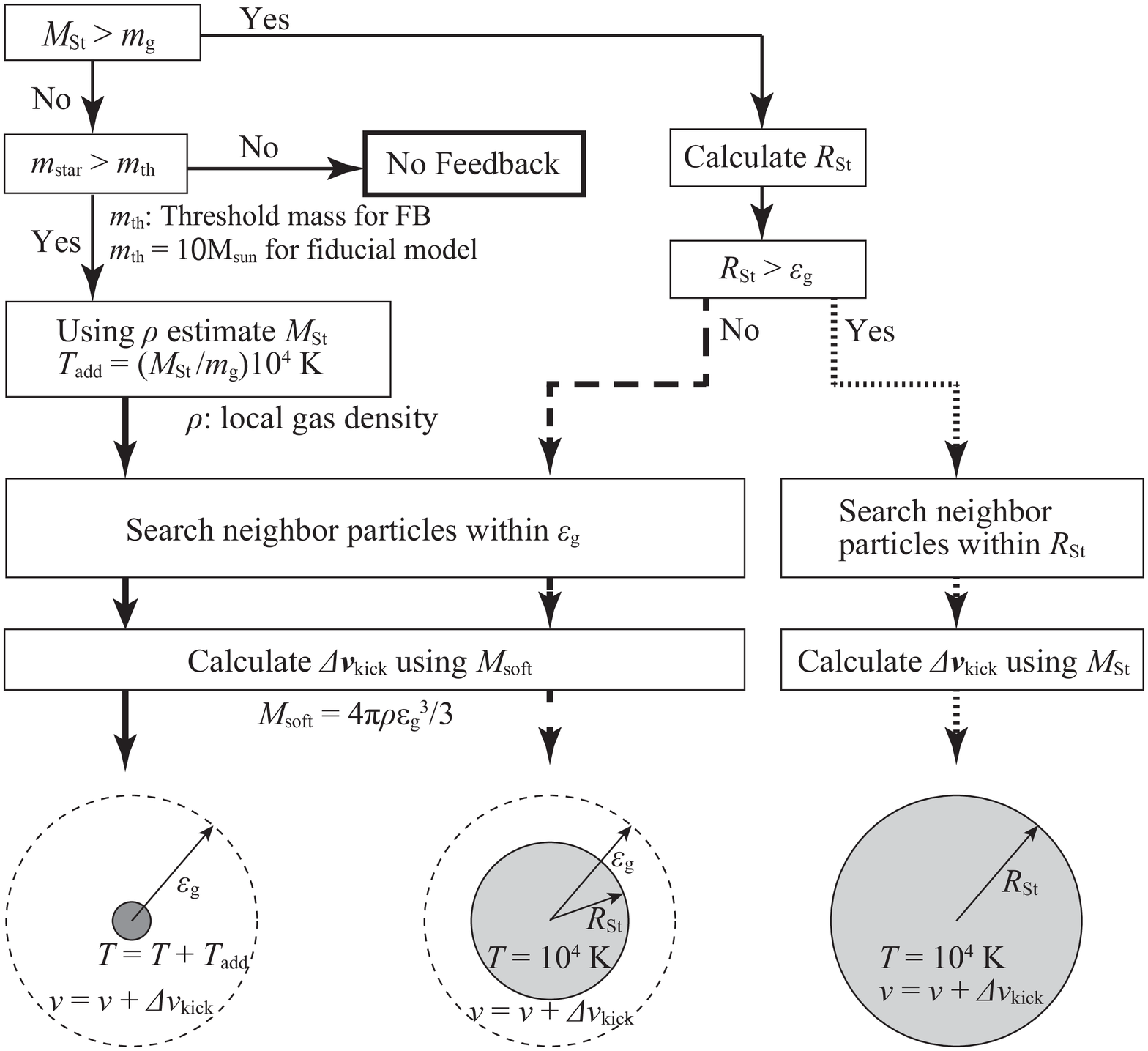}
 \end{center}
\caption{Summary of our feedback model.}\label{fig:FB}
\end{figure*}

\subsection{Other mechanical feedback}\label{sec:other}
In the momentum feedback model of \citet{2013MNRAS.436.1836R}, they assumed $s=2.5$ following \citet{2013MNRAS.432..455D}. This value includes the mechanical feedback due to radiation pressure ($V_{\rm rad}$), protostellar outflow ($V_{\rm ps}$), stellar wind ($V_{\rm ms}$), and supernova ($V_{\rm SN}$). They estimated each velocity as $V_{\rm rad}=190$, $V_{\rm ps}=40$, $V_{\rm SN}=50$, and $V_{\rm ms}=200$\,km\,s$^{-1}$. If we scale the summation by $V_{\rm rad}$, $s=(190+40+50+200)/190\simeq2.5$. In our model, we consider the feedback from protostellar outflow and stellar wind. The corresponding $s=240/190\simeq 1.26$. Because $\langle h\nu\rangle_{\rm i}$ in our model is 1.86 times larger than $h\nu$ in \citet{2013MNRAS.436.1836R}. We therefore define the momentum feedback from protostellar outflow and stellar wind as 
\begin{eqnarray}
\Delta v_{\rm sw}=s'\frac{Q\langle h\nu\rangle_{\rm i}}{M_{\mathrm{H}_{\mathrm{II}}} c} \Delta t,\label{eq:kick_other}
\end{eqnarray}
where $s'\equiv s/1.87=0.673$. We add this velocity kick to gas particles within the Str{\"o}mgren radius in addition to $\Delta v_{\rm dusty}$ irrespective of $r$.
In this study, we ignore the mass loss of massive stars due to stellar wind.

\subsection{Treatment of sub-softening scale}
When the molecular gas is too dense, we sometimes encounter a situation in which the Str{\"o}mgren radius is too small to include at least one gas particle. If we did not add any energy to gas particles in such a case, we would underestimate the strength of the feedback. To solve this issue, we add the momentum feedback to gas particles within the gas softening length using equation~(\ref{eq:kick_other}). Instead of $M_{\rm \HII}$, we adopt the total gas mass within the gas softening length $M_{\rm soft}$.

We also add thermal energy to the nearest gas particle by increasing the temperature of the gas particle, with an upper limit of $10^4$\,K. The temperature is determined to be proportional to the gas mass inside the Str{\"o}mgren radius. 
We also set the minimum mass to give the momentum feedback to prevent a small gas mass from obtaining too strong a kick. The typical minimum mass is set to be the initial gas particle mass. 
In Fig.~\ref{fig:FB}, we summarize the procedures of our scheme. 

We also treat star formation as a sub-softening scale physics. When a gas particle satisfies the star formation condition (threshold density, temperature, and converging flow ), a star particle is created. The mass of the forming star is chosen from a given initial mass function. In our code, however, we do not allow stars whose masses exceeds the local gas mass to form. We set a parameter, $r_{\rm max}$, which defines a radial region where gas particles can contribute to the star formation. The mass of the forming star is limited to the half of the gas mass within $r_{\rm max}$. This assumption naturally relates the mass of the most massive star and the total stellar mass of the host star cluster. The value of $r_{\rm max}$ depends on the resolution of the simulation and must be large enough to form massive stars \citep[see Paper I ][ for more details]{2020arXiv200512906H}. A typical value is in the range of 0.1--0.5\,pc for our targeting resolution. This corresponds to a gas-particle mass of 0.01--0.1\,$M_{\odot}$.

\subsection{Initial conditions}\label{model3}

Using \textsc{ASURA+BRIDGE} including our feedback model, we perform a series of $N$-body simulations of star-cluster formation starting from turbulent molecular clouds. 
We adopt a model similar to that of \citet{2018ApJ...859...68K} as initial conditions, which are homogeneous spherical molecular clouds with turbulent velocity field $k=-4$.
In \citet{2018ApJ...859...68K}, RHD simulations were performed with a sink method and their dusty \HII region model. Although we do not solve radiation transfer, but we adopt an \HII region model based on their model (see Appendix 2). We therefore compare our results with theirs. In their simulation, this model results in the formation of several small clusters, although each stellar particles represents a cluster.

We set up our model following their fiducial model (M1E5R20), with a total gas mass of $M_{\rm g}=10^5M_{\odot}$ and an initial radius of $r_{g}=20$\, pc. We set the initial virial ratio ($\alpha_{\rm vir}=|E_{\rm k}|/|E_{\rm p}|$) to be 1, where $E_{\rm k}$ and $E_{\rm p}$ are kinetic and potential energy, respectively. Thus, this model is initially super-virial. We adopt an initial temperature of 20\,K and a gas-particle mass ($m_{\rm g}$) of $0.1M_{\odot}$.

We take the same parameters as those in model Low of Paper II. The gas softening, $\epsilon_{\rm g}= 14,000$\,au (0.07\,pc). The star formation threshold density, $n_{\rm th} = 5\times10^5$\,cm$^{-3}$. The temperature is 20\,K. The star formation efficiency, $c_{*} = 0.02$. We do not use softening for stars, i.e., $\epsilon_{\rm s}$\,=\,0.0. These parameters are summarized in table~\ref{tb:IC_F}. Three models, K18R20noFB (without feedback from massive stars), K18R20HII (with \HII feedback), and K18R20HIIsw (with \HII feedback and velocity feedback due to stellar wind), are compared. 
To test the dependence on the mass resolution, we run a model (K18R20HIIm001) with a high mass resolution, $m_{\rm g}=0.01 M_{\odot}$, in which the other parameters are the same as those in model K18R20HII.

We also perform the same simulations as for models K18R20HII and K18R20HIIm001 but with softening for stars for comparison. These models (K18R20HII-soft and K18R20HIIm001-soft in table~\ref{tb:IC_F}) apply a softening length equal to that of gas particles. 

To understand the effects of the resolution, we set up model K18R20HII-soft-L, in which a parameter set similar to model M1E5R20 in \citet{2016ApJ...819..137K} is used. In the model in \citet{2016ApJ...819..137K}, a spacial resolution (grid size) of 0.3125\,pc and threshold density for sink formation of $7.8\times 10^3$\,cm$^{-3}$ are adopted. 
For comparison, we use the gas softening length $\epsilon_{\rm g} = 70,000$\,au (0.27\,pc) and star formation density threshold, $n_{\rm th}=3\times 10^3$\,cm$^{-3}$. This set resolves the Jeans mass with 50 SPH particles. As the threshold density is lower, we set $r_{\rm mas} = 0.5$\,pc. For model K18R20HII-soft-L, the other parameters are the same as those of model K18R20HII-soft. 

We set $r_{\rm max}=0.2$\,pc for all models except model K18R20HII-soft-L, where $r_{\rm max}=0.5$\,pc. The minimum mass for feedback is set to be $8\,M_{\odot}$, The minimum value to give the \HII feedback is set to be 0.1\,$M_{\odot}$, but $0.03\,M_{\odot}$ for models with $m_{\rm g}=0.01\,M_{\odot}$ to avoid too large a feedback velocity due to the small included gas mass. If the Str\"{o}mgren radius reaches $30$\,pc, we switch off the \HII feedback from the massive star. Such a large \HII region appears when a massive star escapes from the system due to strong dynamical scattering in simulations without softening. When a massive star runs to the empty region out of the cloud, the estimated Str\"{o}mgren radius increases too much due to the excessively low density of gas. This is an artificial effect caused by assuming an isolated molecular cloud as an initial condition. To avoid such an artificial effect, we need to set a maximum Str\"{o}mgren radius similar to the size of the entire system.

In addition, we set up four models referring to \citet{2020MNRAS.497.3830F}. In these models, the initial virial ratio of gas is 0.25 (sub-virial) and the others are the same as those of \citet{2018ApJ...859...68K}. 
As the star-formation efficiency depends on the initial virial ratio of molecular clouds, this model is expected to form more stars. We note that \citet{2020MNRAS.497.3830F} used an AMR code with a sink method and radiation transfer model different from those used in \citet{2018ApJ...859...68K}. 
Model F20HIIsw-soft has softening for stars, model F20HIIsw has no softening, models F20HII has no stellar wind feedback, and model F20HIIswm001 has a high mass resolution (see table~\ref{tb:IC_F}).

\subsection{Integration}
Here, we summarize the parameters used for the integration with \textsc{ASURA+BRIDGE}. 
The stars are integrated with the sixth-order Hermite scheme  with individual timesteps or \textsc{PeTar} (see Section 2.1 and Paper II). We adopt a timestep for the Bridge scheme of $\Delta t=200$\,yr.  It is determined from the local dynamical (free-fall) time, mass resolution and softening length of the gas \citep[see ][ for more details]{2021arXiv210105934F}. In the presence of feedback, we also need to consider the timescale of the mechanical feedback. In our simulations, the feedback velocity is smaller than 100\,km\,s$^{-1}$ and our stepsize is short enough to resolve this motion.

We set the accuracy parameter for the Hermite individual timestep \citep{1992PASJ...44..141M,2008NewA...13..498N} to 0.2 and 0.01 for the sixth-order Hermite scheme and \textsc{PeTar}, respectively. 
As \textsc{PeTar} in \textsc{ASURA} cannot treat gravitational softening, we use the sixth-order Hermite scheme for all runs with gravitational softening for stars. For runs without softening, we first use the sixth-order Hermite scheme and switch to \textsc{PeTar} after the number of star particles becomes sufficiently large, as the P$^3$T scheme is based on a tree code, and therefore, the overhead of the tree code is too large for small number of particles.

\begin{table*}
  \tbl{Models and results for cluster-forming clouds}{%
  \begin{tabular}{lcccccccccccc}
      \hline
      Name  & $M_{\rm g}$  & $m_{\rm g}$ & $R_{\rm g}$& $n_{\rm ini}$ & $\alpha_{\rm vir}$& $t_{\rm ff,ini}$  & $\epsilon_{\rm s}$ & $\epsilon_{\rm g}$ & $n_{\rm th}$ & $r_{\rm max}$ & \multicolumn{2}{c}{Feedback} \\
       & $(M_{\odot})$ & $(M_{\odot})$ & (pc) &  (cm$^{-3}$) & (Myr) &  & (pc) & (pc) & (cm$^{-3}$) & (pc) & \HII & SW \\ 
      \hline
      K18R20noFB & $1\times 10^5$ & $0.1$ & 20 & $86$ & 1 & $4.7$  & 0.0 & $0.07$ & $5\times 10^5$ & 0.2 &  N & N \\
      K18R20HII & $1\times 10^5$ & $0.1$ & 20 & $86$ & 1 & $4.7$ & 0.0 & $0.07$ & $5\times 10^5$ & 0.2 & Y & N \\
      K18R20HIIm001 & $1\times 10^5$ & $0.01$ & 20 & $86$ & 1 & $4.7$  & 0.0 & $0.07$ & $5\times 10^5$ & 0.2 & Y & N \\
      K18R20HIIsw & $1\times 10^5$ & $0.1$ & 20 & $86$ & 1 & $4.7$  & 0.0 & $0.07$ & $5\times 10^5$ & 0.2 & Y & Y \\
      F20HII & $1\times 10^5$ & $0.1$ & 20 & $86$ & 0.25 & $4.7$  & 0.0 & $0.07$ & $5\times 10^5$ & 0.2 & Y & N \\
      F20HIIsw & $1\times 10^5$ & $0.1$ & 20 & $86$ & 0.25 & $4.7$  & 0.0 & $0.07$ & $5\times 10^5$ & 0.2 & Y & Y \\
      F20HIIswm001 & $1\times 10^5$ & $0.01$ & 20 & $86$ & 0.25 & $4.7$  & 0.0 & $0.07$ & $5\times 10^5$ & 0.2 & Y & Y \\
      \hline
      K18R20HII-soft & $1\times 10^5$ & $0.1$ & 20 & $86$ & 1 & $4.7$  & $0.07$ & $0.07$ & $5\times 10^5$ & 0.2 & Y & N \\
      K18R20HIIm001-soft & $1\times 10^5$ & $0.01$ & 20 & $86$ & 1 & $4.7$  & $0.07$ & $0.07$ & $5\times 10^5$ & 0.2 & Y & N \\
      K18R20HII-soft-L & $1\times 10^5$ & $0.1$ & 20 & $86$ & 1 & $4.7$ & $0.27$ & $0.27$ & $3\times 10^3$ & 0.5 & Y & N \\
      F20HIIsw-soft & $1\times 10^5$ & $0.1$ & 20 & $86$ & 0.25 & $4.7$  & $0.07$ & $0.07$ & $5\times 10^5$ & 0.2 & Y & Y \\
      \hline
    \end{tabular}}\label{tb:IC_F}
\begin{tabnote}
Here, `\HII' indicates \HII region model including radiation pressure model. `SW' indicates stellar wind model. `soft' indicates `with softening' for stars, and `L' indicates low resolutions of softening for stars and gas.
\end{tabnote}
\end{table*}

\section{Star cluster formation simulations with feedback}

\begin{figure*}
 \begin{center}
 \includegraphics[width=7.cm]{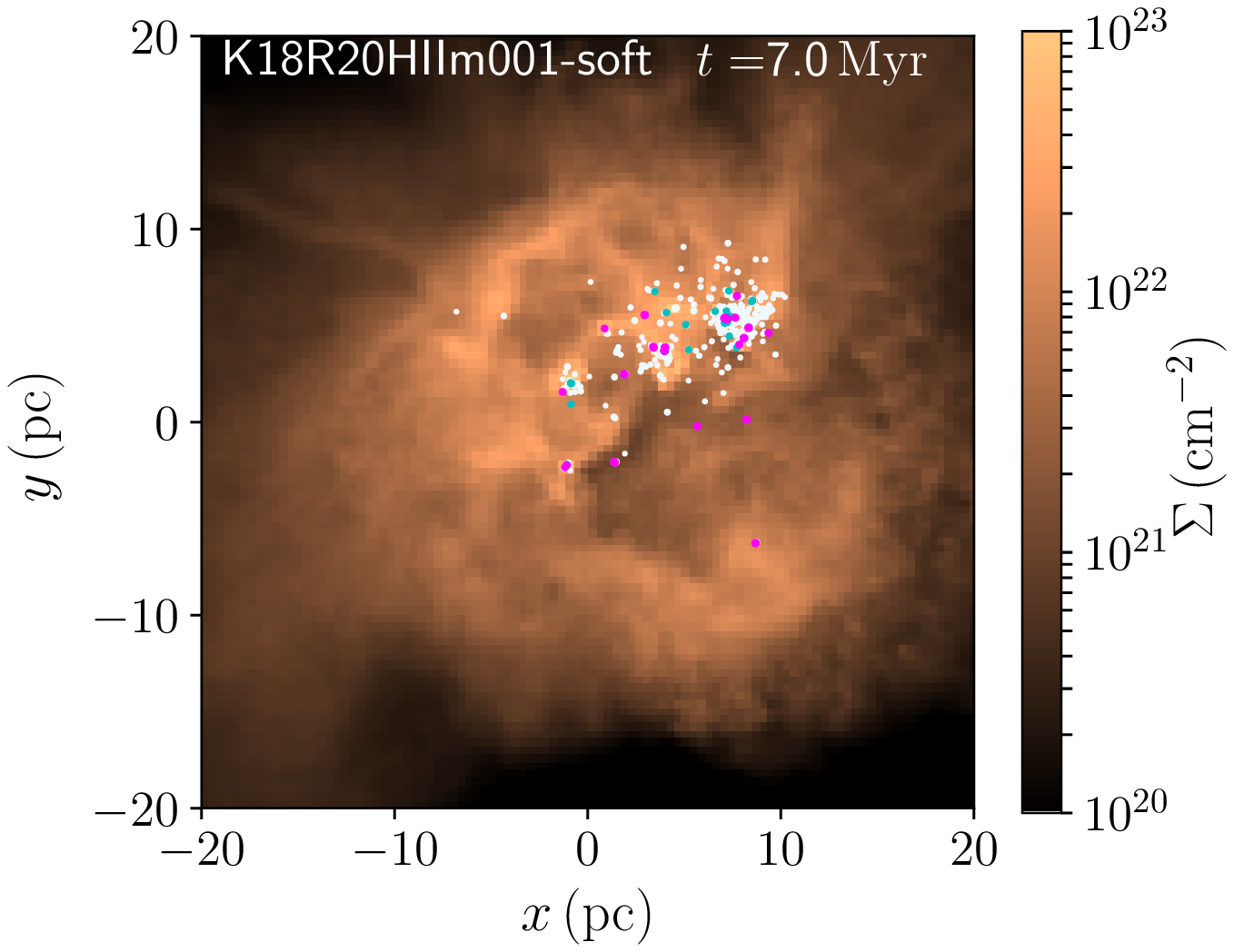}
  \includegraphics[width=7.cm]{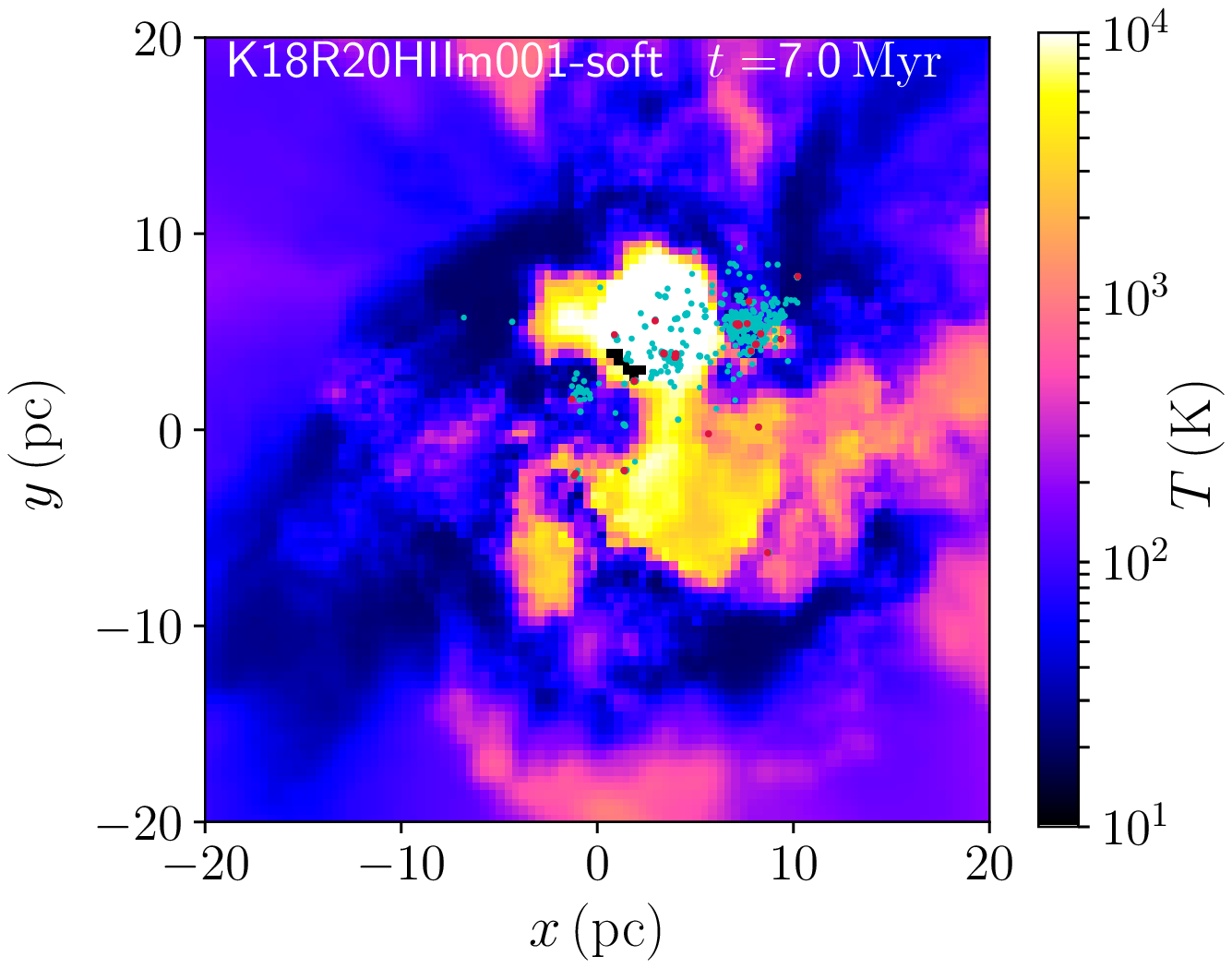}\\
 \includegraphics[width=7.cm]{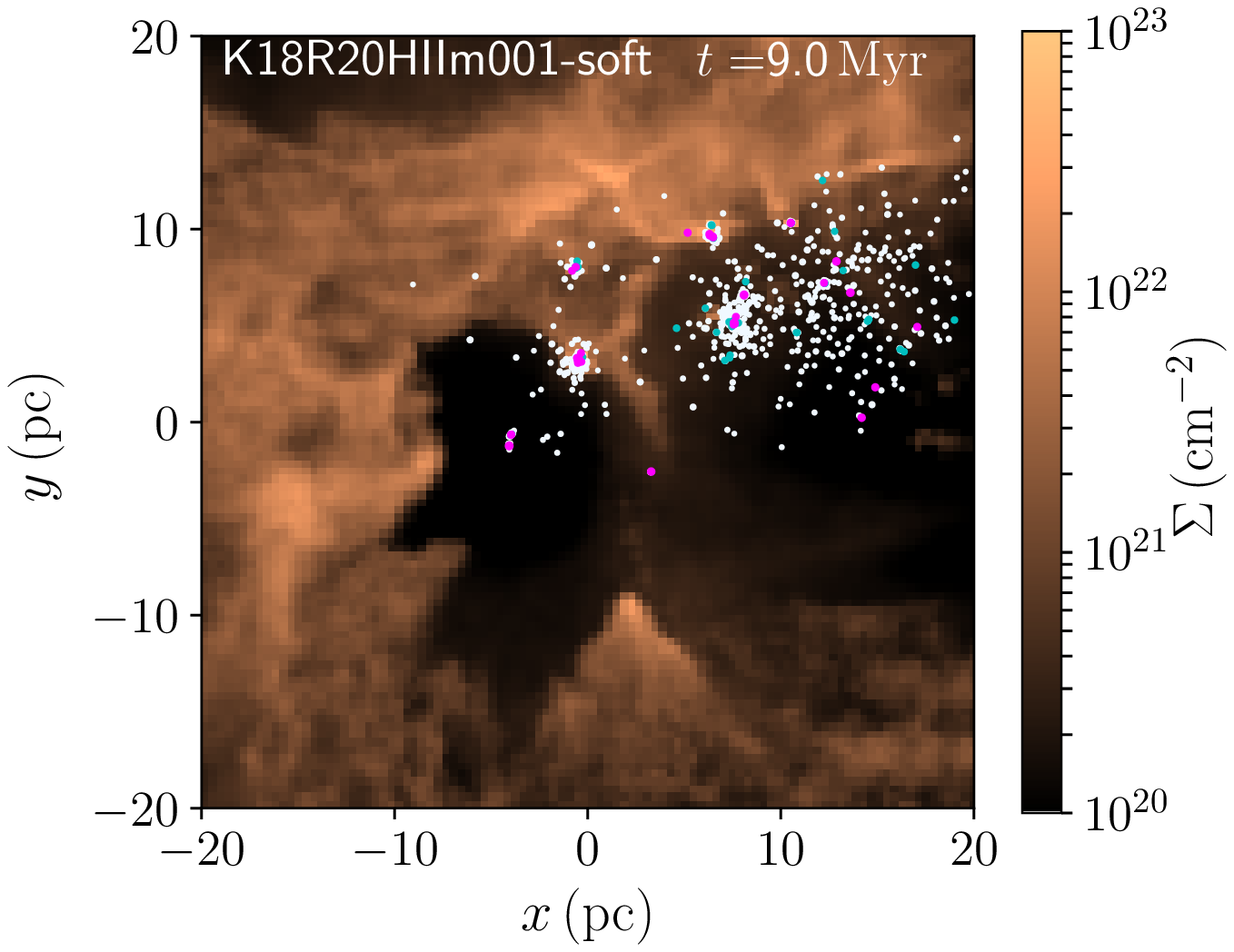}
  \includegraphics[width=7.cm]{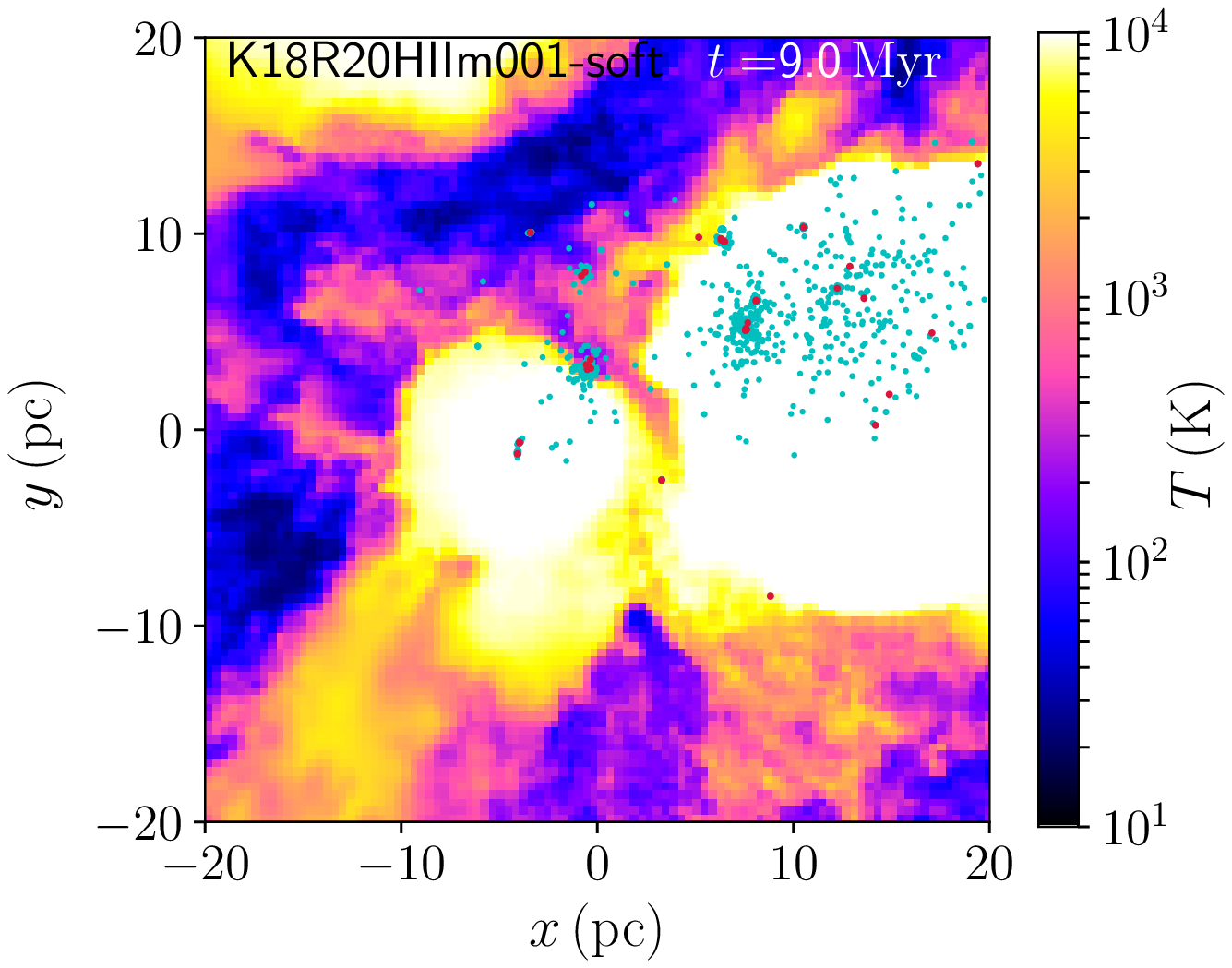}\\
 \includegraphics[width=7.cm]{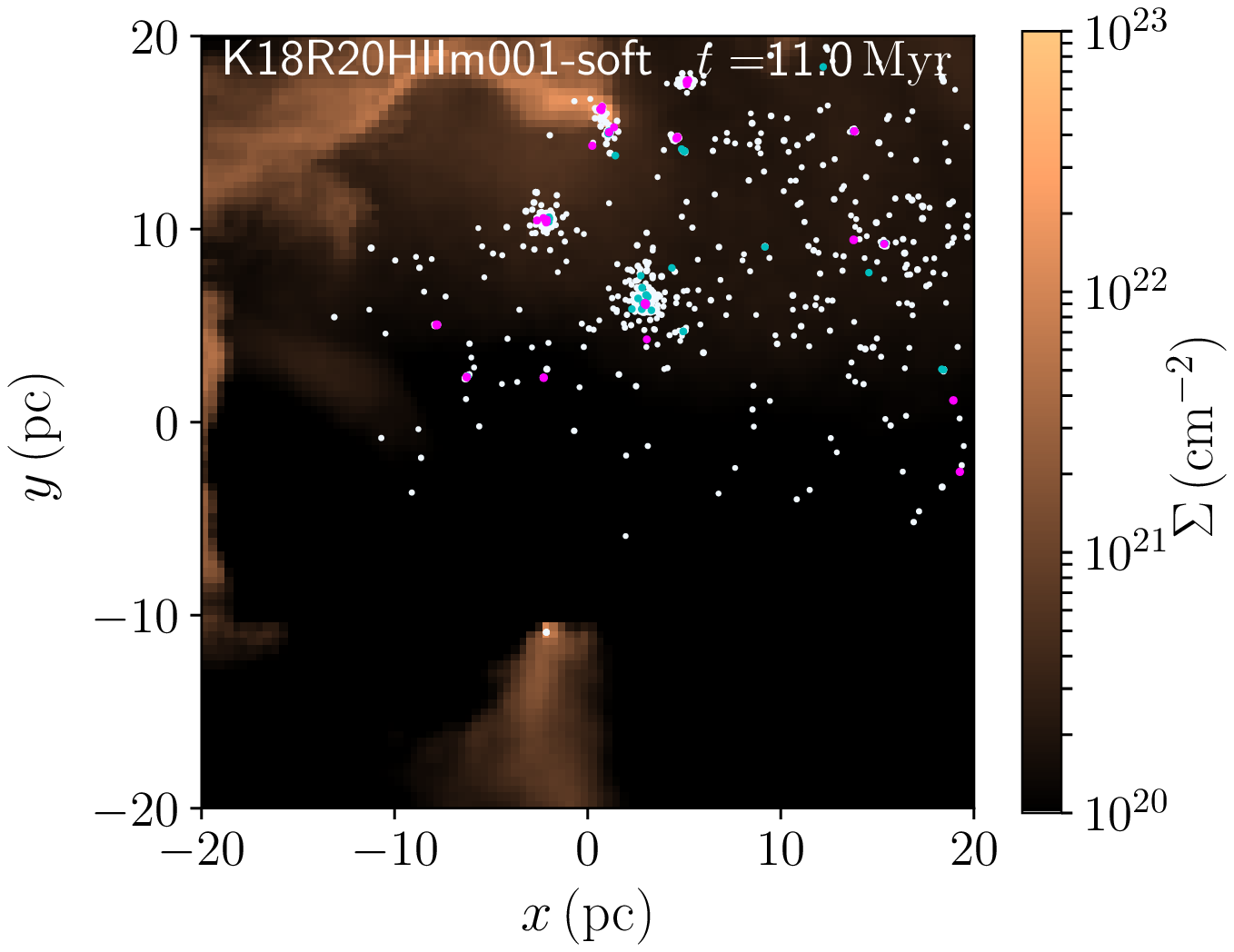}
  \includegraphics[width=7.cm]{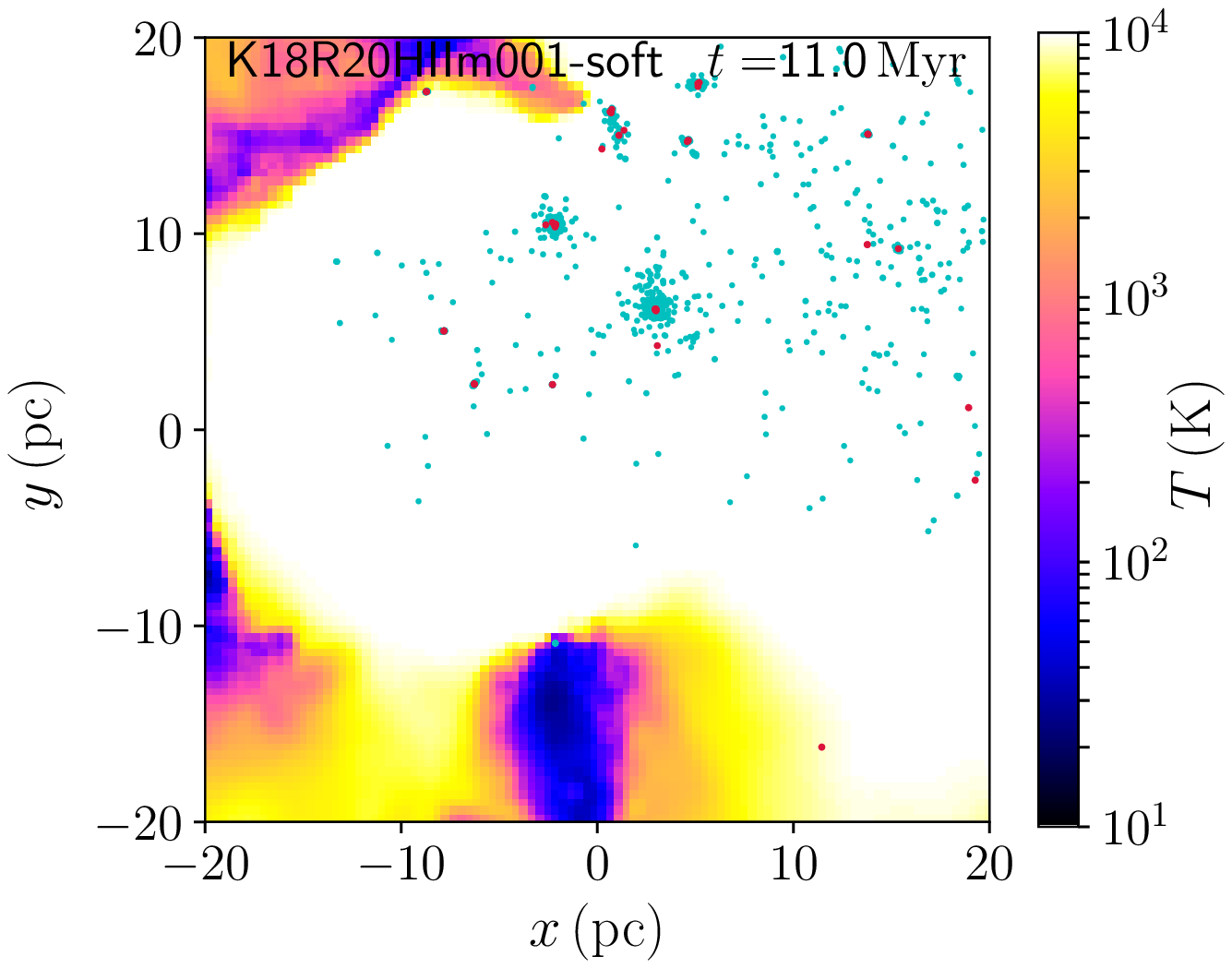}\\
 \end{center}
\caption{Time evolution of the distribution of gas and stars for models K18R20m001HII-soft. Left: white, blue, and magenta dots indicate stars with masses of 1--8, 8--20, and $>20 M_{\odot}$. Color scale indicates gas surface density. Right: blue and red dots indicate stars with masses of 1--20 and $>20 M_{\odot}$, respectively. Color scale indicates the gas temperature. }\label{fig:snapshot_Kim18m001-soft}
\end{figure*}

\begin{figure}
 \begin{center}
  \includegraphics[width=7.8cm]{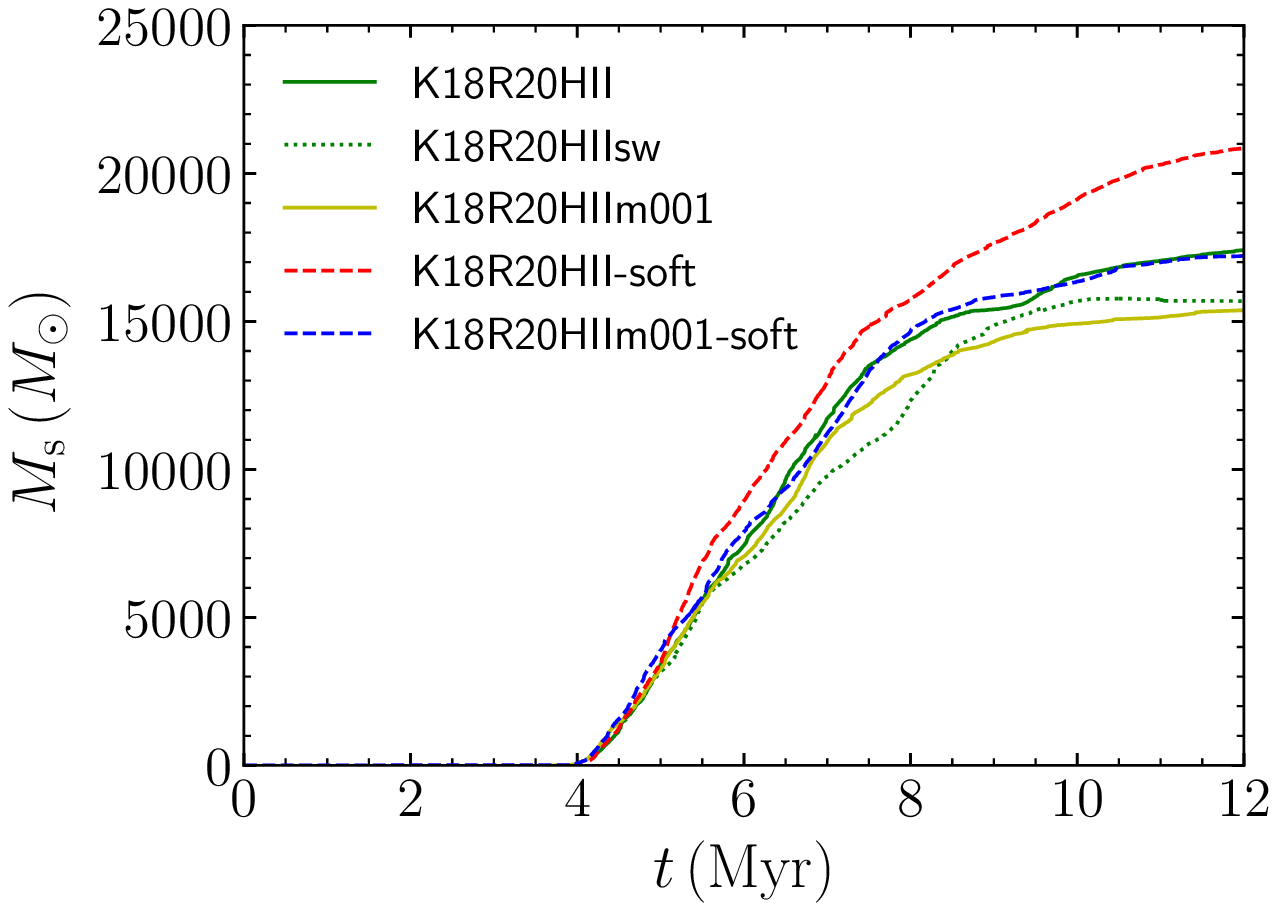}
 \end{center}
\caption{Stellar mass evolution of K18R20 models.}\label{fig:stellar_mass_ev_Kim18_r20_B}
\end{figure}

\begin{figure*}
 \begin{center}
 \includegraphics[width=7.cm]{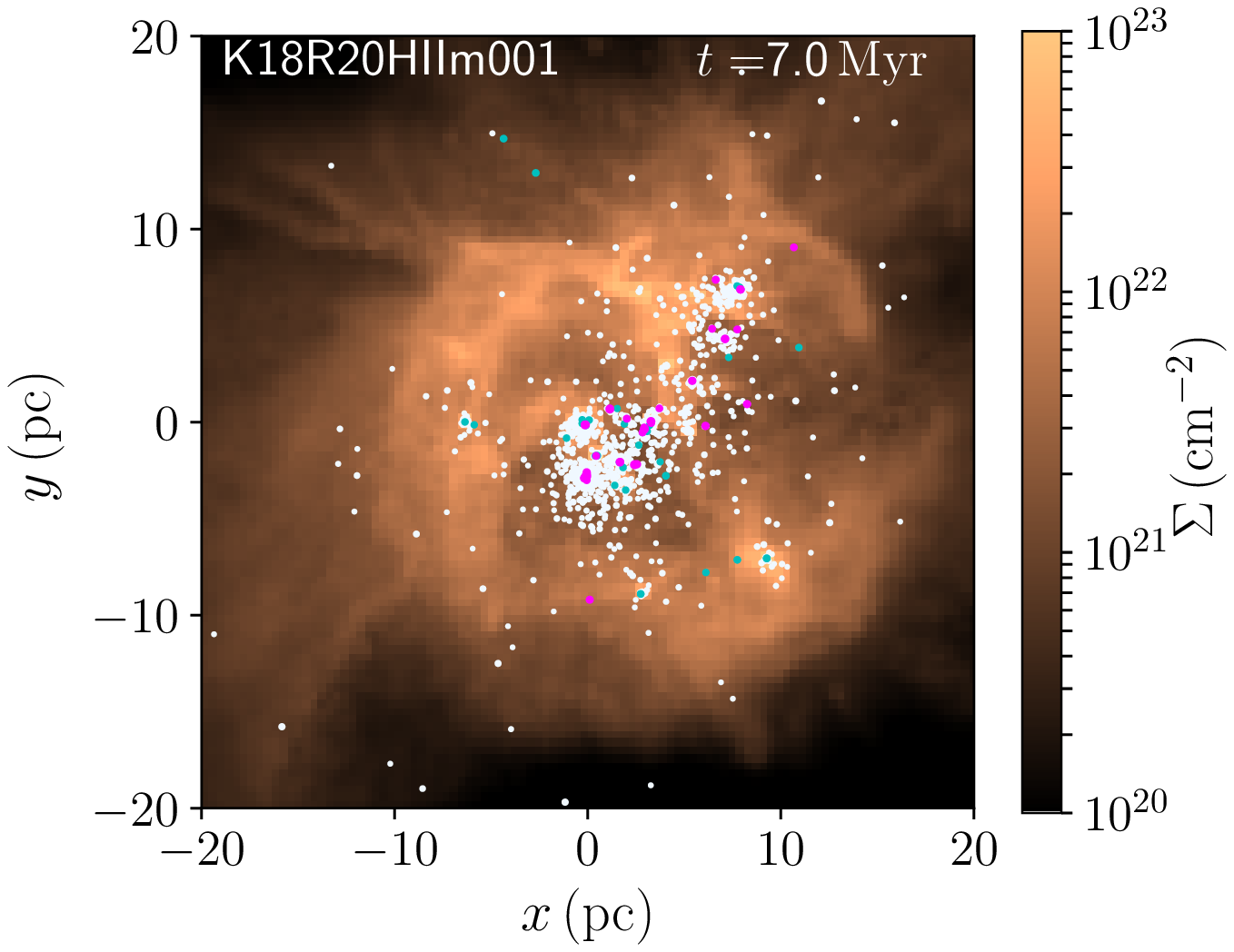}
  \includegraphics[width=7.cm]{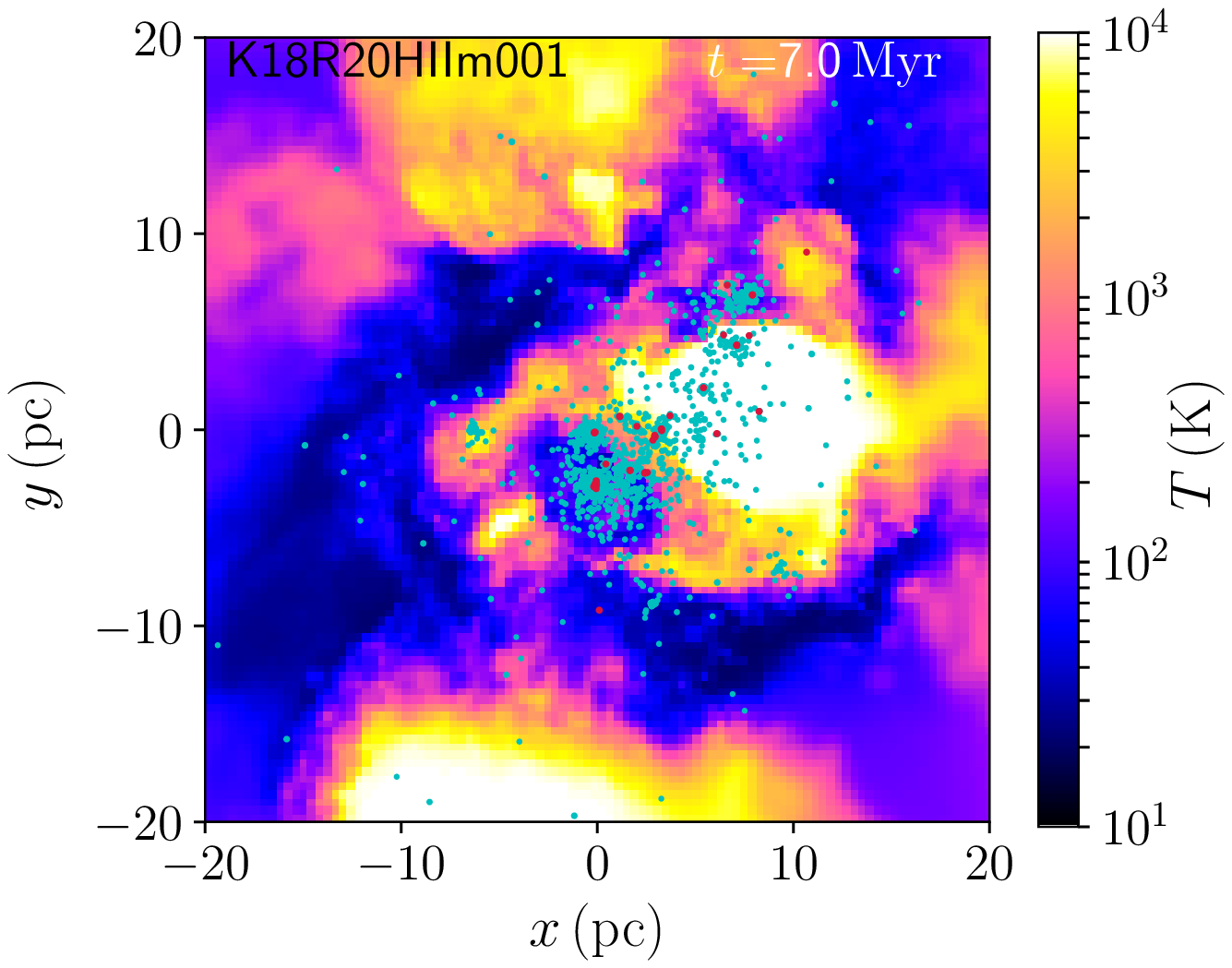}\\
 \includegraphics[width=7.cm]{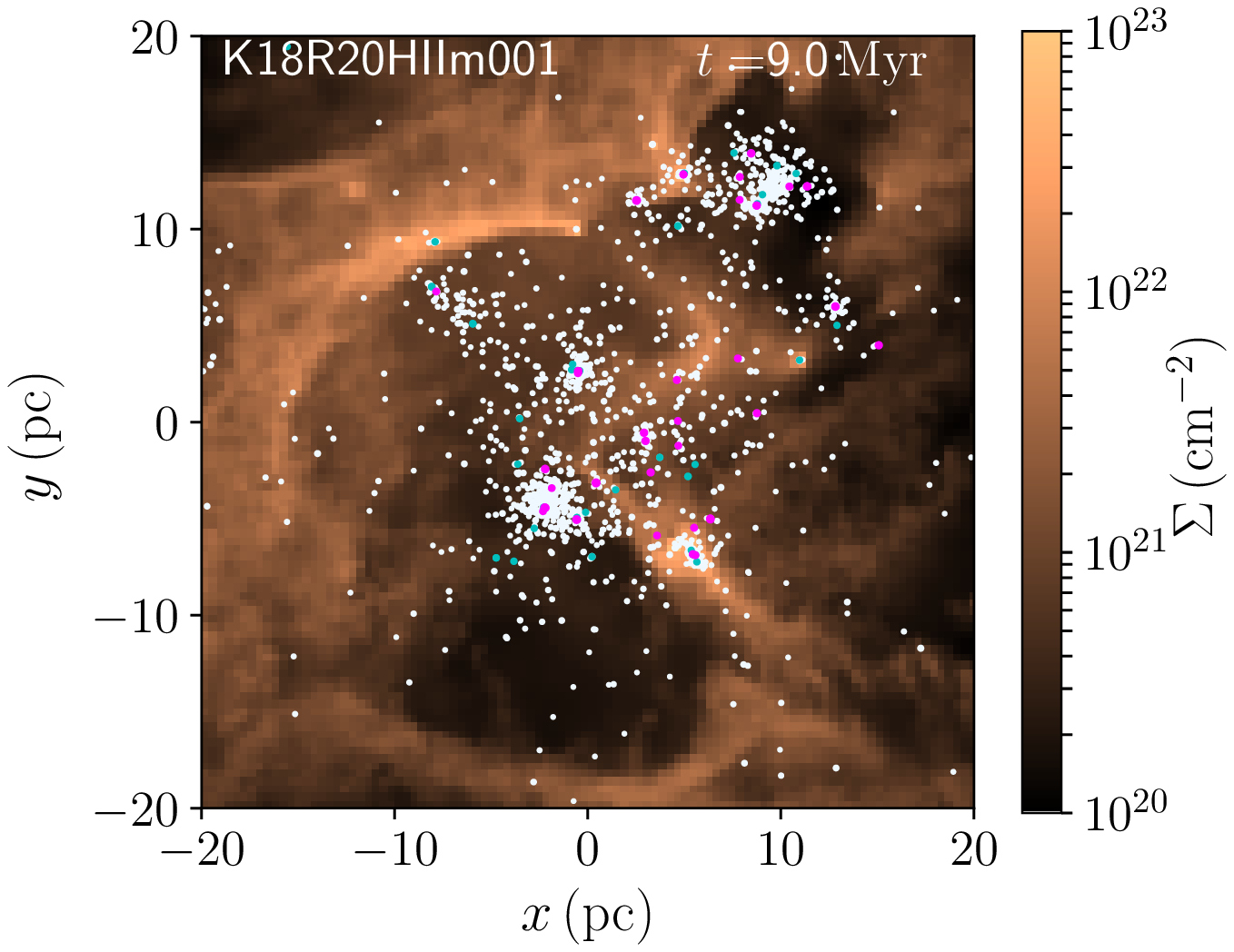}
  \includegraphics[width=7.cm]{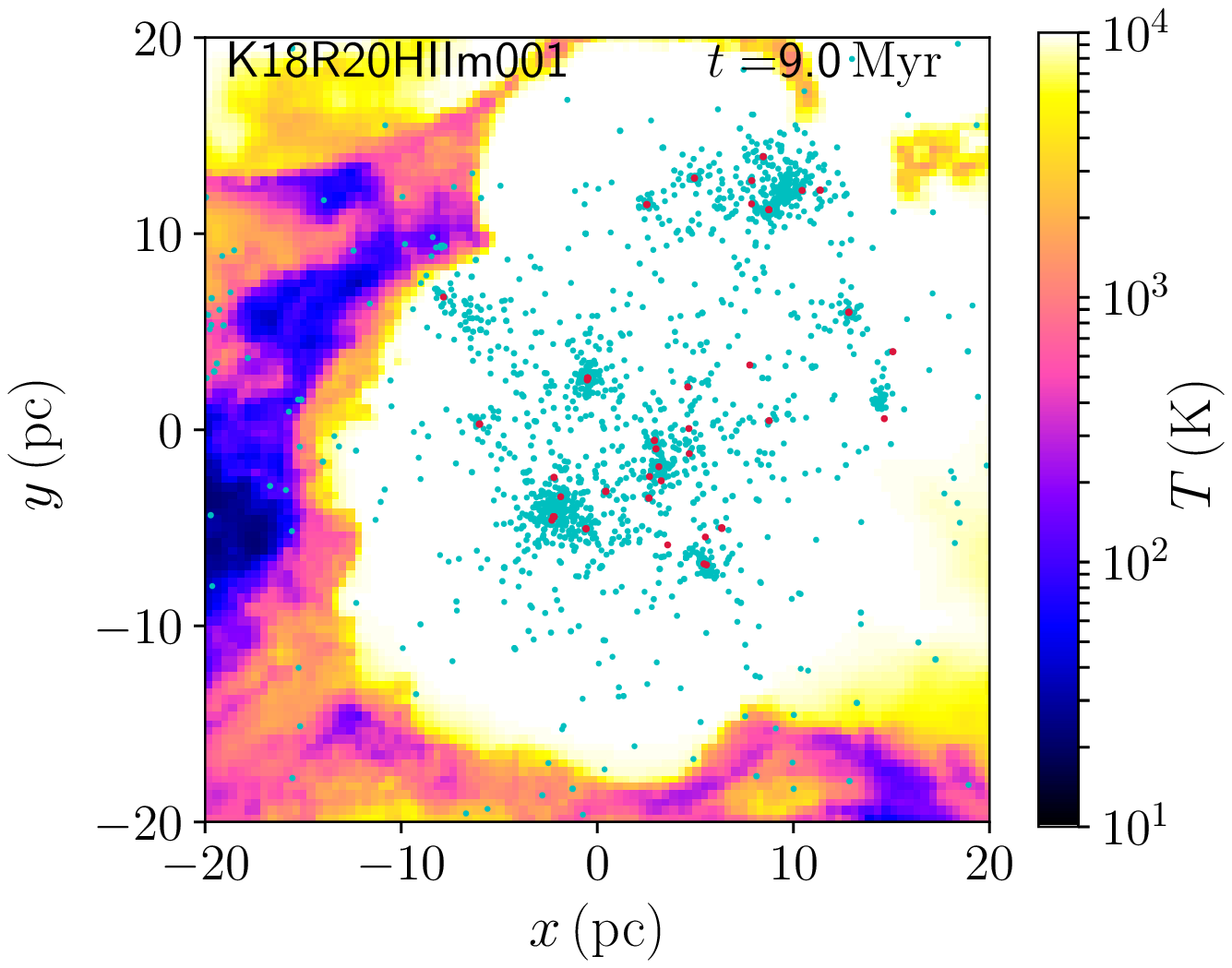}\\
 \includegraphics[width=7.cm]{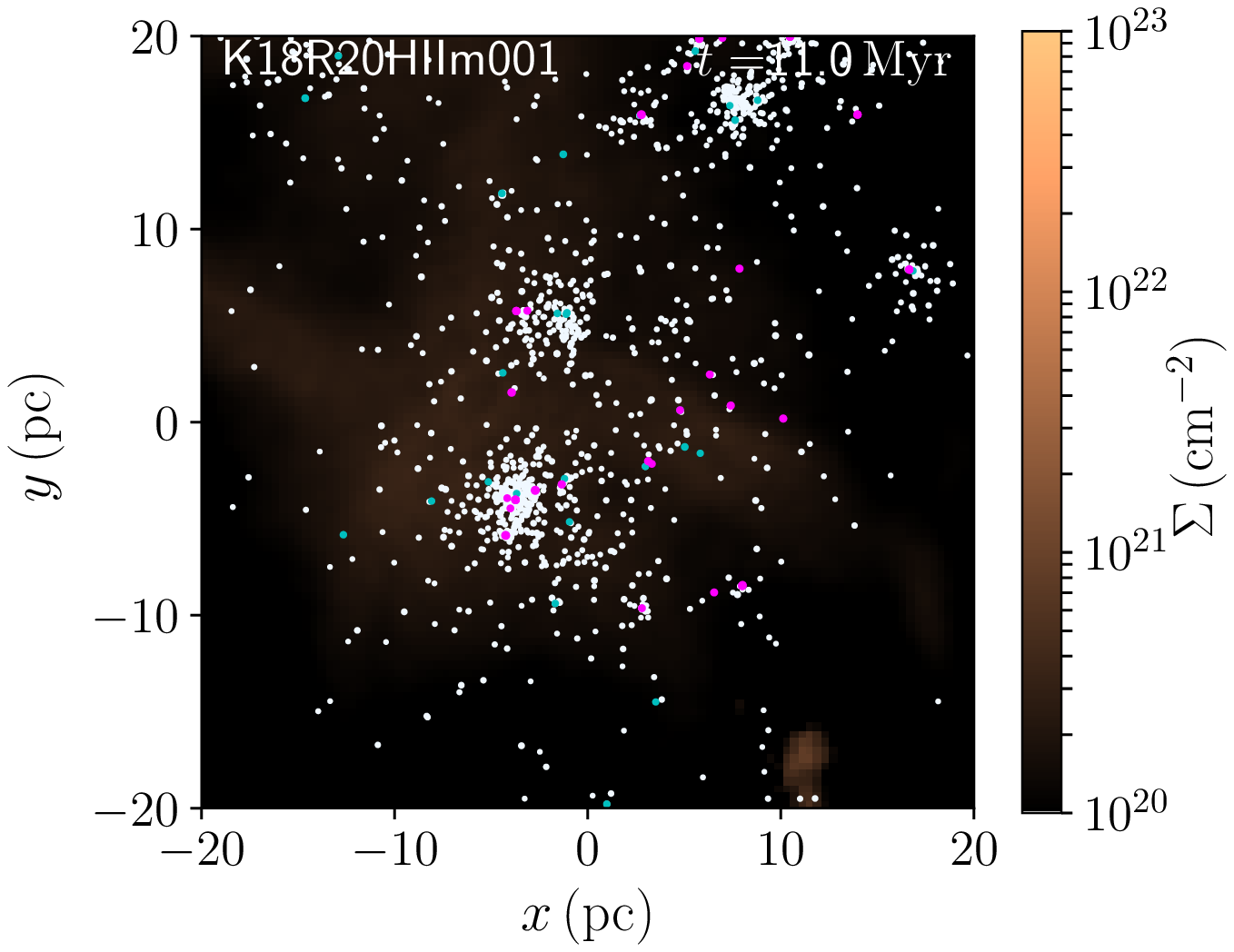}
  \includegraphics[width=7.cm]{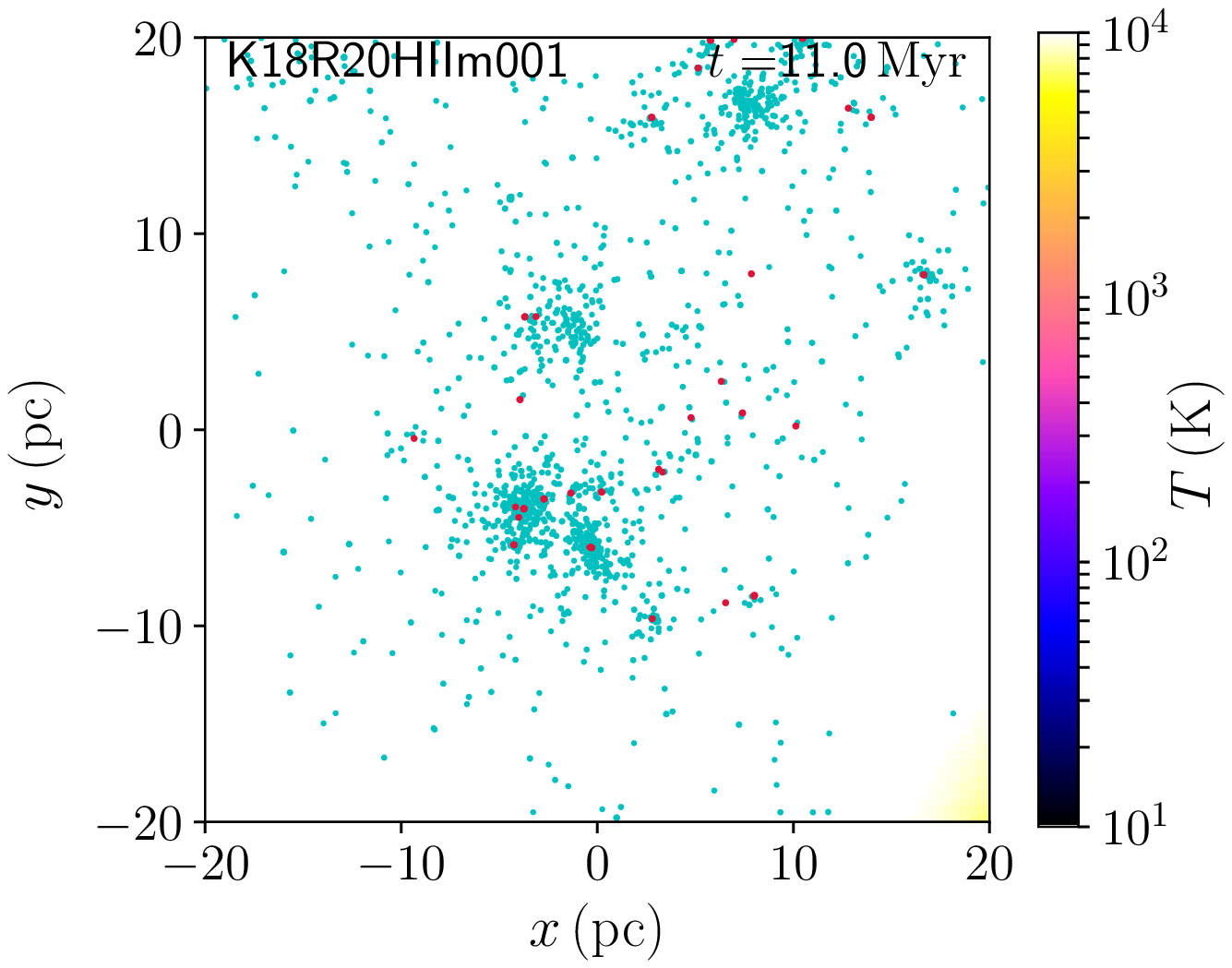}\\
 \end{center}
\caption{Same as Fig.~\ref{fig:snapshot_Kim18m001-soft} but for models K18R20m001HII. }\label{fig:snapshot_Kim18m001}
\end{figure*}

\subsection{Super-virial case}

In Fig.~\ref{fig:snapshot_Kim18m001-soft}, we present snapshots of models Kim18R20HII-soft. The time evolution of the total stellar mass formed in these simulations are shown in Fig.~\ref{fig:stellar_mass_ev_Kim18_r20_B}. The star formation starts at around the initial free-fall time ($\sim 5$\,Myr). Once massive stars form, stellar feedback is switched on. In the early phase, however, the feedback energy is insufficient to stop the star formation, and therefore, the star formation in the system continues. Star formation occurs not only in the center of the entire system but in several clumps. They finally evolve to several clusters.

Although stellar feedback does not immediately stop the star formation, it slows down it after the formation of a few thousands stars. 
At $\sim7$\,Myr (see Fig.~\ref{fig:snapshot_Kim18m001-soft}), the \HII regions expand to the outside of the cluster. At this moment, several clumps that includes some massive stars exist in the molecular cloud. At 9\,Myr, the gas inside the star-forming region is almost blown away. However, some clumps still continues star formation. Thus, the formation time of the formed clumps has a small dispersion. These individual clumps remain bound, but the separations between clumps increases (see the snapshots at 11\,Myr). 
We will discuss the details of the formed clumps in Section 3.4. 

In Fig.~\ref{fig:snapshot_Kim18m001}, we show the snapshot of model K18R20HIIm001, in which gravitational softening for stars is not adopted. In this simulation, \HII regions form out of the central star forming region (see snapshot at 7\,Myr). These \HII regions are formed by escaping massive stars, known as runaway/walkaway stars. In Fig.~\ref{fig:stellar_mass_ev_Kim18_r20_B}, we present the time evolution of the total stellar mass of this model. The star formation is slightly suppressed in the case without softening compared to that with softening. 
The difference is larger in the case with lower mass resolution (models K18R20HII and K18R20HII-soft). 

\begin{figure*}
 \begin{center}
  \includegraphics[width=7.8cm]{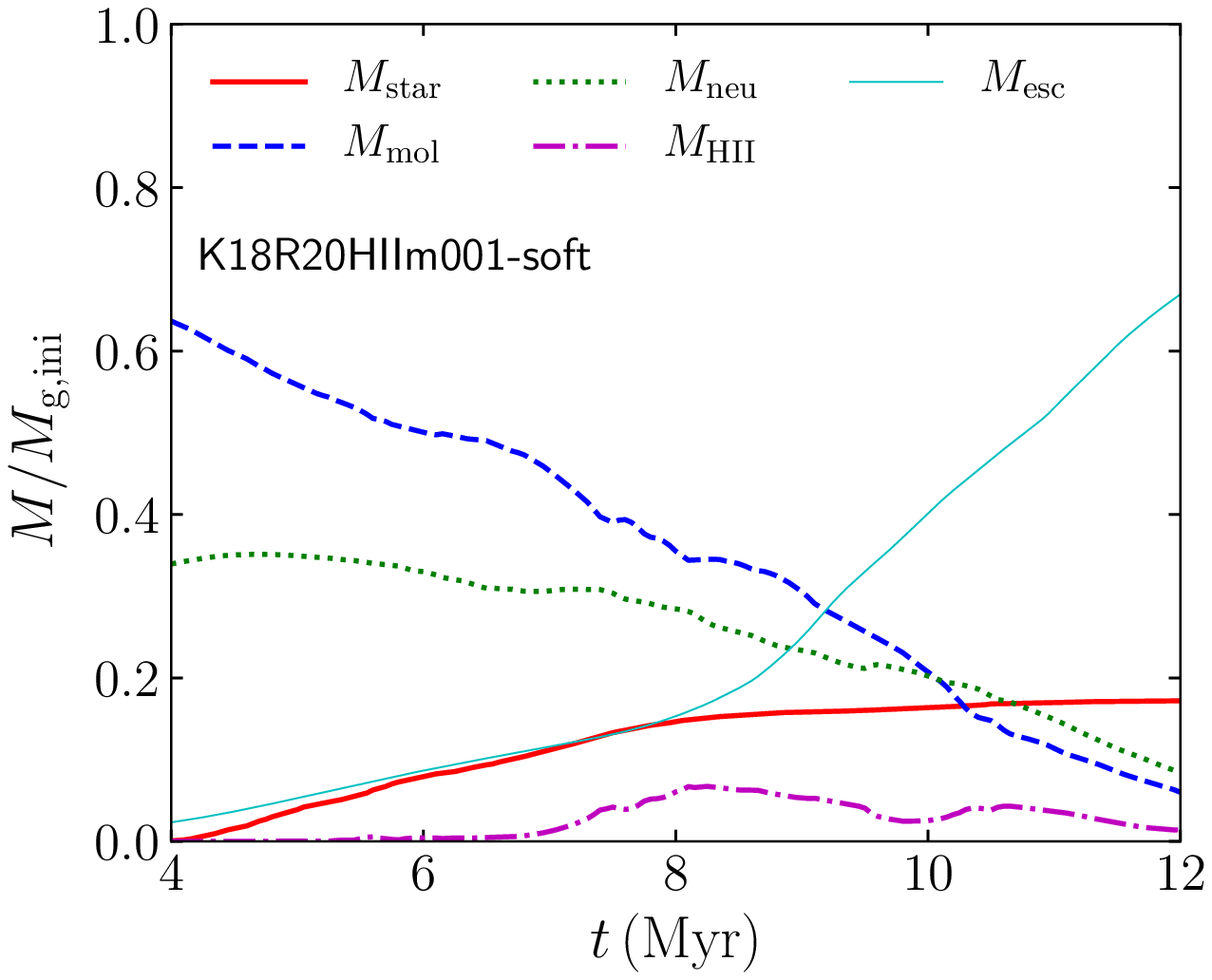}
  \includegraphics[width=7.8cm]{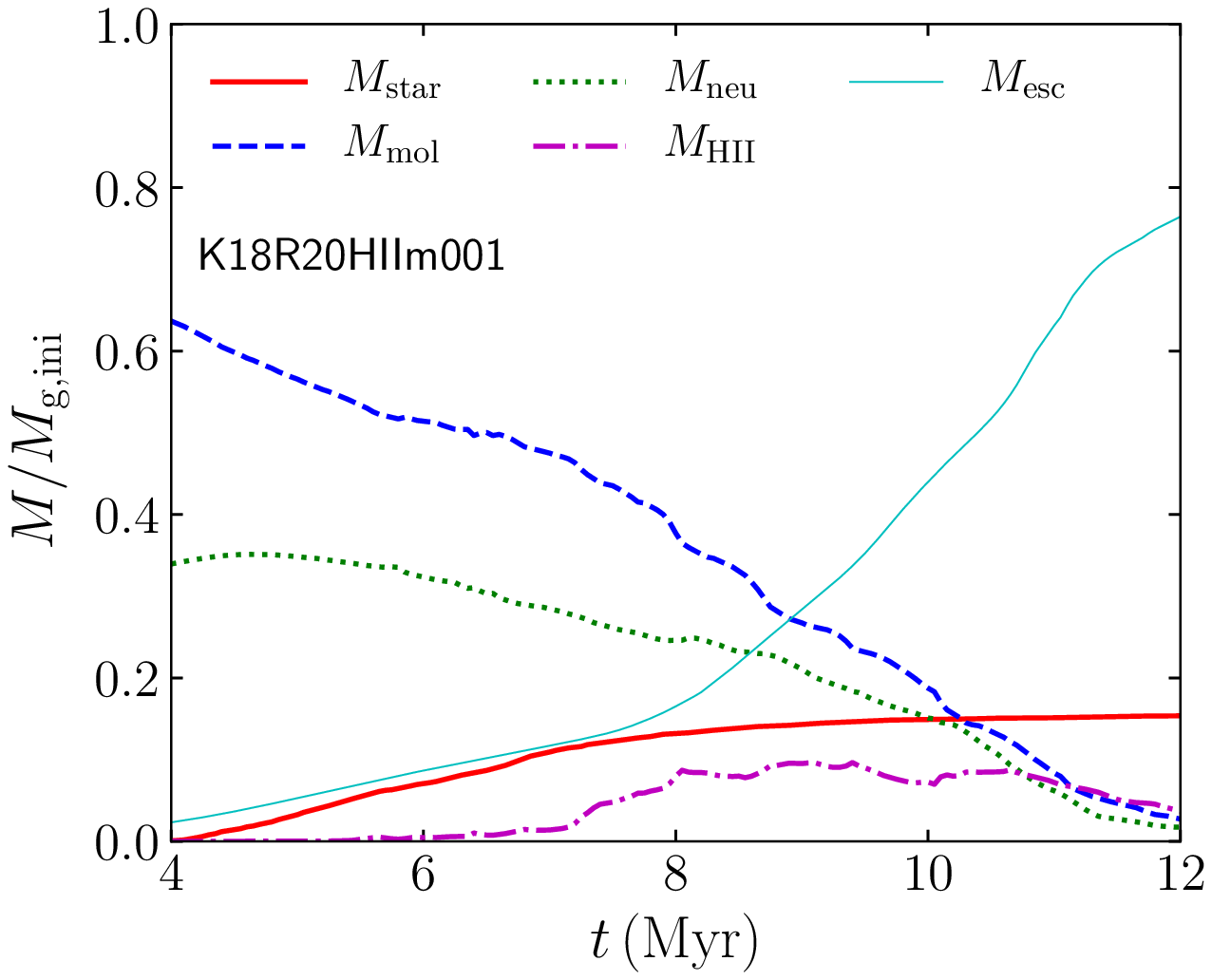}
 \end{center}
\caption{Mass fraction of gas and stars for models K18R20HIIm001-soft (left) and K18R20HIIm001 (right). Here, we define gas with $T<100$\,K and $n>100$\,cm$^{-3}$ as molecular gas ($M_{\rm mol}$), that with $T>8000$\,K as ionized gas ($M_{\rm \HII}$), and the rest as neutral gas ($M_{\rm neu}$). Gas particles reach further than 40\,pc are classified as escaped gas ($M_{\rm esc}$). }\label{fig:gas_frac_Kim18}
\end{figure*}

In Fig.~\ref{fig:gas_frac_Kim18}, we present the mass fractions of gas and stars as a function of time for models K18R20HIIm001-soft and K18R20HIIm001. As we do not solve the chemical network, we assume that gas with $T<100$\,K and $n>100$\,cm$^{-3}$ is molecular gas ($M_{\rm mol}$), that with $T>8000$\,K is ionized gas ($M_{\rm \HII}$), and the rest is neutral gas ($M_{\rm neu}$). We count all gas particles reach further than 40\,pc as escaped gas ($M_{\rm esc}$) irrespective of their densities and temperatures. Before star formation starts, more than 60\,\% is high-density and low-temperature gas (molecular gas) and $\sim35$\,\% is neutral gas. After 7\,Myr, the amount of ionized gas starts to increases. However, the maximum is only $\sim10$\,\%. When the feedback becomes strong enough ($\gtrsim 8$\,Myr), the fraction of escaped gas dramatically increases. The escaped gas fraction finally reaches $\sim70$\,\% at the end of the simulation. On the other hand, the neutral and molecular gas continuously decreases. Molecular gas is converted to stars or ionized by the feedback. The fraction of ionized gas within 40\,pc decreases later in the simulation. Comparing the models with and without softening, the fraction of ionized gas is slightly larger in the case without softening.

\begin{figure*}
 \begin{center}
  \includegraphics[width=7.8cm]{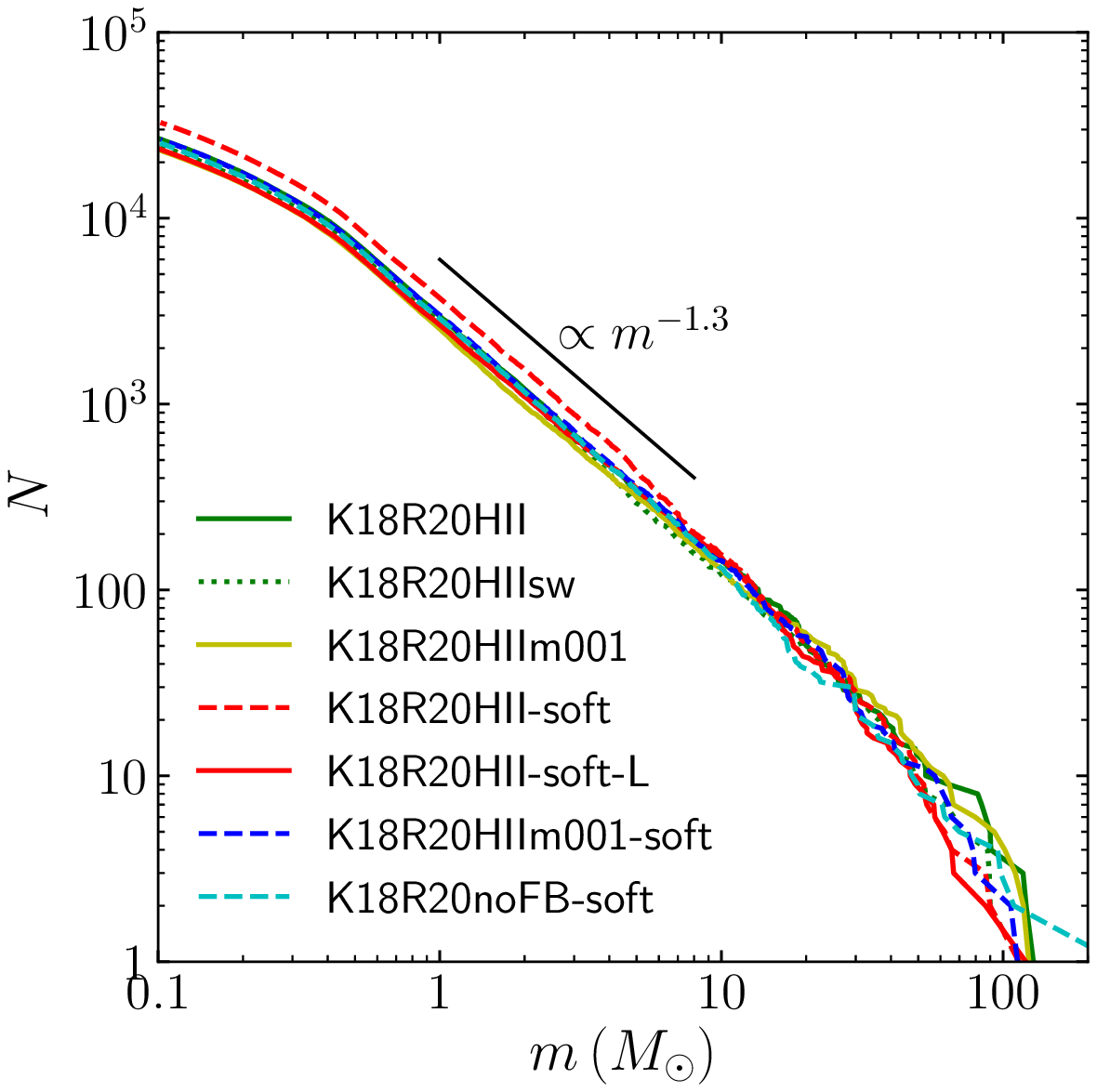}
  \includegraphics[width=7.8cm]{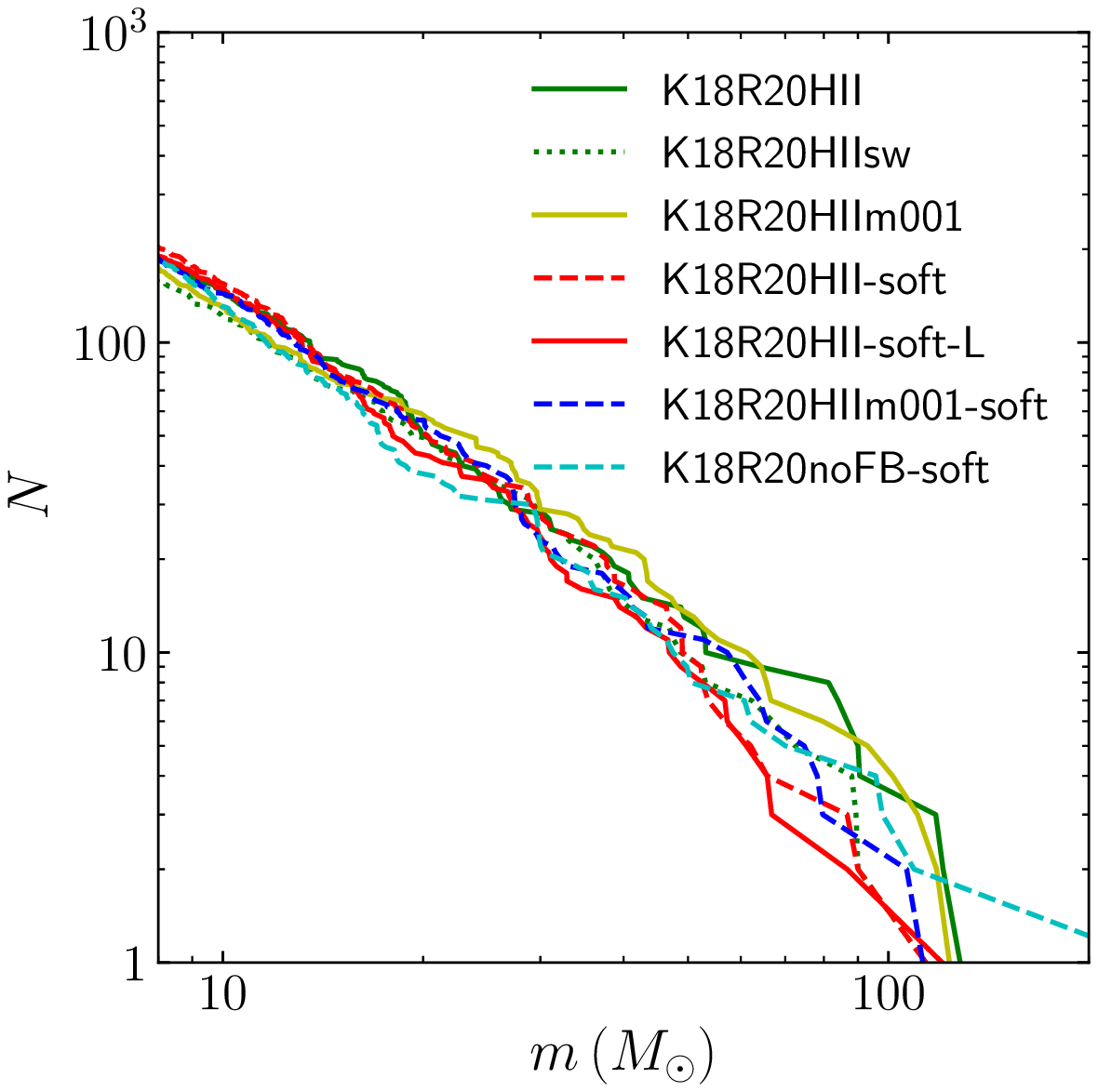}
 \end{center}
\caption{Mass functions obtained at $t=12$\,Myr. Left: entire mass function. Right: mass function of massive stars. \label{fig:MF_Kim18}}
\end{figure*}

We further investigate the cause of the difference in the cases with/without softening for stars. 
In Fig.~\ref{fig:MF_Kim18}, we show the stellar mass functions at the end of the simulation ($t=12$\,Myr). The mass function is almost identical between different models, but in the massive end, we observe some differences. In our star-by-star formation scheme, we assume the IMF of \citet{2001MNRAS.322..231K} and allow stars to form up to $150\,M_{\odot}$. However, the massive end of the mass functions of the formed stars is truncated. The truncation is caused by the limited amount of mass in the star-forming regions. When the gas mass within $r_{\rm max}$ is not sufficient to form a mass chosen form the given mass function, we choose a sufficiently small mass for the forming star. Such a truncation is also seen in the mass functions of observed open clusters \citep{2006MNRAS.365.1333W}. If we switch off the feedback (see model K18R20noFB-soft), the truncation mass grows to higher mass compared with the cases with feedback. 

\begin{figure}
 \begin{center}
  \includegraphics[width=7.8cm]{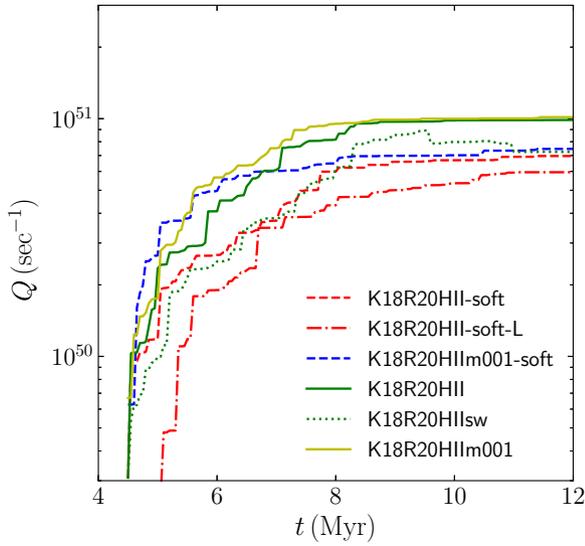}
 \end{center}
\caption{Evolution of the total photon counts from all massive stars ($>8 M_{\odot}$).}\label{fig:Q_ev_Kim18}
\end{figure}

This limit works in the super-virial models, in which stars form in small stellar clumps. The amount of gas in clumps cannot be sufficient to form massive stars up to the upper mass limit of the mass function, and therefore the mass of the most massive star in clumps is lower than the upper mass limit of the given mass function. With stellar softening, massive stars formed in clumps cannot be kicked out due to the interactions with binaries and stay inside clumps. The massive stars in clumps ionize the gas and terminate the star formation before forming more massive stars. However, without stellar softening, massive stars can be ejected from clumps, and as a consequence, star formation can continue for longer. During the more extended star formation period, clumps can accumulate more gas and form more massive stars. This difference alters the shape of the mass function at the massive end.
Therefore, more ionizing photons are emitted from the massive stars in the models without softening. This is confirmed in the time evolution of $Q$ (see Fig.~\ref{fig:Q_mass} and equation~\ref{eq:Q}) shown in Fig.~\ref{fig:Q_ev_Kim18}, where we count $Q$ from all massive stars formed. 

\begin{figure}
 \begin{center}
  \includegraphics[width=7.8cm]{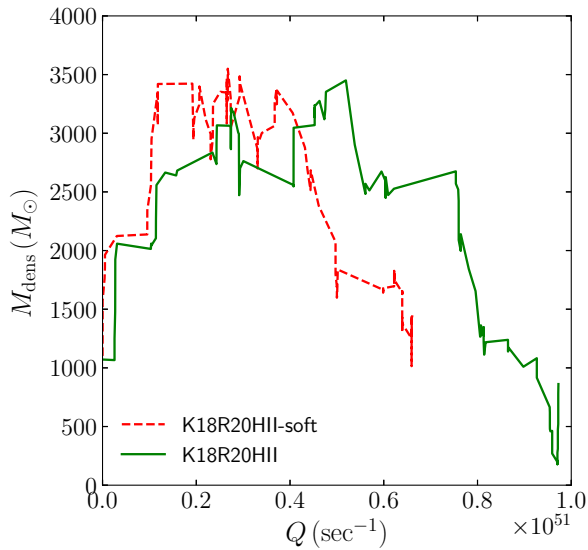}
 \end{center}
\caption{Relation between the total ionization photon counts per second from all massive stars ($>8 M_{\odot}$) and the total mass of dense cold gas ($n>10^6$\,cm$^{-3}$ and $T<100$\,K).}\label{fig:Q_Mdens_Kim18}
\end{figure}

To understand the relation between feedback and star formation in the cases with/without softening, we present the total mass of cold ($<100$\,K) and dense ($>10^6$\,cm$^{-3}$) as a function of the total ionizing photon counts from all massive stars in Fig.~\ref{fig:Q_Mdens_Kim18}. Here, we only compare models K18R20HII-soft and K18R20HII, because the star formation is not significantly different in other models with a higher mass resolution. As shown in this figure, dense cold gas forms massive stars and the ionizing photon counts increase, and then the photon counts start to decrease due to the ionization and quench of the star formation. In the model without softening, the total amount of dense gas is initially smaller than that in the model with softening, and maintains small value almost all the time. However, the dense gas survives longer after the formation of massive stars in the case with softening. This is because some massive stars escape from the star-forming regions as runaways due to dynamical interactions. Such runaways ionize less dense regions in the system outside of star-forming clumps.
Therefore, the star formation continues longer in the models with softening, and as a consequence more low-mass stars form.

\begin{figure}
 \begin{center}
  \includegraphics[width=7.8cm]{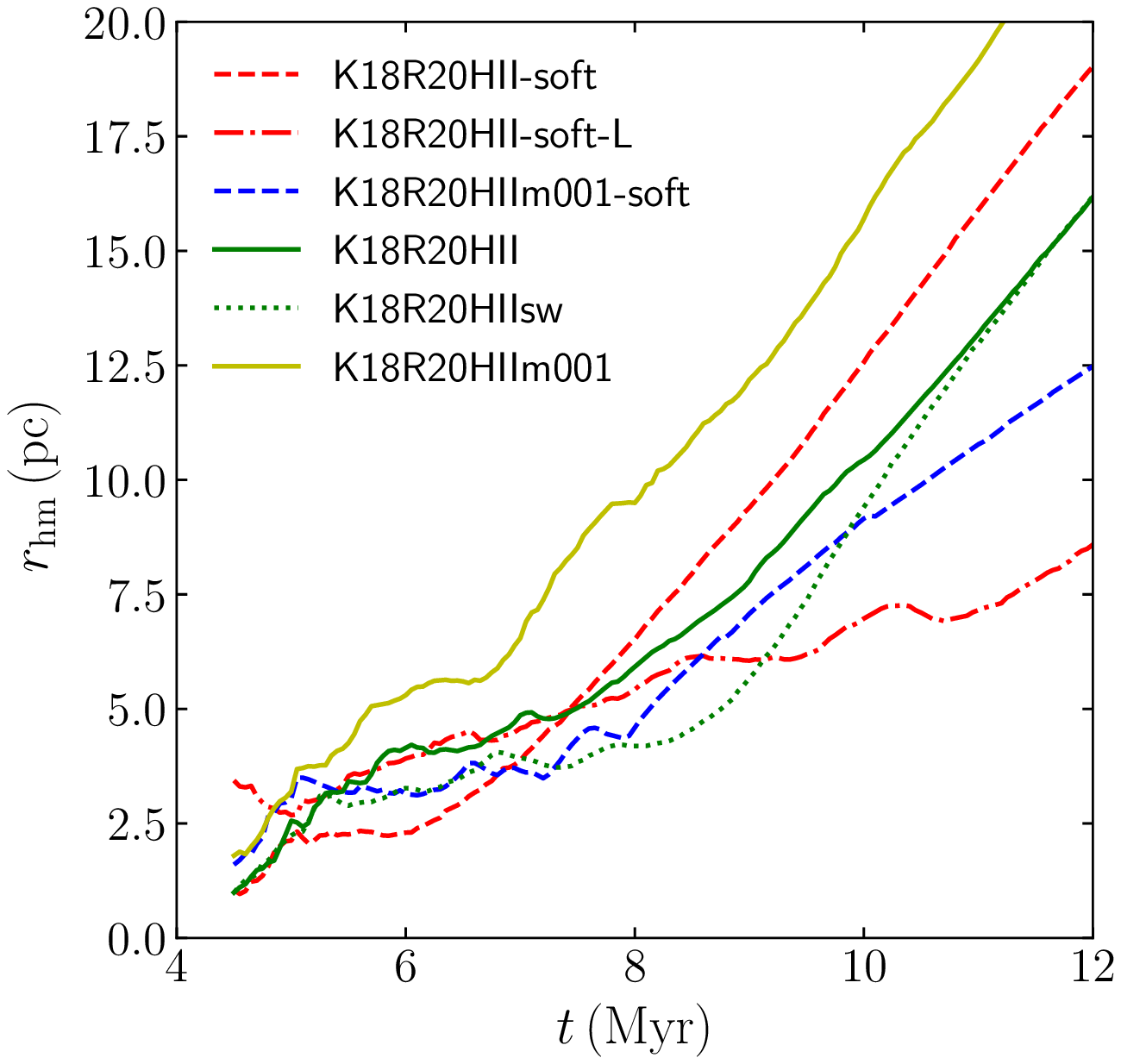}
 \end{center}
\caption{Evolution of the half-mass radius of the entire stellar system for K18R20 models.}\label{fig:r_hm_Kim18}
\end{figure}

In Fig.~\ref{fig:r_hm_Kim18}, we present the evolution of the half-mass radius of the entire stellar system. The  system is unbound and continues to expand after the gas expulsion, whereas individual clusters are bound (see Fig.~\ref{fig:snapshot_Kim18m001}). The details of these clusters are discussed in section~\ref{clusters}. The entire system is more compact with stellar softening before the gas expulsion. 
In the model K18R20HIIsw with stellar wind feedback, the stellar wind slightly suppresses the star formation, but the effect is small.

\subsection{Sub-virial case}

\begin{figure*}
 \begin{center}
 \includegraphics[width=7.cm]{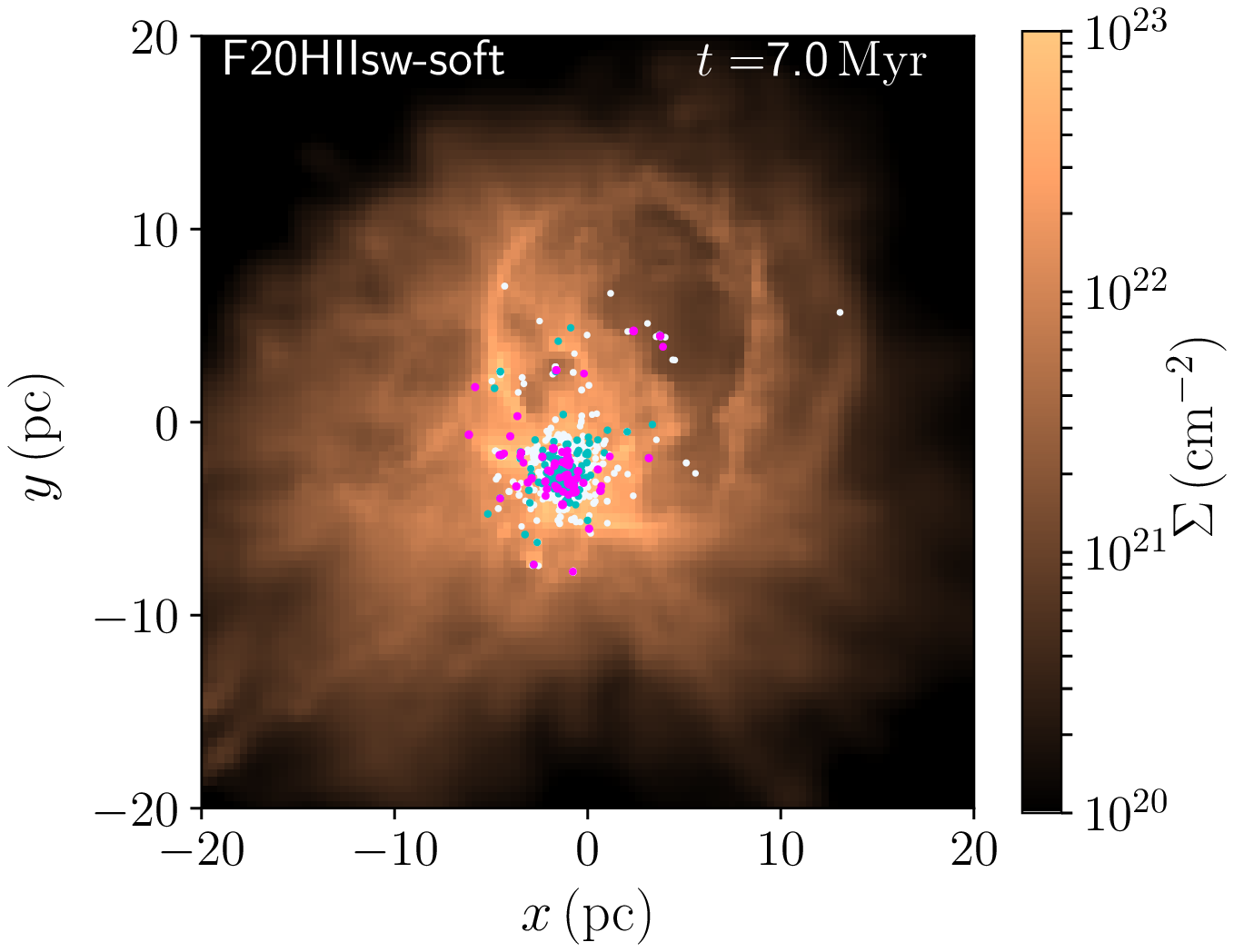}
  \includegraphics[width=7.cm]{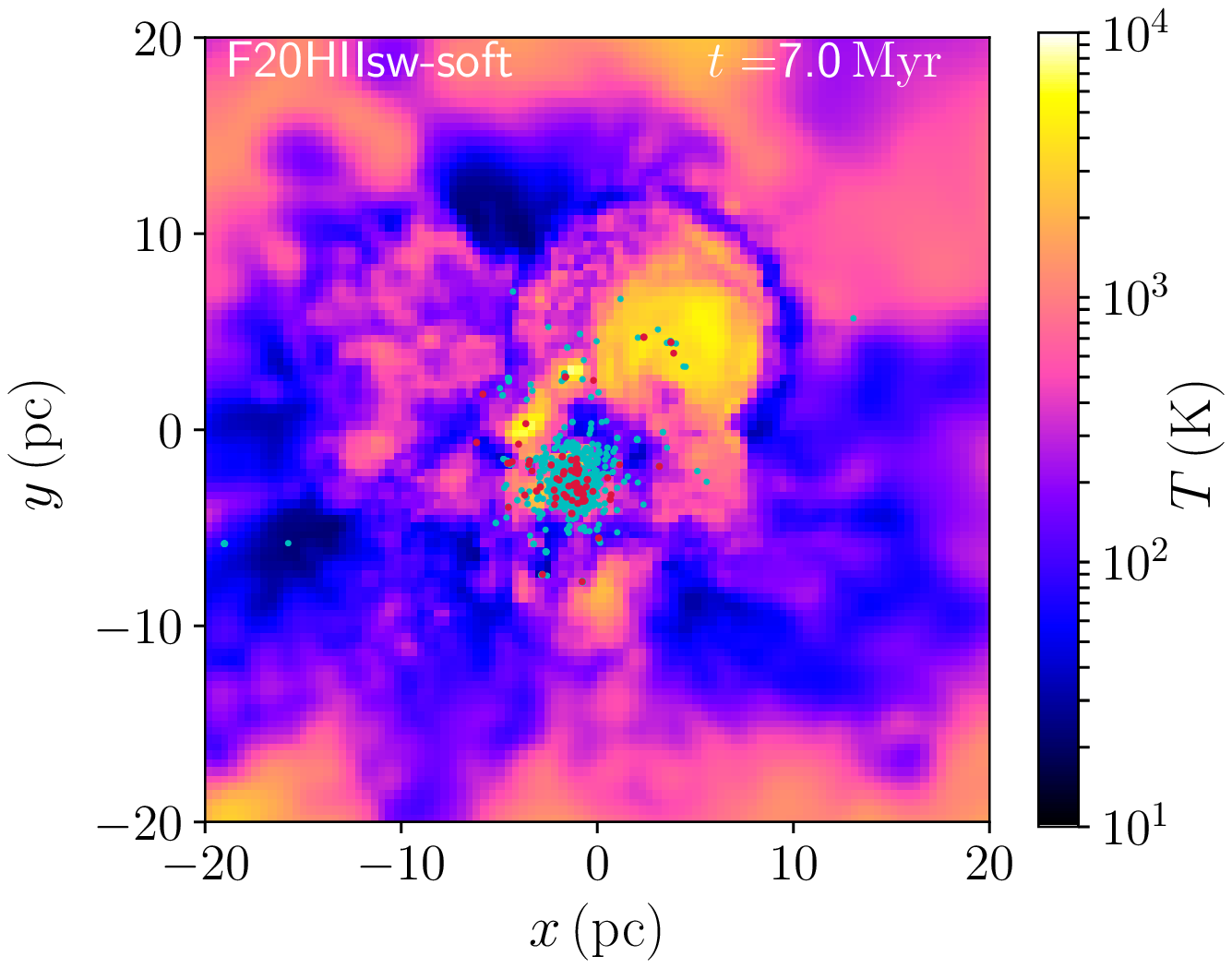}\\
 \includegraphics[width=7.cm]{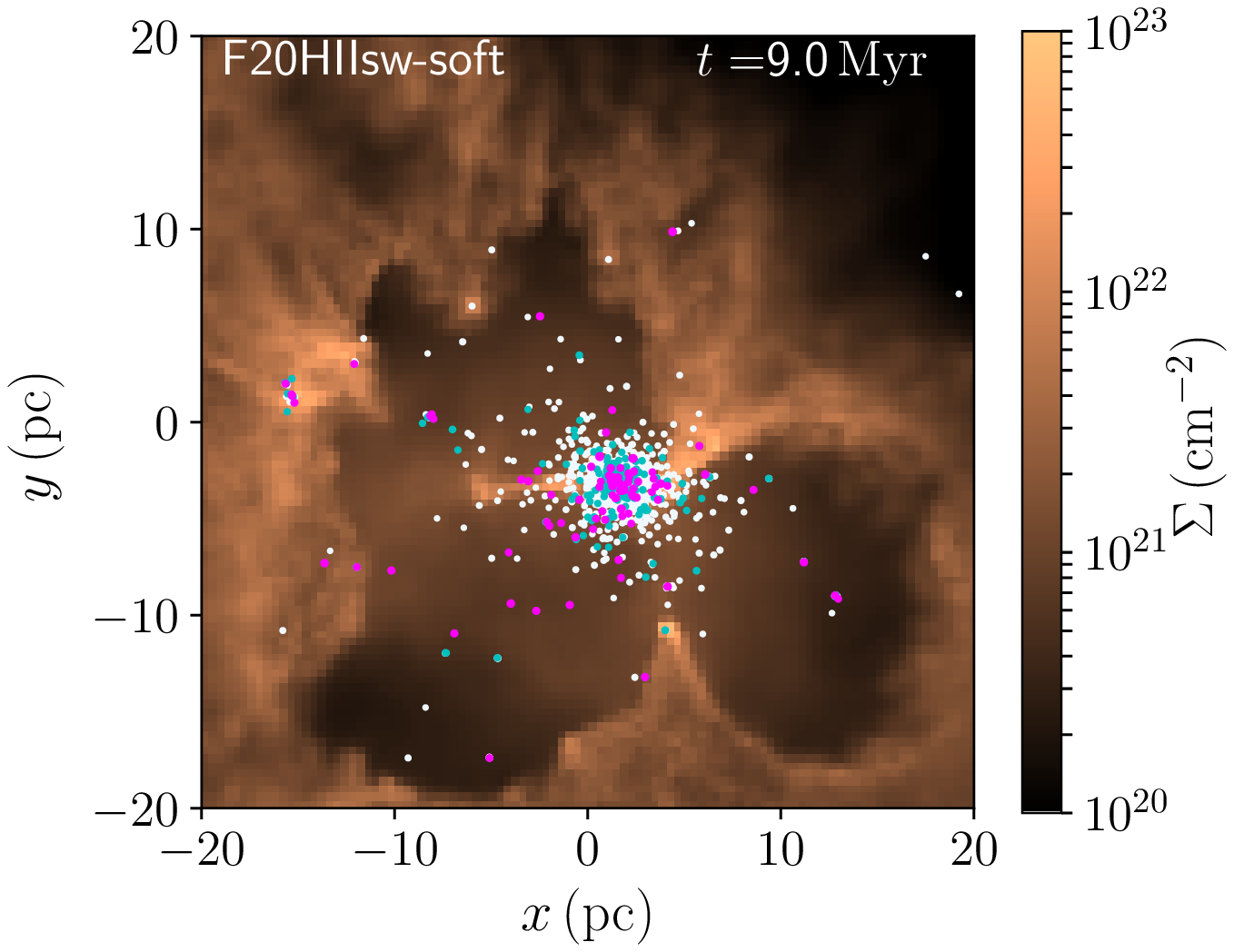}
  \includegraphics[width=7.cm]{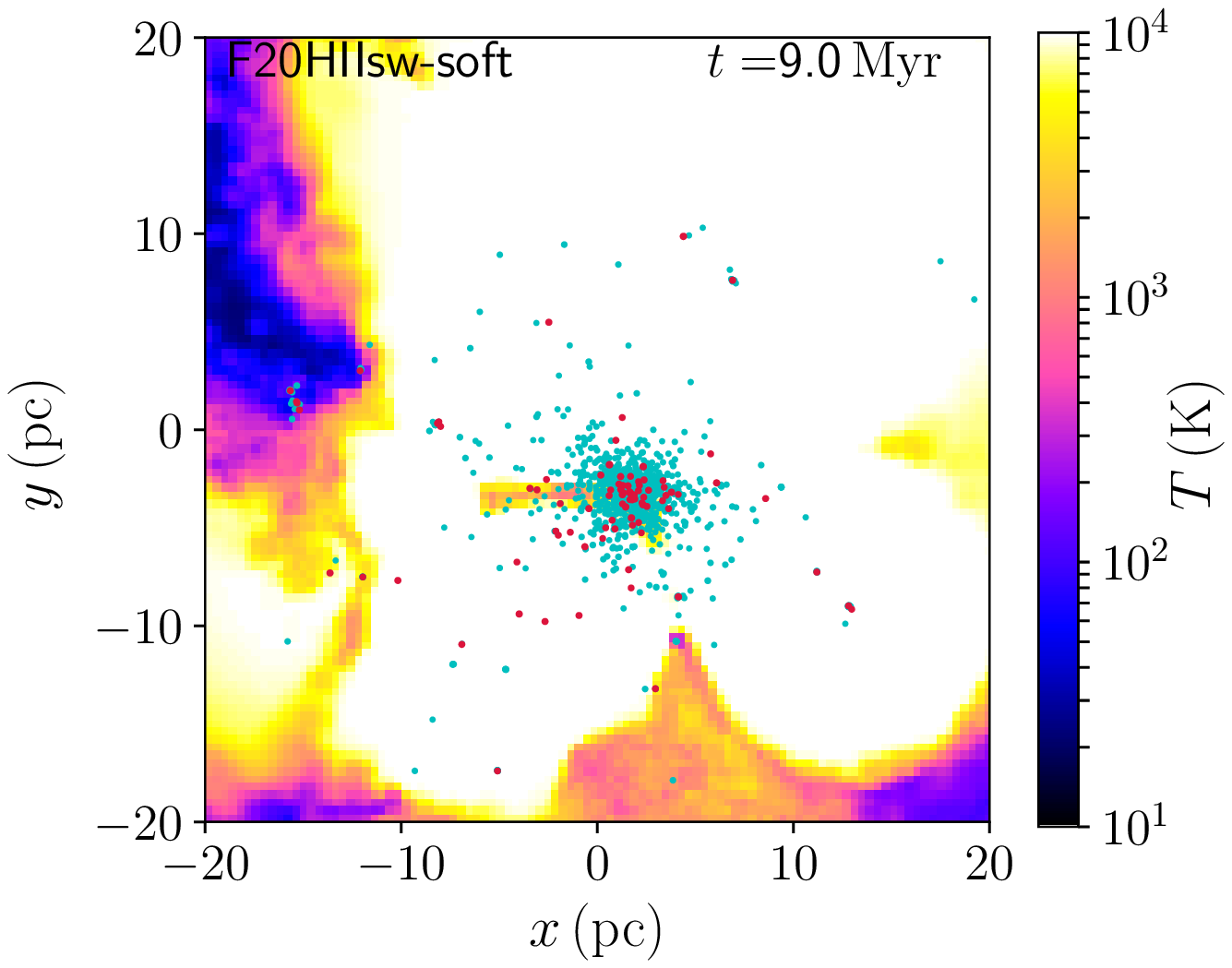}\\
 \end{center}
\caption{Same as Fig.~\ref{fig:snapshot_Kim18m001-soft} but for model F20HIIsw-soft. }\label{fig:snapshot_F20sw-soft}
\end{figure*}

\begin{figure*}
 \begin{center}
 \includegraphics[width=7.cm]{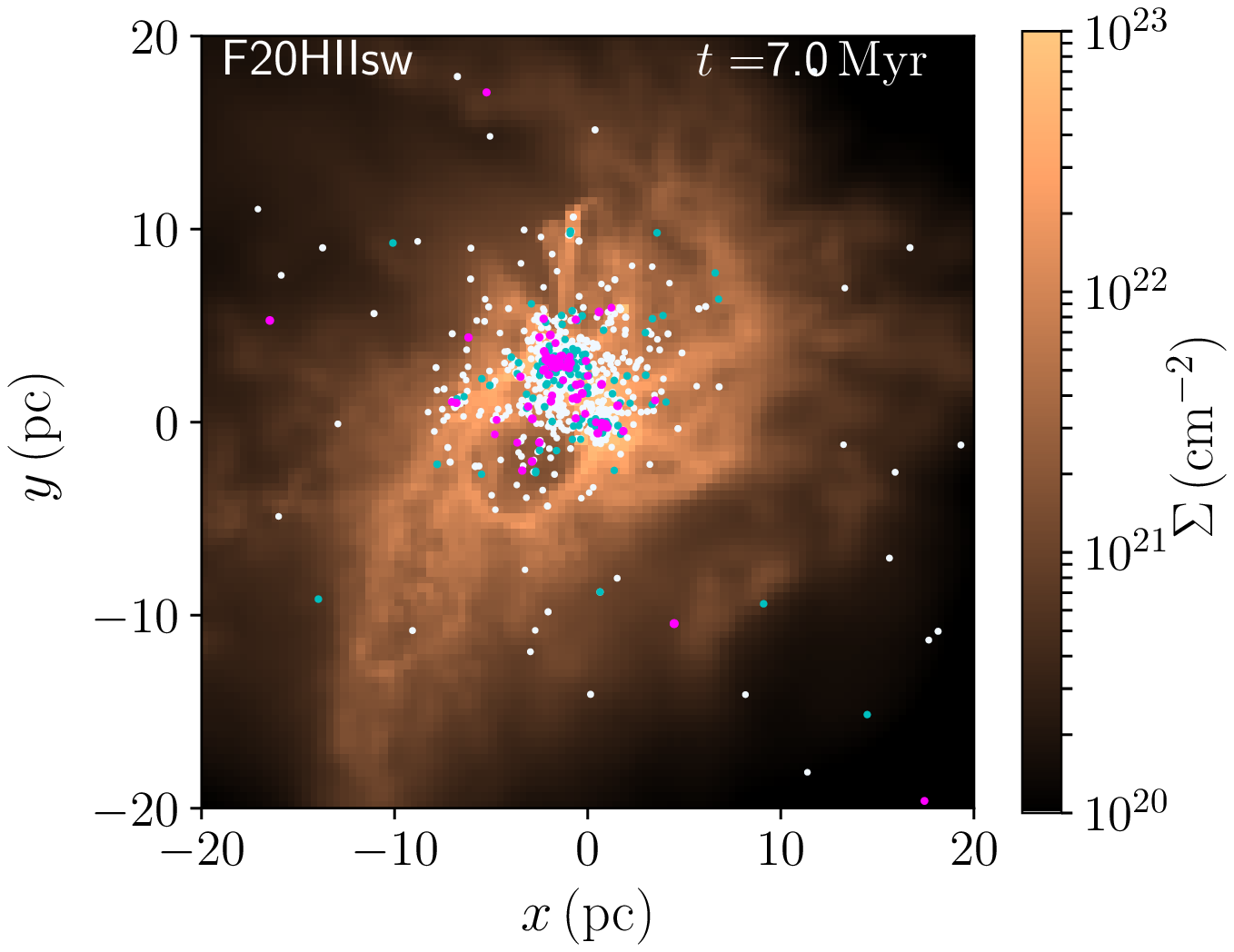}
  \includegraphics[width=7.cm]{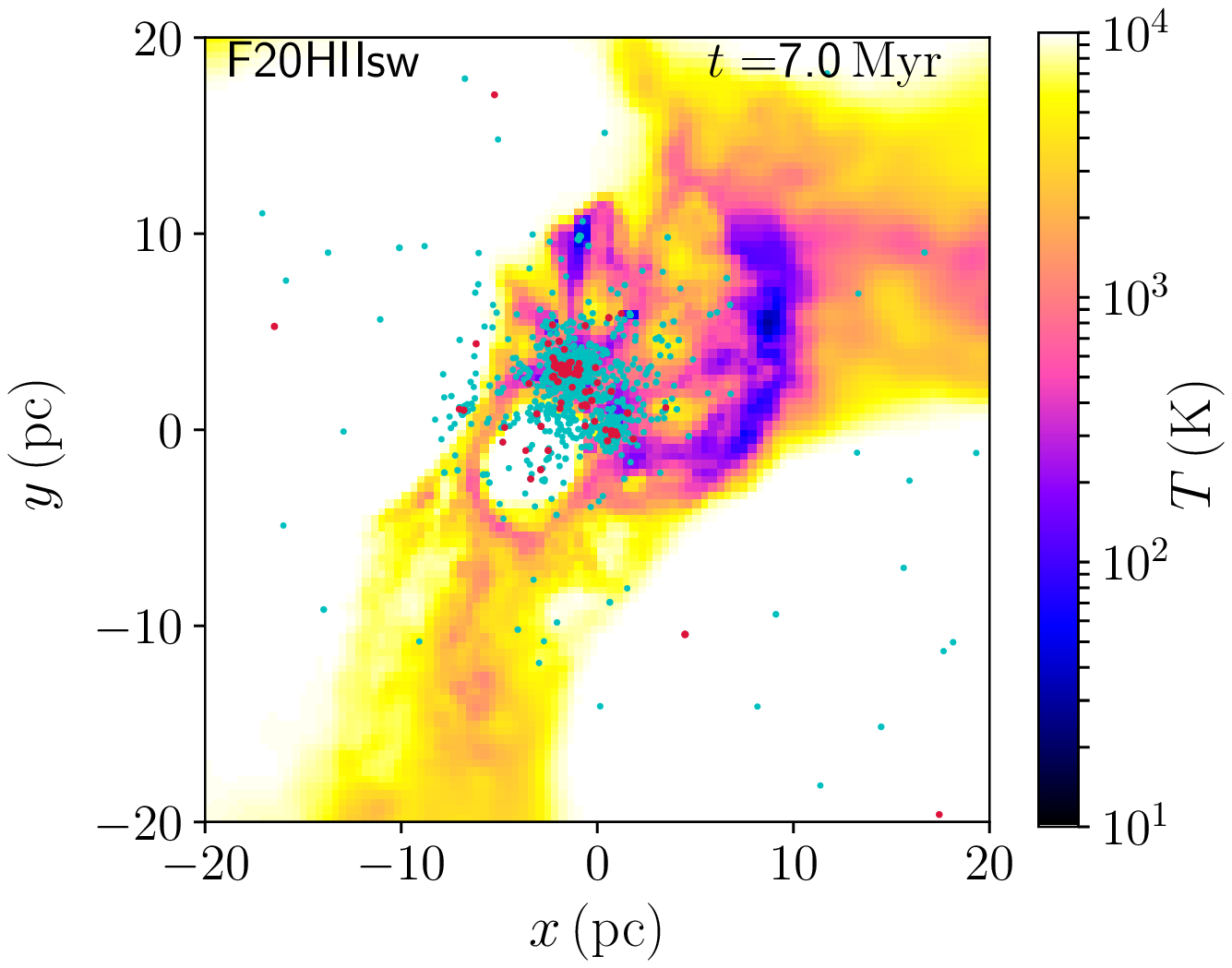}\\
 \includegraphics[width=7.cm]{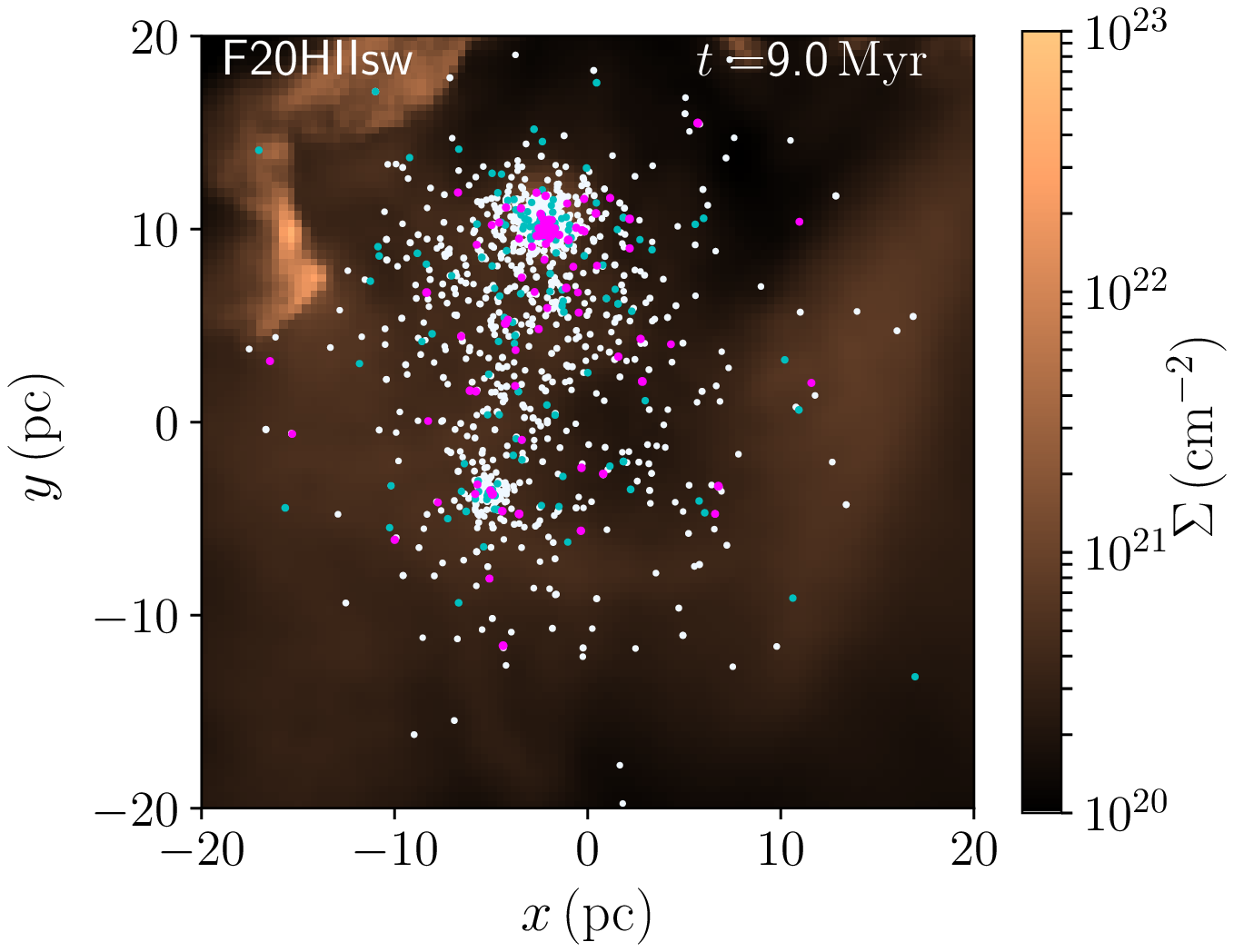}
  \includegraphics[width=7.cm]{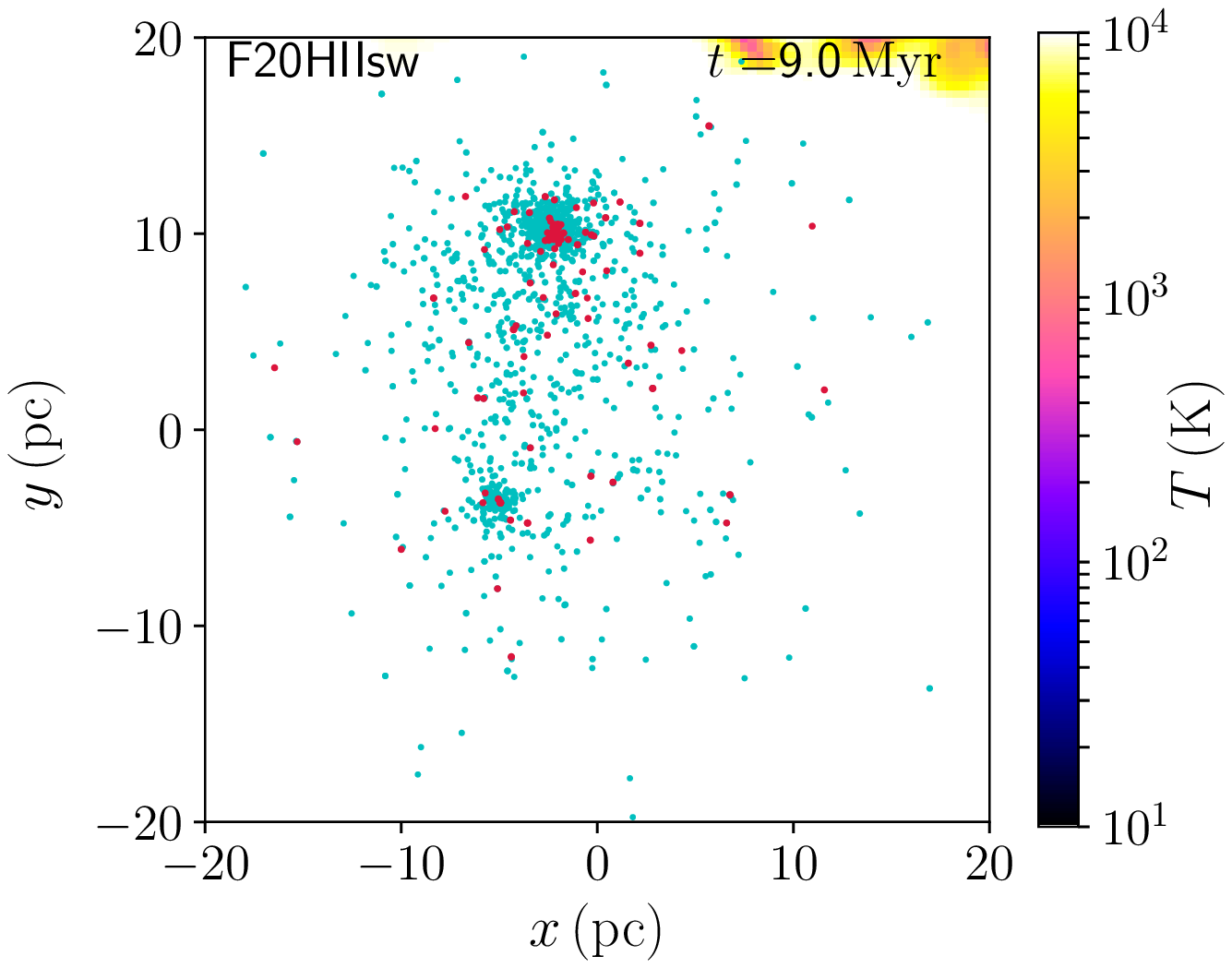}\\
 \end{center}
\caption{Same as Fig.~\ref{fig:snapshot_Kim18m001-soft} but for model F20HIIsw.}\label{fig:snapshot_F20sw}
\end{figure*}

In Figs.~\ref{fig:snapshot_F20sw-soft} and \ref{fig:snapshot_F20sw}, we present snapshots of models F20HIIsw-soft and F20HIIsw. These models are initially sub-virial. Therefore, the stellar distribution is less clumpy, and one massive cluster forms. 
The evolution of the total stellar mass is shown in Fig.~\ref{fig:stellar_mass_ev_Fuk20}. The star formation starts at $\sim4$\,Myr, which is similar to the super-virial case (K18R20 models). However, the total stellar mass is smaller with softening for stars. This is opposite to the trend in the super-virial case. We further perform additional simulations with a different random seed for the initial turbulence (models F20HIIsw-s2 and F20HIIsw-soft-s2) and confirm that this trend is not a result of random seeds (see Fig.~\ref{fig:stellar_mass_ev_Fuk20}). 

In Fig.~\ref{fig:gas_frac_Fuk20}, we present the time evolution of the mass fraction of gas and stars for models with and without softening for stars. The trend of the evolution in which the molecular gas decreases as stars form is similar to that of K18R20 models. The fraction of escaped gas is much smaller than the super-viral cases. The neutral gas fraction does not decrease much. 
Comparing the cases with/without softening for stars, the fractions of ionized and escaped gas are larger in the case without softening than the soften case.

We show the stellar mass function at $t=9$\,Myr in Fig.~\ref{fig:MF_Fuk20} and the time evolution of the total ionizing photon counts in Fig.~\ref{fig:Q_ev_Fuk20}. Similar to the super-virial case, models without stellar softening result in the formation of a larger number of massive stars and, as a consequence, more ionizing photons compared with the models with softening.  However, the difference is smaller than that in the super-virial case; with a different random seed for the initial turbulent velocity field, the formed mass functions show no clear difference between the cases with and without the stellar softening. In the sub-virial case, one massive cluster forms in the system, and therefore the given mass function can be fully sampled up to the end of IMF.

In Fig.~\ref{fig:Q_Mdens_Fuk20}, we present the relation between the ionizing photon counts and the amount of cold dense gas for models F20HIIsw and F20HIIsw-soft. The initial evolution is similar for both models. In both cases, the amount of cold dense gas starts to decrease after the photon counts reached $\sim 5 \times 10^{51}$\,s$^{-1}$. In contrast with the super-virial case (models K18R20HII and K18R20HII-soft), the amount of dense gas increases again in the case without softening and reaches a higher peak later on. With softening, it continues to decrease after the first peak. The second peak in the case without stellar softening is due to the feedback in the outskirts of the star-forming region. As shown in Fig.~\ref{fig:snapshot_F20sw}, some massive stars are ejected from the central star forming region and ionize the gas when they reach the outskirts of the cloud, in which the gas density is low enough and the \HII region can expand. The formed \HII regions push the gas toward the cloud center, and the gas in the potential well increases. This causes the second peak of the dense and cold gas mass in Fig.~\ref{fig:Q_Mdens_Fuk20} and results in a higher star formation efficiency than in the model with softening.

In Fig.~\ref{fig:r_hm_Fuk20}, we present the evolution of the half-mass radius of the formed stars. Before the gas expulsion, the half-mass radius is roughly constant. The value is similar to the results in \citet{2020MNRAS.497.3830F} (1.5\,pc). After gas expulsion, the half-mass radius starts to increase, but the central part of the formed cluster remains bound. 
In \citet{2020MNRAS.497.3830F}, each stellar particles was treated as a star cluster and a cluster mass function was demonstrated. In our simulation, however, a massive cluster formed rather than a bunch of clusters.

We also present the results with a higher mass resolution (model F20HIIswm001) and without stellar-wind feedback (model F20HII) in Figs.~\ref{fig:stellar_mass_ev_Fuk20}, \ref{fig:MF_Fuk20}, and \ref{fig:Q_ev_Fuk20}. With a higher mass resolution, the total stellar mass is slightly smaller than that with a lower mass resolution, as is also seen in the super-virial case. Without stellar-wind feedback, the total stellar mass increases, but it is within the variations due to the random seed for the initial condition.

\subsection{Randomness effects}
\begin{figure}
 \begin{center}
  \includegraphics[width=7.8cm]{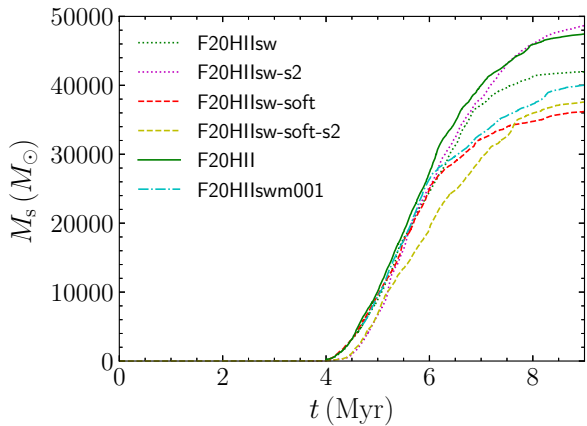}
 \end{center}
\caption{Stellar mass evolution of F20 models.}\label{fig:stellar_mass_ev_Fuk20}
\end{figure}

\begin{figure*}
 \begin{center}
  \includegraphics[width=7.8cm]{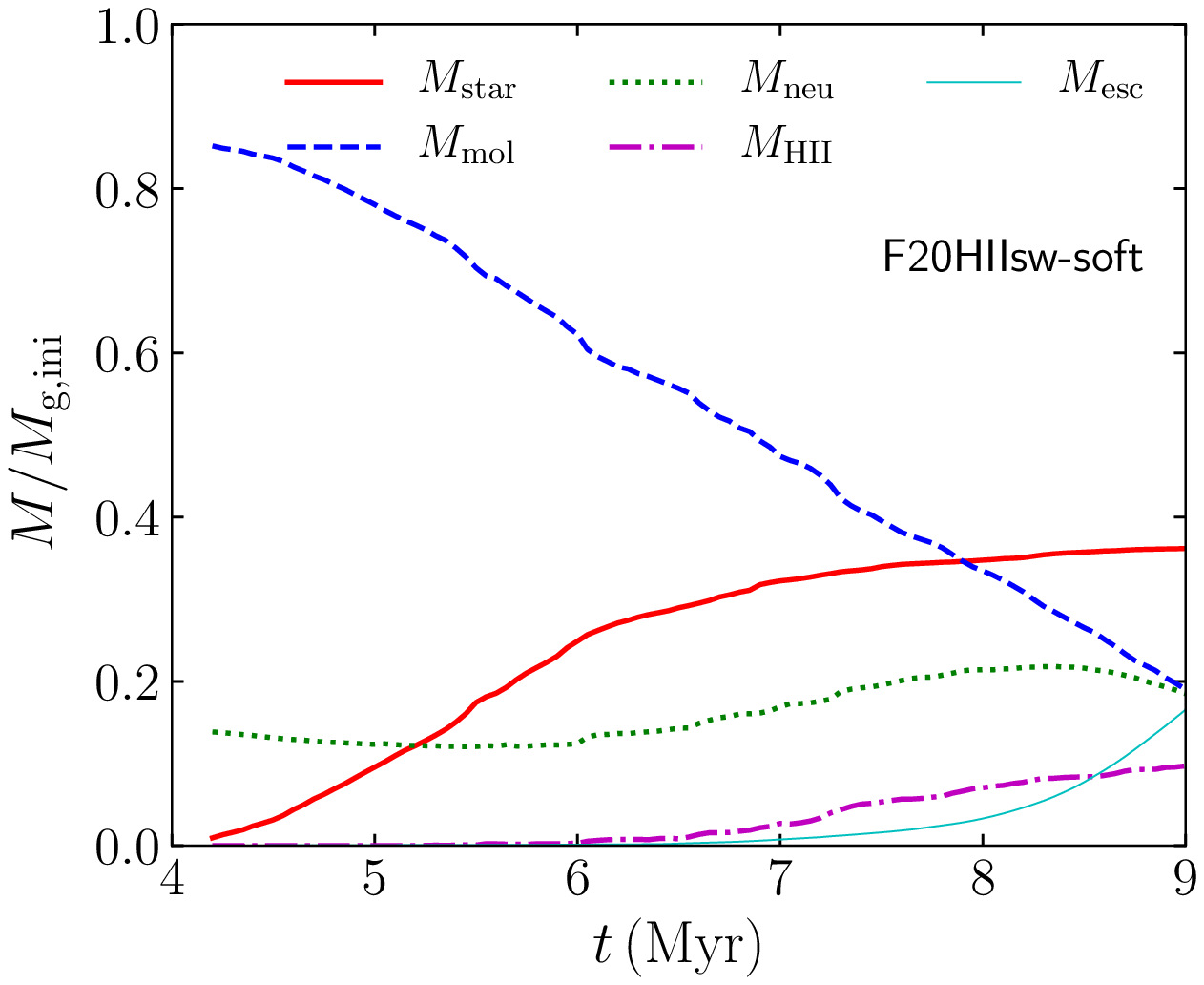}
  \includegraphics[width=7.8cm]{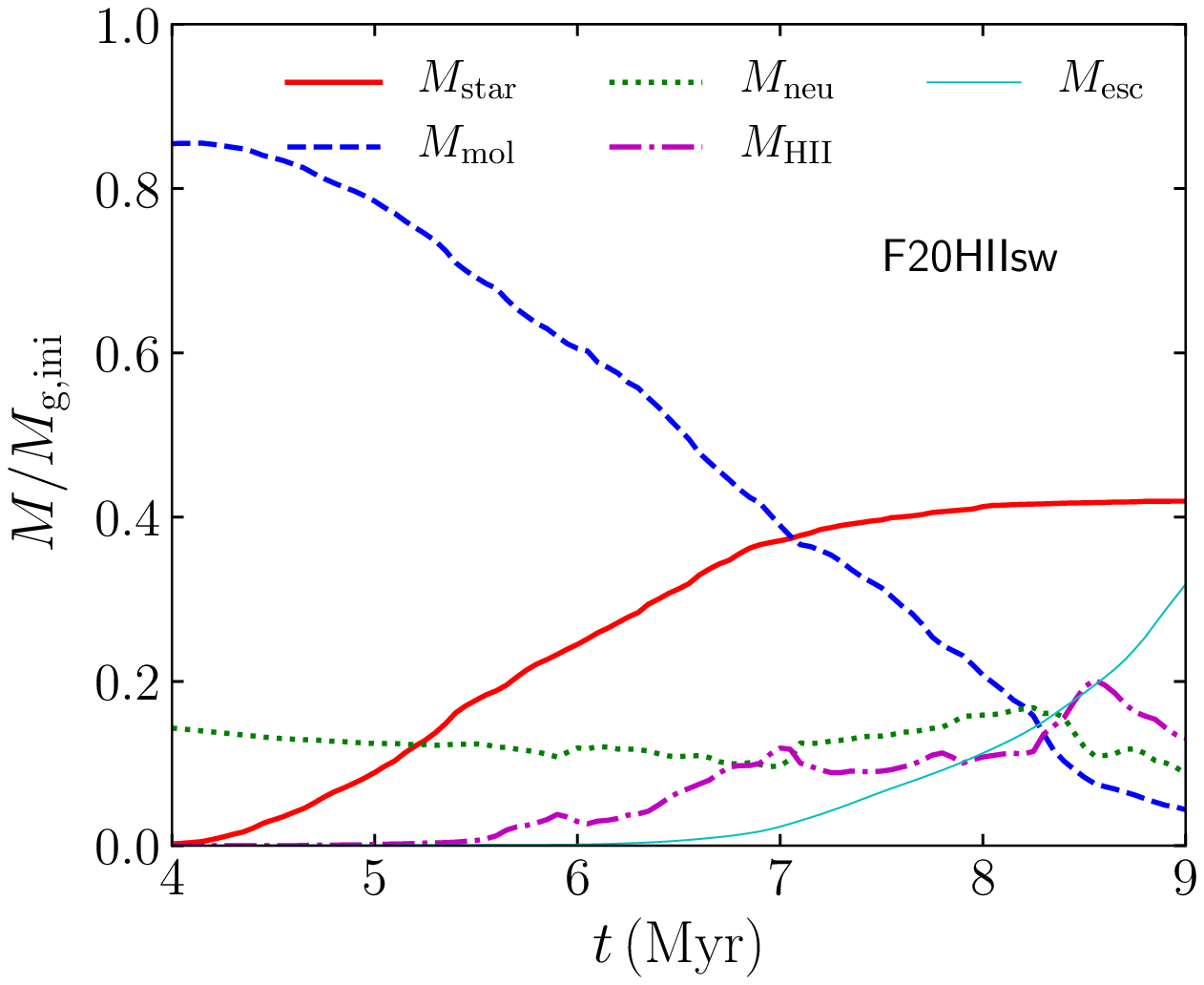}
 \end{center}
\caption{Same as Fig.~\ref{fig:gas_frac_Kim18} but for models F20HIIsw and F20HIIsw-soft.}\label{fig:gas_frac_Fuk20}
\end{figure*}

\begin{figure}
 \begin{center}
  \includegraphics[width=7.8cm]{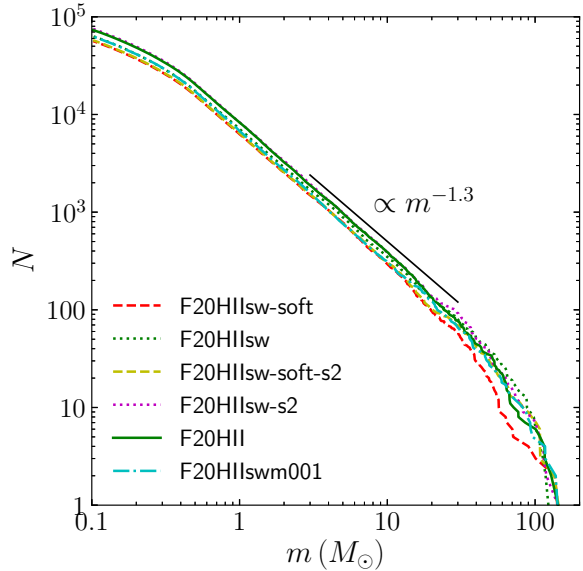}
 \end{center}
\caption{Same as the left panel of Fig.~\ref{fig:MF_Kim18} but for F20 models at $t=9$\,Myr.\label{fig:MF_Fuk20}}
\end{figure}

\begin{figure}
\begin{center}
  \includegraphics[width=7.8cm]{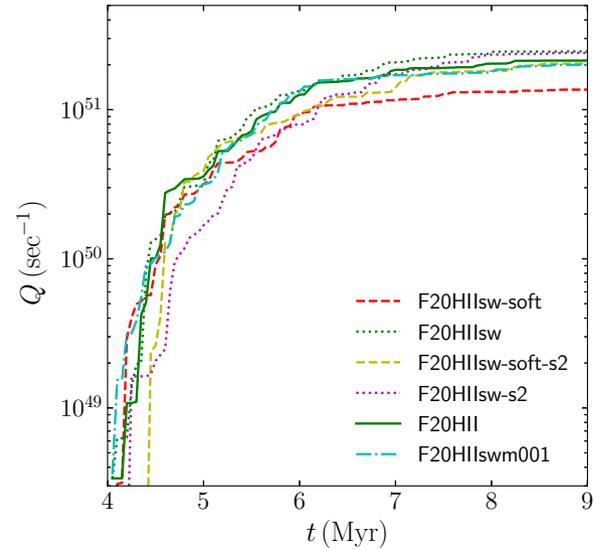}
 \end{center}
\caption{Same as Fig.~\ref{fig:Q_ev_Kim18} but for F20 models at $t=9$\,Myr.\label{fig:Q_ev_Fuk20}}
\end{figure}

\begin{figure}
 \begin{center}
  \includegraphics[width=7.8cm]{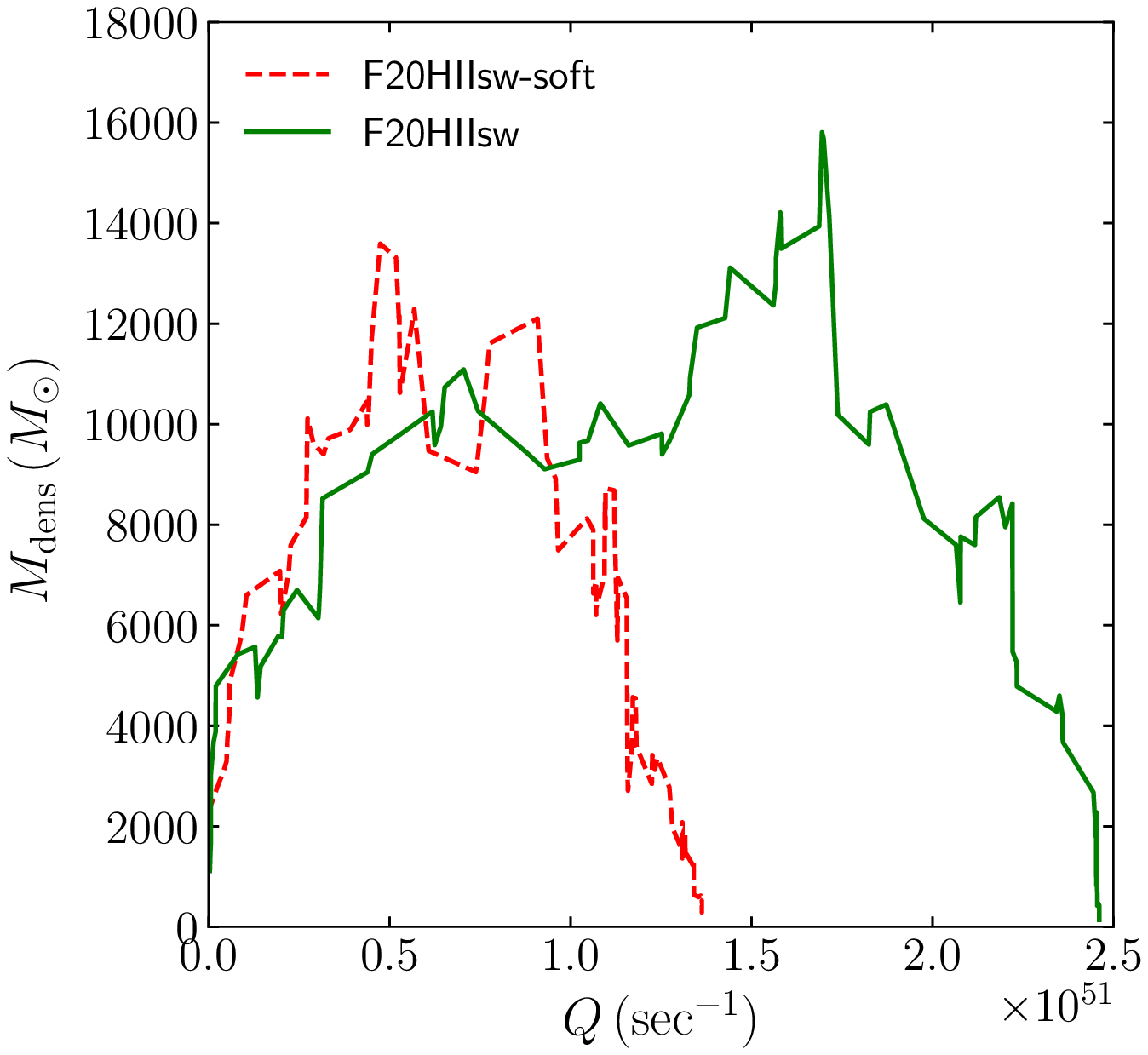}
 \end{center}
\caption{Same as Fig.~\ref{fig:Q_Mdens_Kim18} but for F20 models. \label{fig:Q_Mdens_Fuk20}}
\end{figure}

\begin{figure}
 \begin{center}
  \includegraphics[width=7.8cm]{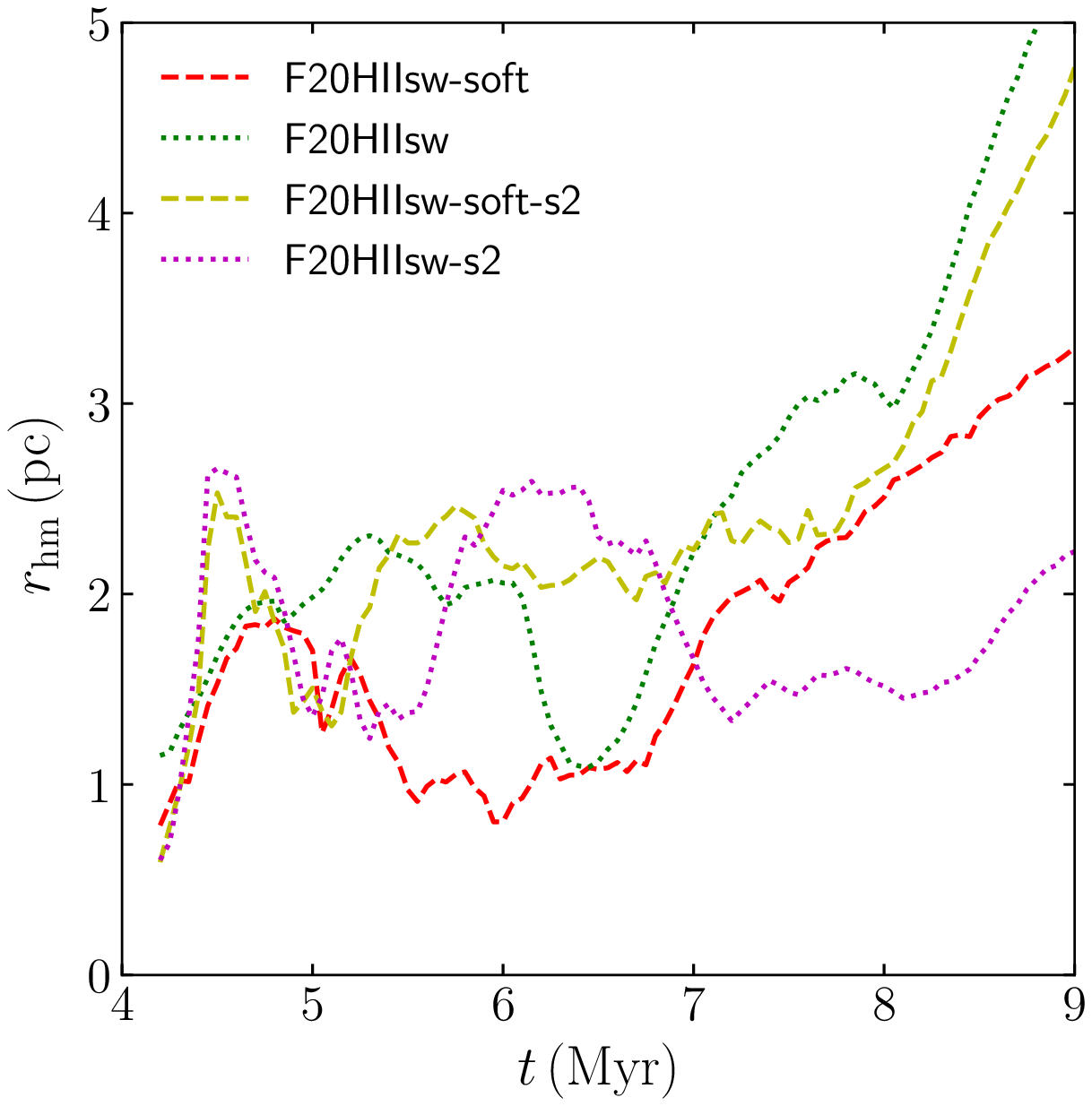}
 \end{center}
\caption{Same as Fig.~\ref{fig:r_hm_Kim18} but for F20 models.}\label{fig:r_hm_Fuk20}
\end{figure}

Here, we briefly discuss the effect of the randomness of these simulations. 
The structure of the turbulent molecular clouds and the distribution of the forming star clusters depend on the random seed for the initial turbulent velocity fields. For models F20HIIsw and F20HIIsw-soft (Seed 1), we perform another runs with different random seed for the initial condition. These are models F20HIIsw-s2 and F20HIIsw-soft-s2 (Seed 2). The results are shown in Figs.~\ref{fig:stellar_mass_ev_Fuk20}, \ref{fig:MF_Fuk20}, \ref{fig:Q_ev_Fuk20}, and \ref{fig:r_hm_Fuk20}. Compared with Seed 1, the difference in the total stellar mass between models with/without softening is smaller for Seed 2. However, the difference in the high-mass end in the mass function is not clear for Seed 2.

The formation of massive stars is another randomness in our simulations. The timing and the positions of the formation of massive stars is somehow random because we draw a mass when we form a star. The feedback process can also be affected by the randomness of the formation of runaway stars as a result of three-body encounters. The outcome of three-body encounters is highly chaotic, and therefore the mass and velocity of stars scattered from the center to the outskirts of clusters can also randomly change in each run.

\subsection{Comparison with previous works}
As described in section \ref{model3}, we adopt models similar to those of \citet{2018ApJ...859...68K} and \citet{2020MNRAS.497.3830F}.
Hereafter, we compare our results with these studies, in which they performed simulations using grid-based codes, sink methods, and super-particles for stars with softening. 

\begin{figure}
 \begin{center}
  \includegraphics[width=7.8cm]{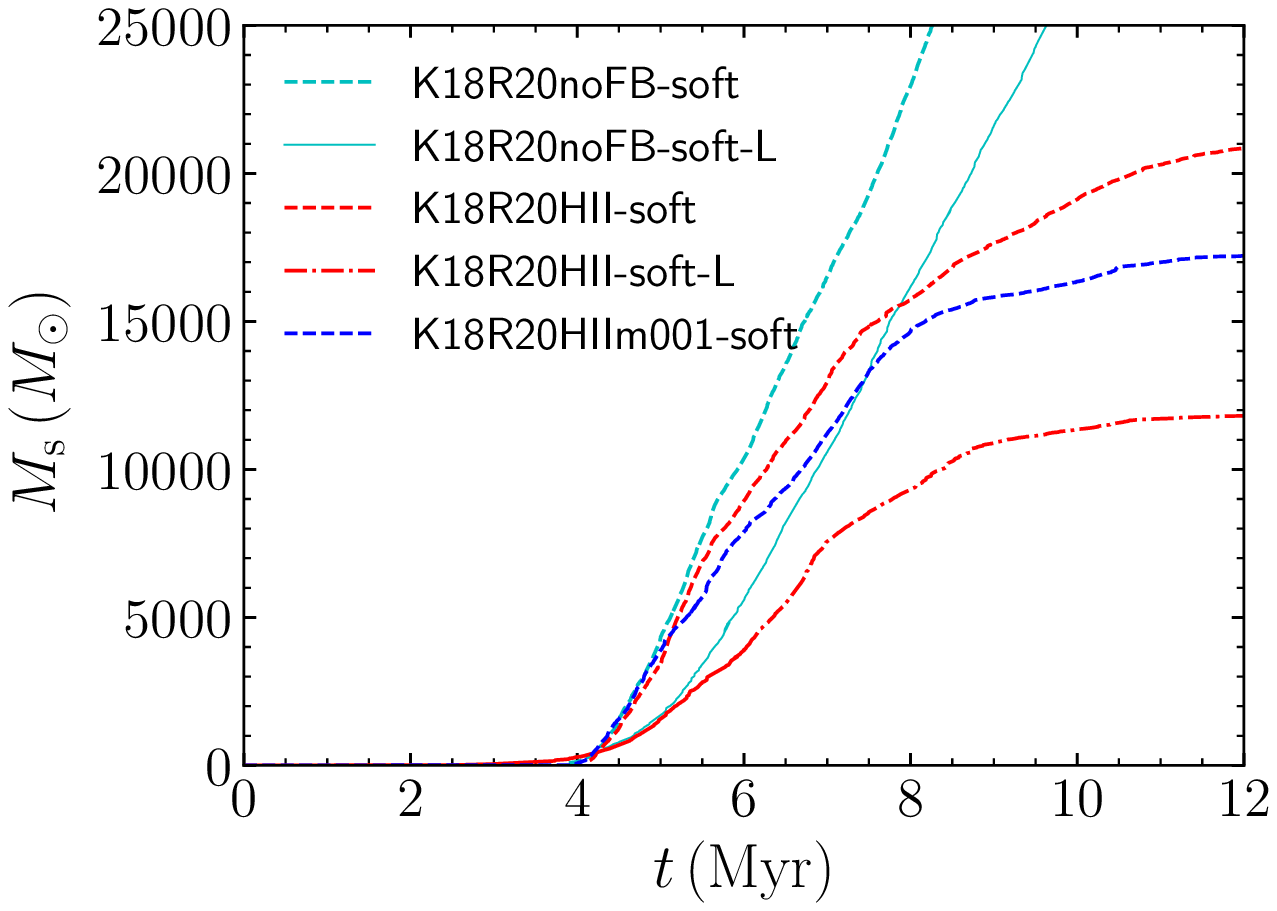}
 \end{center}
\caption{Stellar mass evolution of K18R20 models.}\label{fig:stellar_mass_ev_Kim18_r20}
\end{figure}

Fig.~\ref{fig:stellar_mass_ev_Kim18_r20} shows the stellar mass evolution of our models. Compared to those in \citet{2018ApJ...859...68K}, the star formation proceeds for longer in our simulations, and the final stellar mass is larger (see models K18HII-soft and K18HIIm001-soft). This may be due to the difference in softening lengths adopted in our simulation and in theirs. We additionally perform a simulation with a larger softening length, which is similar to the grid size adopted in \citet{2018ApJ...859...68K}, named as model K18R20HII-soft-L. The final stellar mass at 12\,Myr ($\sim 1.2\times 10^4 M_{\odot}$) is smaller than model K18R20HII-soft, and it is similar to that obtained in \citet{2018ApJ...859...68K}.  

\citet{2018ApJ...859...68K} also performed their simulations with higher and lower resolutions, and confirmed that the final stellar mass evolution converges. In our simulation, in contrast, the star formation proceeds slightly more rapidly with a run with smaller softening for gas (similar to higher spatial resolution). This results in the formation of more stars in our simulation (model K18R20HII).
We can confirm that star formation is more efficient for a smaller softening in models without feedback (models K18R20noFB-soft and K18R20noFB-soft-L; see Fig.~\ref{fig:stellar_mass_ev_Kim18_r20}). Thus, the rapid increase in forming stars with a smaller softening length causes the slightly higher star-formation efficiency in our simulation. 

We also compare our results (F20 models) with that of \citet{2020MNRAS.497.3830F}. Our models even including the stellar wind form roughly twice as many stars as theirs ($\sim 2\times 10^4 M_{\odot}$). The difference comes from the difference in the treatment of stellar particles and the related mass function. As shown in Fig.~\ref{fig:MF_Fuk20}, the high-mass end of the mass function of the formed stars is truncated in our simulations. In \citet{2020MNRAS.497.3830F}, in contrast, the mass function is assumed to continue up to $150 M_{\odot}$ following the Chabrier mass function \citep{2003PASP..115..763C}. In their model, they used the photon production rate per unit mass, $s_{*} = 8.0\times 10^{46}$\,s$^{-1}$\,$M_{\odot}^{-1}$. As the final stellar mass in their simulation was $\sim 2\times 10^4 M_{\odot}$, the ionizing photon count was estimated to be $1.8\times 10^{51}$ per second. This is comparable to ours in the end (see Fig.~\ref{fig:Q_ev_Fuk20}). Thus, the number of photons necessary to ionize the entire system is similar in both simulations, but the total stellar masses to reach that point are different.

Thus, we confirm that our feedback models give results similar to previous simulations performed using radiation hydrodynamics codes. We also find positive and negative effects of the stellar feedback on the star formation when we correctly treat the dynamical evolution of star clusters. We identify a small dependence of the total stellar mass on the gas-particle mass, although we might underestimate the feedback for the \HII region model. When the gas density is high, there are only a few gas particles within the Str{\"o}mgren radius. If the gas mass within the Str{\"o}mgren radius estimated from the gas density is for 2.5 gas particles, we consider the feedback energy for only two particles and ignore the feedback energy for 0.5 particles. This treatment underestimates the feedback energy, especially the mass resolution of gas is low.

\subsection{Formation and evolution of star clusters}\label{clusters}
We further compare the star clusters formed in our simulations with observations and previous theoretical works. 
We detect clusters from the snapshots for models K18R20HII and K18R20HIIm001 at 5, 6, 7, 9, and 11\,Myr using \textsc{HOP} \citep{1998ApJ...498..137E} in \textsc{AMUSE}. \textsc{HOP} is an algorithm that detects clumps using local stellar densities. We use the same parameter set as that used in \citet{2019MNRAS.486.3019F}. After 8\,Myr, most clumps are almost gas-free. We therefore select dynamically bound particles and count them as cluster members at 9 and 11\,Myr. In contrast, clumps are embedded in the gas at an earlier time. We count all particles chosen by \textsc{HOP} as cluster members ($<$\,8\,Myr). The minimum number of stars for the detection is 100. As the expected mean stellar mass is $\sim 0.5 M_{\odot}$, clusters of $\lesssim 50 M_{\odot}$ are ignored in our analysis. 
We then calculate the half-mass radius ($r_{\rm h, cl}$) and half-mass density ($\rho_{\rm h, cl}$) of the detected clusters. There is no clear difference in the mass and radius distributions of the two models, although the mass resolution for gas is different. 

\begin{figure*}
 \begin{center}
 \includegraphics[width=7.8cm]{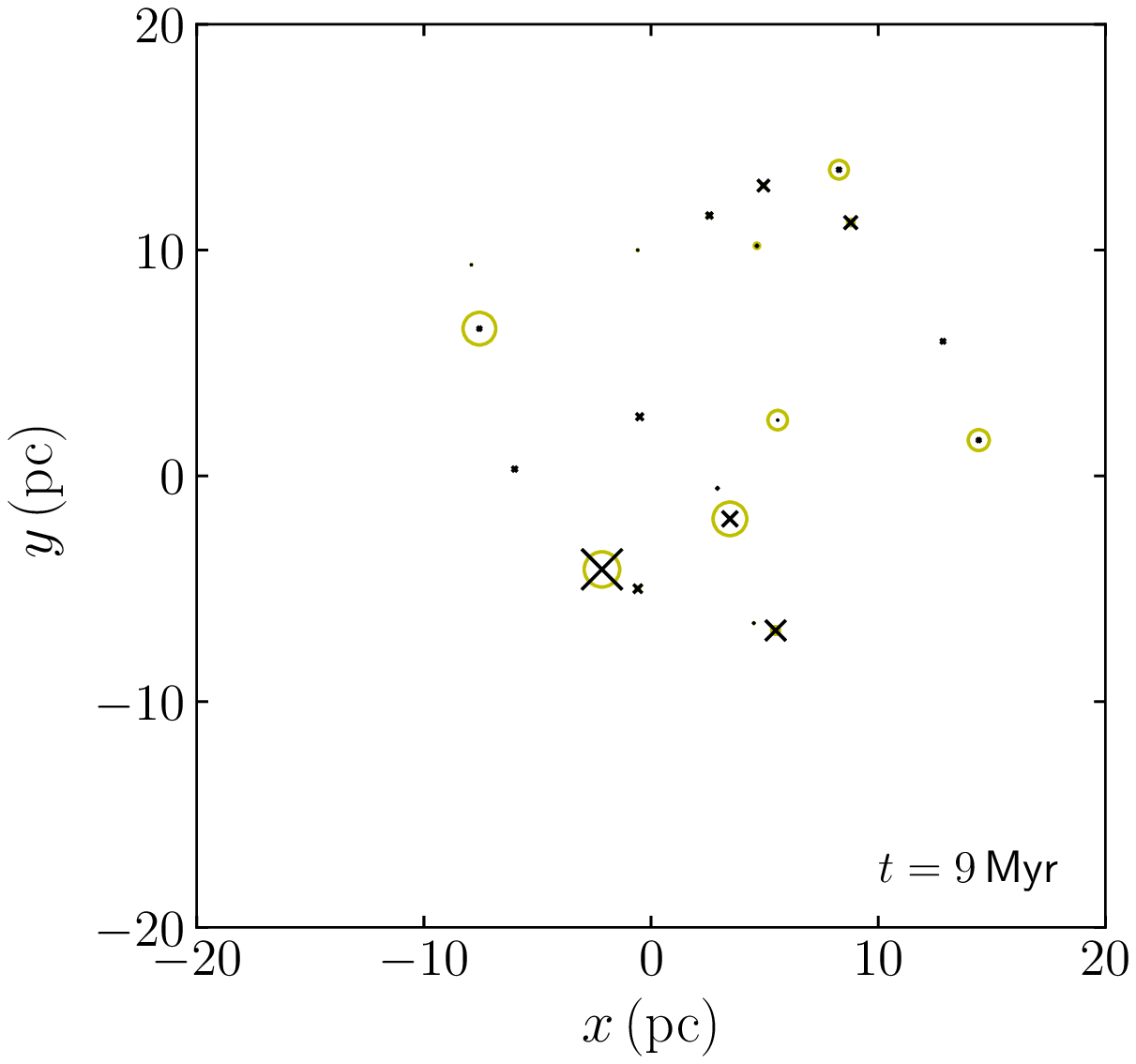}
  \includegraphics[width=7.8cm]{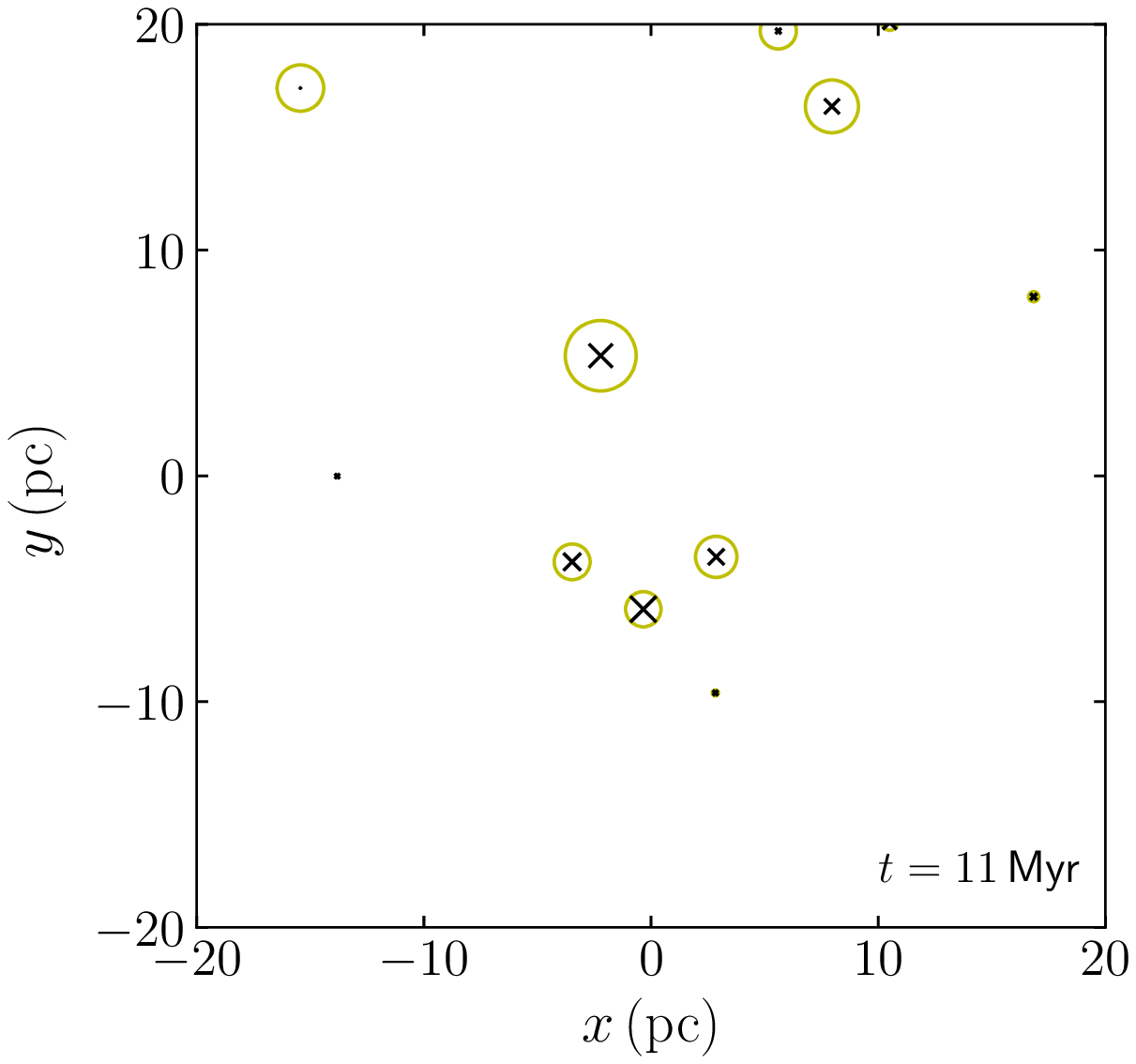}
 \end{center}
\caption{Cluster distribution of model K18R20HIIm001 at $t=9$ (left) and $11$ (right) Myr. Yellow circles represent the half-mass radii. Black crosses indicate the center-of-mass positions. The sizes of crosses represent the cluster masses.}\label{fig:clump_dist_Kim18}
\end{figure*}

In Fig.~\ref{fig:clump_dist_Kim18}, we present the spatial distribution of the detected clusters at 9 and 11\,Myr for model K18R20HIIm001. We indicate the half-mass radii and masses by the sizes of the circles and crosses. We confirm that our clump-finding method can detect major clumps (see also Fig.~\ref{fig:snapshot_Kim18m001}). Note that some clumps are found to have a much higher mean mass that than expected from the given mass function ($\sim0.5 M_{\odot}$). This may be due to the clump-finding methods. In the following figures, we present clumps with a mean stellar mass larger than $0.8\,M_{\odot}$ with small symbols.

\begin{figure}
 \begin{center}
  \includegraphics[width=7.8cm]{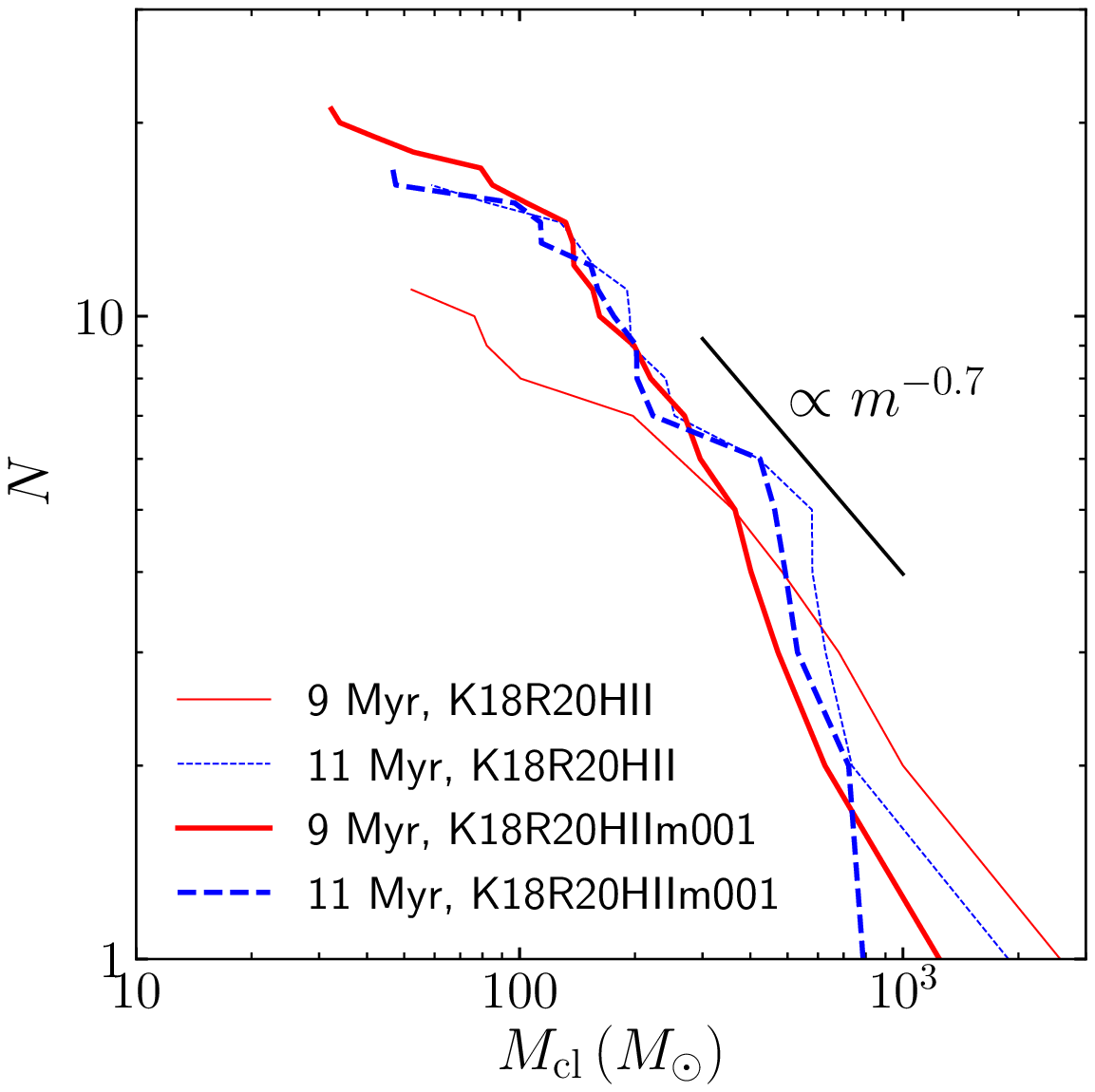}
 \end{center}
\caption{Cumulative cluster mass function of K18R20 models.}\label{fig:CMF_Kim18}
\end{figure}

In Fig.~\ref{fig:CMF_Kim18}, we present the cluster mass functions obtained from our simulations. The mass of the most massive cluster is $\sim 1000 M_{\odot}$. 
\citet{2015MNRAS.449..726F} performed a series of $N$-body simulations starting from initial conditions constructed from the results of SPH simulations of molecular clouds. They first performed SPH simulations of molecular clouds for one initial free-fall time. Subsequently, they stopped the SPH simulations and formed stellar particles assuming a star-formation efficiency dependent on the local gas density. They further performed $N$-body simulations using star particles only assuming an instantaneous gas expulsion. Although their method simplified the gas expulsion process, they gave a relation between the mass of the most massive cluster ($M_{\rm cl, max}$) as a function of initial cloud mass ($M_{\rm g}$):
\begin{equation}
    M_{\rm cl, max} = 6.1 M_{\rm g}^{0.51}.
\end{equation}
Applying this relation to our results, we expect the maximum mass of a cluster to be $\sim2000 M_{\odot}$. This mass is consistent with our results. 

\citet{2015MNRAS.449..726F} obtained a power of their obtained cluster mass function as $\alpha = -1.7$. In Fig.~\ref{fig:CMF_Kim18}, we also present this relation. As our number of obtained clusters is only $\sim 20$ per simulation, it is difficult to fit a power law to our cluster mass function, but the power is apparently shallower than $-2$.

\begin{figure*}
 \begin{center}
  \includegraphics[width=7.8cm]{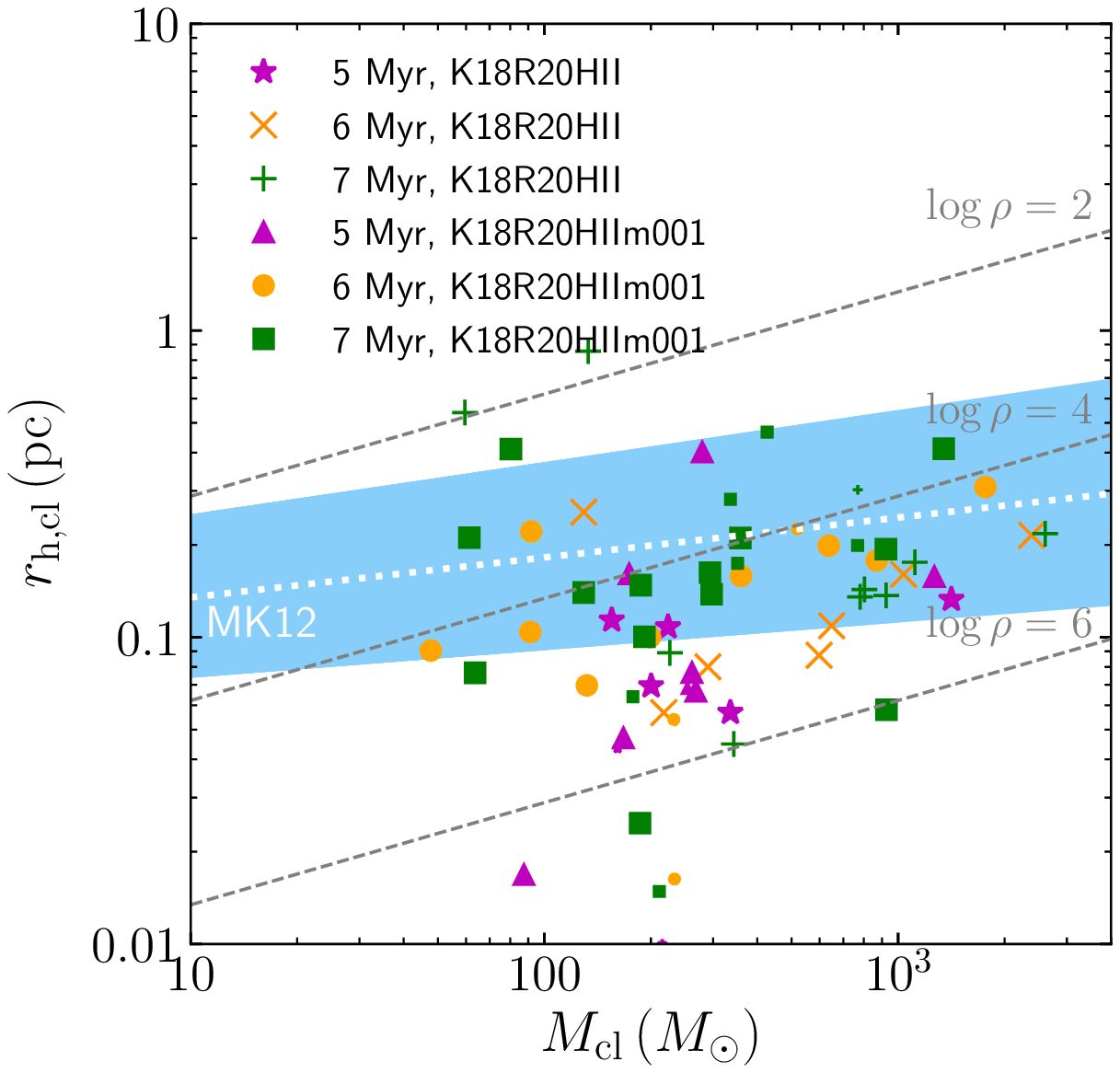}
  \includegraphics[width=7.8cm]{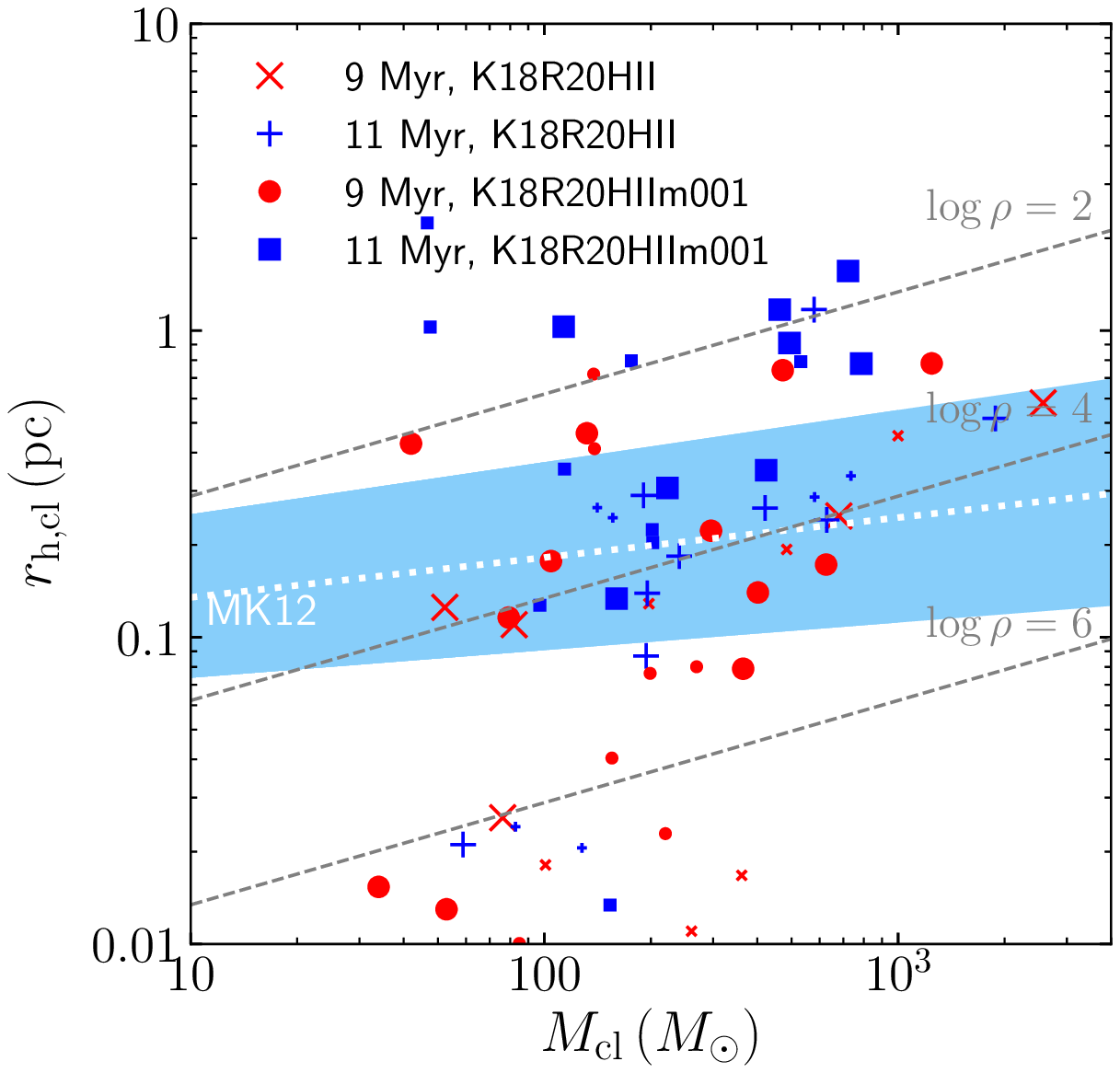}
 \end{center}
\caption{Relation between the mass and half-mass radius of embedded clusters (left) and bound clusters (right) detected in models K18R20HII and K18R20HIIm001. Light blue shaded region and white dotted line indicate equation (\ref{eq:MK12rad}). Small symbols indicate clusters with mean stellar mass of $>0.8\,M_{\odot}$.}\label{fig:mass_radius_Kim18}
\end{figure*}

We also show the relation between cluster mass and half-mass radius in Fig.~\ref{fig:mass_radius_Kim18}. At $\lesssim 7$\,Myr, the clusters are still embedded in the gas (see Fig.~\ref{fig:snapshot_Kim18m001}). Therefore, we did not check if the selected stars bound to the cluster or not. At this earlier time, the obtained clusters have radii much smaller than those typical of currently observed open clusters \citet[see figure~2 of][]{2010ARA&A..48..431P}. After gas expulsion due to the stellar feedback ($t\gtrsim 9$\,Myr), the half-mass radius expands. The radius after gas expulsion is consistent with the densest open clusters shown in \citep[figure~2 in ][]{2010ARA&A..48..431P}. For $t=9$ and 11\,Myr, the masses of bound stars are shown. 

\begin{figure*}
 \begin{center}
  \includegraphics[width=7.8cm]{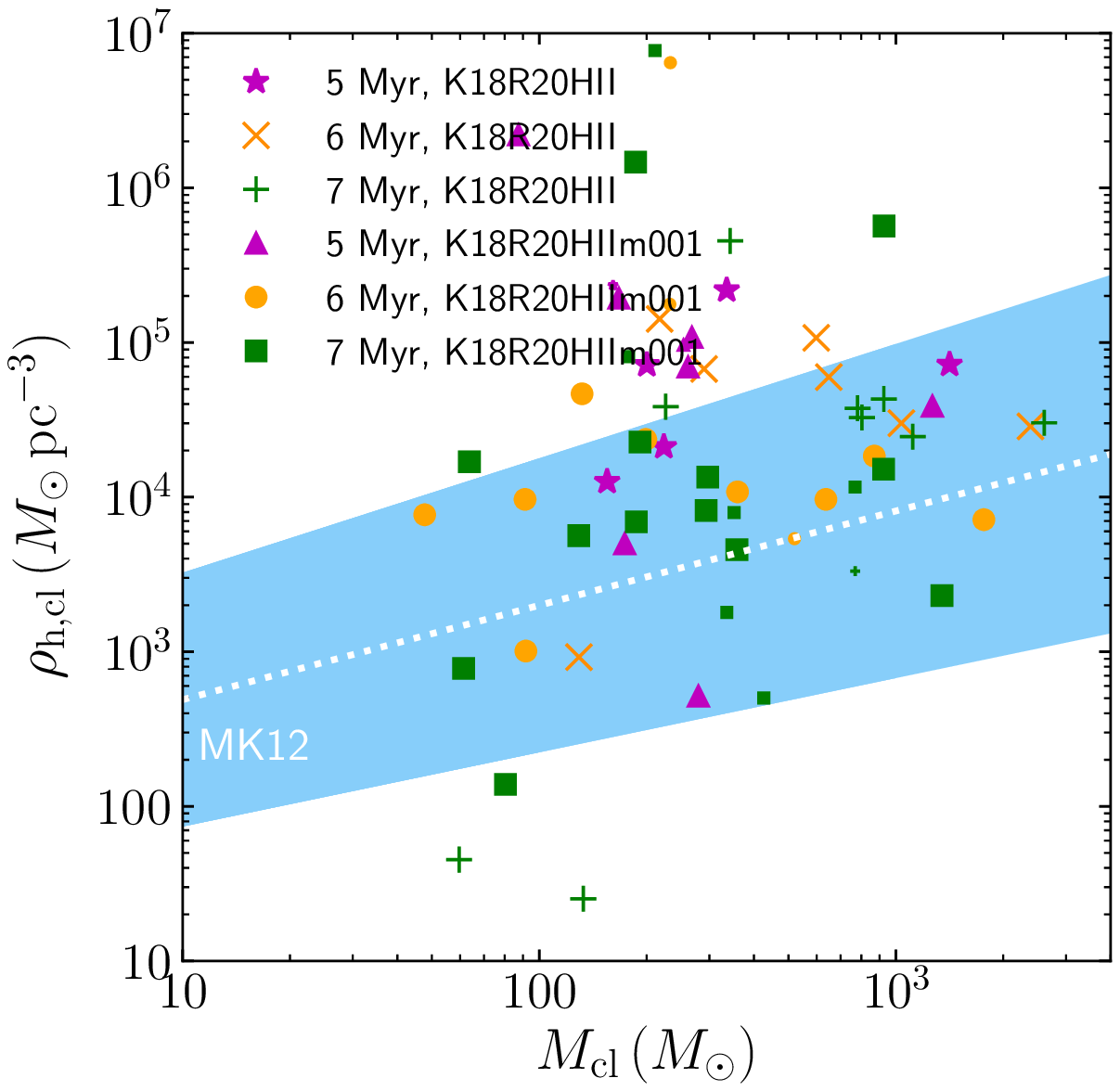}
  \includegraphics[width=7.8cm]{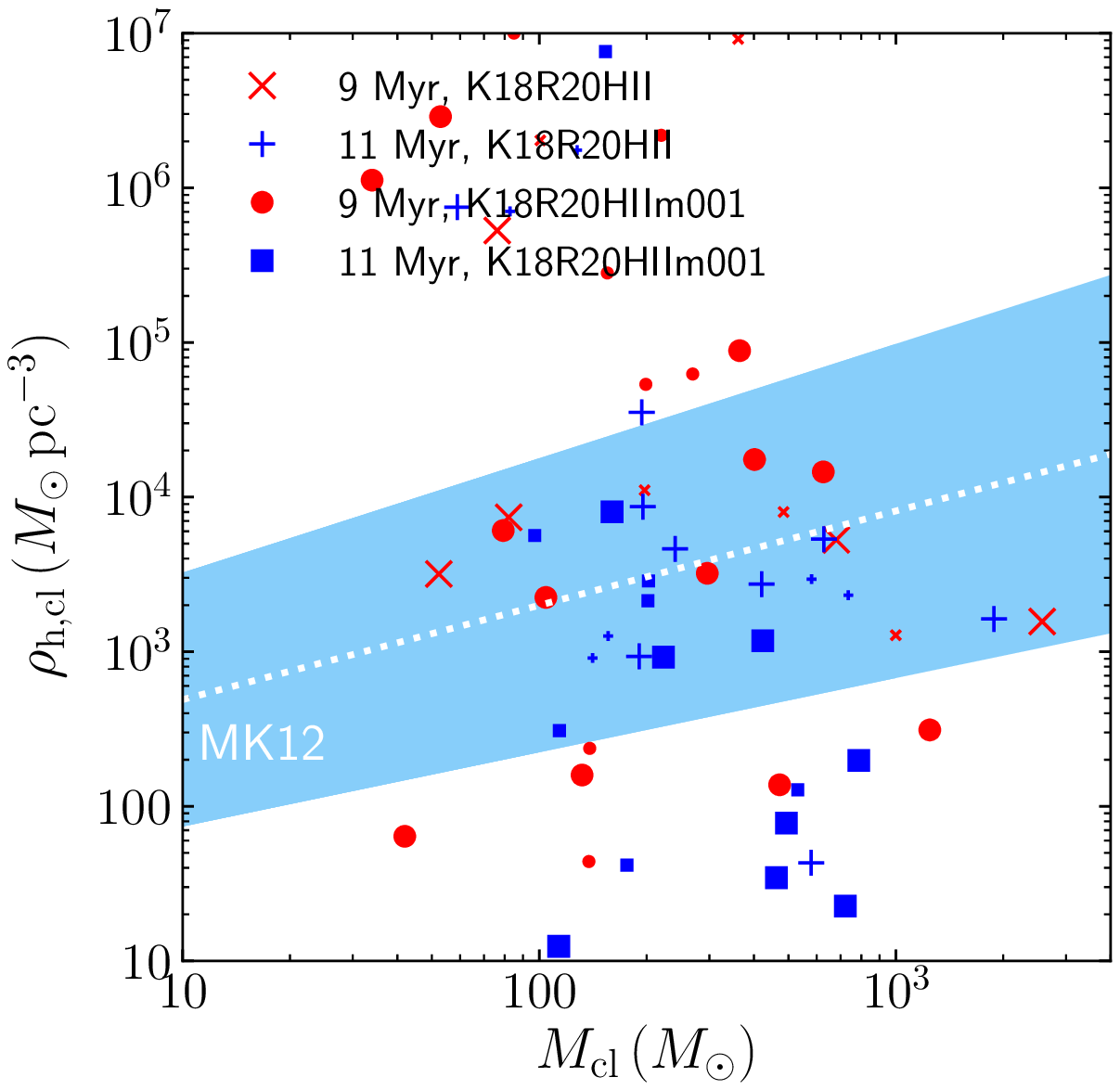}
 \end{center}
\caption{Relation between the mass and half-mass density of embedded clusters (left) and bound clusters (right) detected in models K18R20HII and K18R20HIIm001. Light blue shaded region and white dotted line indicate equation (\ref{eq:MK12dens}).  Small symbols indicate clusters with mean stellar mass of $>0.8\,M_{\odot}$.}
\label{fig:mass_density_Kim18}
\end{figure*}

\citet{1995ApJ...447L..95K} suggested that open clusters were much denser at the moment of formation and the dynamical evolution including binaries decreased the density down to that of currently observed clusters. \citet{2012A&A...543A...8M} estimated the relation between the initial mass and half-mass radius (or density) of star clusters using an inverse dynamical population synthesis, in which the initial mass and density of currently observed star clusters were estimated using `dynamical operators.' They estimated the initial cluster mass and radius as follows:
\begin{equation}
    \frac{r_{\rm h, cl}}{{\rm pc}} = 0.10^{+0.07}_{-0.04} \times \left( \frac{M_{\rm cl}}{M_{\odot}} \right)^{0.13\pm0.04}
    \label{eq:MK12rad}
\end{equation}
and the initial density and mass follows:
\begin{equation}
    \log_{10} \left( \frac{\rho_{\rm h, cl}}{M_{\odot}\,{\rm pc}^{-3}}\right) = 0.61\pm 0.13\times \log _{10} \left( \frac{M_{\rm cl}}{M_{\odot}}\right) + 2.08\pm 0.69.
    \label{eq:MK12dens}
\end{equation}
We show these relations in Fig.~\ref{fig:mass_radius_Kim18} and the mass--density plot in Fig.~\ref{fig:mass_density_Kim18}. Our results are roughly consistent with their results or even more dense when they are still embedded in gas. We also find a large scatter in the relation between the cluster mass and radius (or density). 

We also compare our results with a similar simulation, but assuming instantaneous star formation and gas expulsion performed by \citet{2016ApJ...817....4F}. The mass and radius obtained in both simulations are similar ($100<M_{\rm cl}<10^3 M_{\odot}$ and $0.1<r_{\rm hm, cl}<1$\,pc). Thus, the simplified star formation and gas expulsion models adopted by \citet{2016ApJ...817....4F} are sufficient to study cluster mass function and dynamical evolution of clusters after gas expulsion.

\begin{figure}
 \begin{center}
  \includegraphics[width=7.8cm]{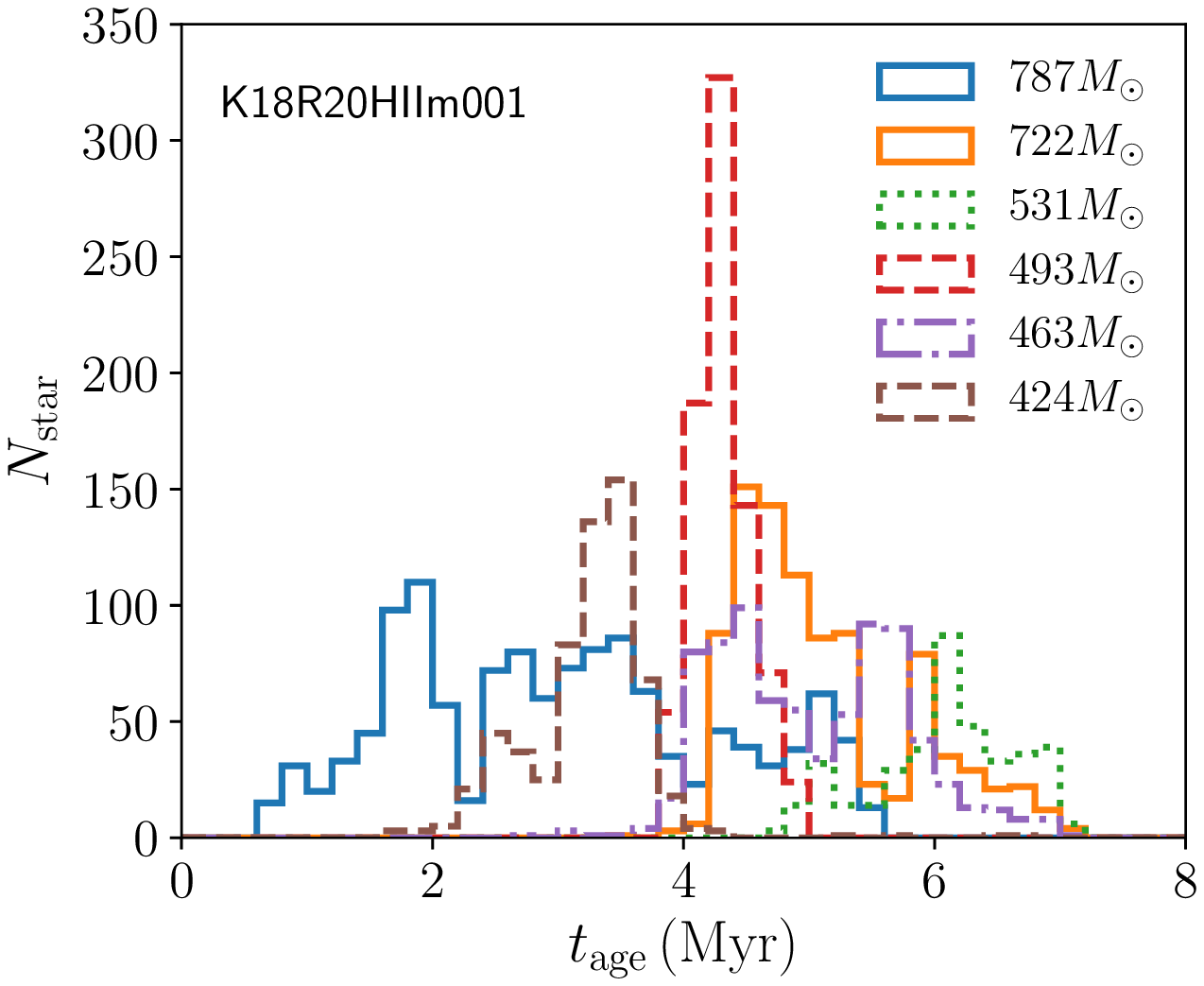}
 \end{center}
\caption{Stellar age distribution of the six most massive clusters at $t=11$\,Myr in model K18R20HIIm001 .}\label{fig:cluster_age_dist_Kim18}
\end{figure}

We further investigate the stellar ages in our formed clusters. 
In Fig.~\ref{fig:cluster_age_dist_Kim18}, we present the age distribution of stars in the six most massive clusters detected at $t=11$\,Myr for model K18R20HIIm001. The age distribution varies for each cluster. One of the biggest clusters continues star formation for $\sim 5$ Myr, while another forms most of its stars within $\sim1$\,Myr. There are also ones with double and triple peaks in the age distribution. These are a result of hierarchical formation \citep{2012ApJ...753...85F,2015PASJ...67...59F}. Multiple peaks of stellar ages in one cluster have also been observed in the Orion Nebula Cluster \citep{2017A&A...604A..22B}.

\begin{figure}
 \begin{center}
  \includegraphics[width=7.8cm]{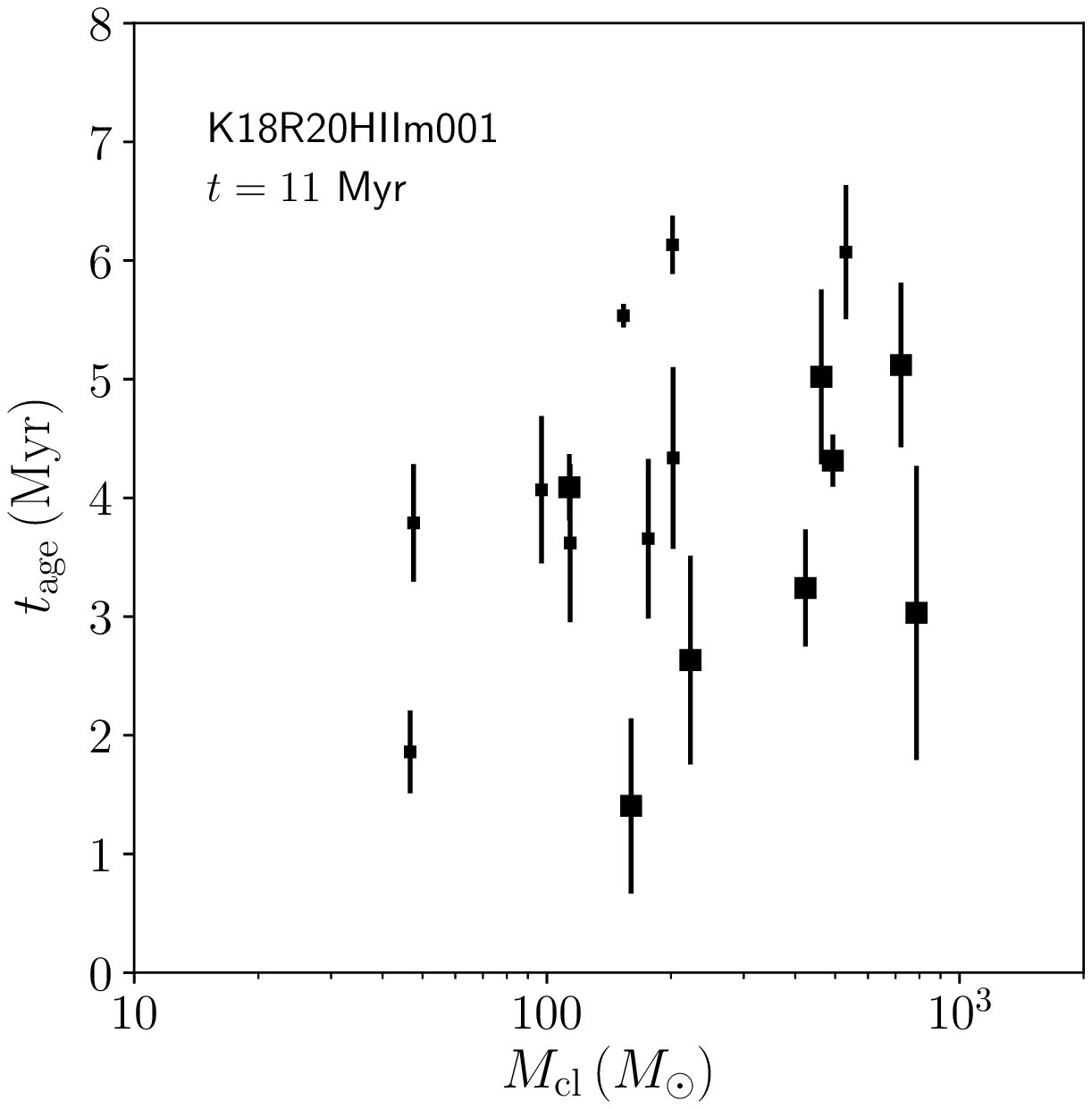}
 \end{center}
\caption{Age (mean stellar age) and age spread among cluster members for clusters detected at $t=11$\,Myr as a function of cluster mass for model K18R20HIIm001. Small symbols indicate clusters with mean stellar mass of $>0.8\,M_{\odot}$.}\label{fig:cluster_age_mass_Kim18}
\end{figure}

We also show the relation between cluster mass and age with the dispersion (standard deviation) in Fig.~\ref{fig:cluster_age_mass_Kim18}. We find a trend that more massive clusters form earlier, although the scatter is large. This is probably related to our initial condition, which is an initially homogeneous sphere. The gas cloud first gravitationally collapses before star formation starts in the center of the cloud, which is the densest region. Hence, the most massive clusters form around the cloud center (see Fig.~\ref{fig:snapshot_Kim18m001}). In contrast, small clumps can form in the outskirts of the cloud, even after the feedback starts to eject the gas from the most massive clusters.

\begin{figure}
 \begin{center}
  \includegraphics[width=7.8cm]{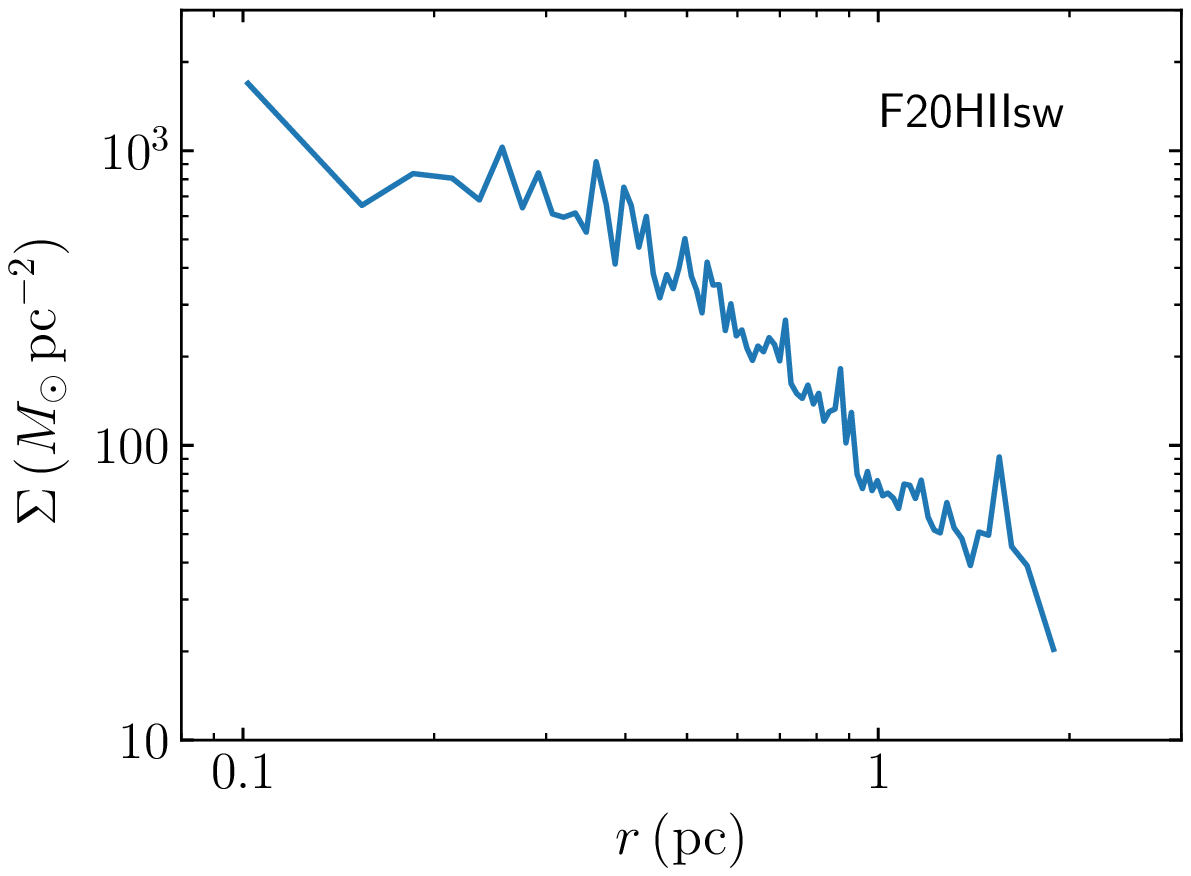}
 \end{center}
\caption{Surface density profile of the most massive cluster at $t=9$\,Myr in model F20HIIsw.}\label{fig:density_prof_F20}
\end{figure}

\begin{figure}
 \begin{center}
  \includegraphics[width=7.8cm]{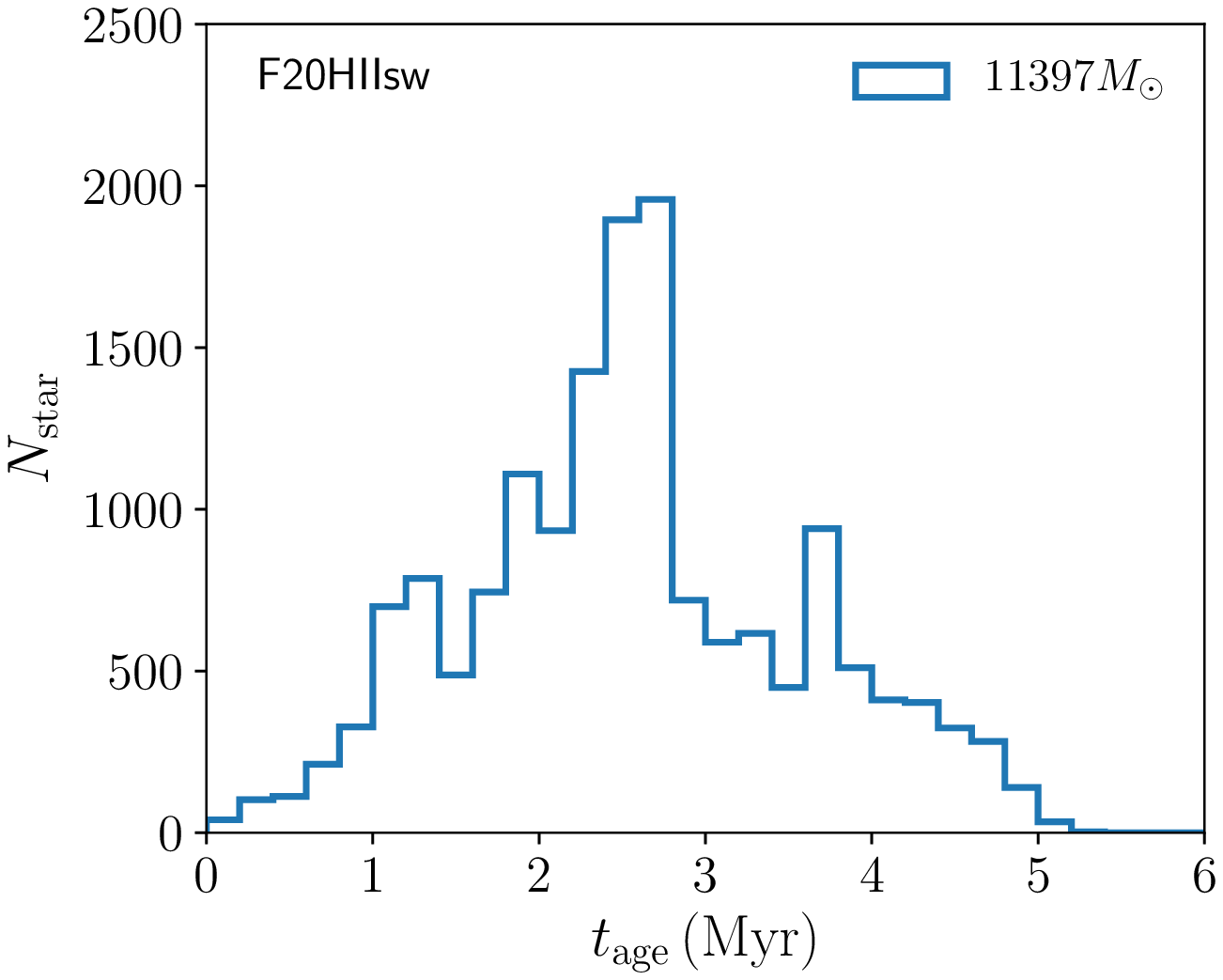}
 \end{center}
\caption{ Stellar age distribution of the most massive cluster at $t=9$\,Myr in model F20HIIsw .}\label{fig:cluster_age_dist_F20}
\end{figure}

 We also detect the clusters formed in the model F20HIIsw. As shown in figure~\ref{fig:snapshot_F20sw}, one massive cluster was formed 
in our sub-virial models. The bound mass of the most massive cluster at $t=9$\,Myr is $12000 M_{\odot}$, which is comparable to the masses of young massive clusters such as NGC 3603 and Trumpler 14 \citep{2010ARA&A..48..431P}. In Fig.~\ref{fig:density_prof_F20}, we present the surface density profile of this cluster. The core radius calculated using the method of \citet{1985ApJ...298...80C} is $0.33$\,pc, which is larger than that of NGC 3603 (0.15\,pc) and Trumpler 14 (0.14\,pc). The half-mass radius of the cluster in our simulation is 0.61\,pc, while the effective radii of NGC 3603 and Trumpler 14 are 0.7 and 0.5\,pc, respectively \citep{2010ARA&A..48..431P}. In Fig.~\ref{fig:cluster_age_dist_F20}, we show the age distribution of the cluster at $t=9$\,Myr. There are a peak at $\sim 2.5$\,Myr, a few small sub-peaks and an elongated distribution.

\citet{2015MNRAS.447..728B} studied the hierarchical formation of young massive clusters using $N$-body simulations without softening, although the gas potential is static in their simulations. Their results suggest that either a monolithic or a hierarchical formation within 1 Myr is preferable for NGC 3603, which is a young massive cluster in the Milky Way. Our result also supports the formation of young massive clusters in a short period.

We also investigate the fraction of mass remaining in bound clusters at the end of the simulations. For the model K18R20HIIm001 (the super-virial case), 5,500 $M_{\odot}$ stars belong to bound clusters. This mass is 37\,\% of the total stellar mass formed in the simulation. For the model F20HIIsw (sub-virial case), 15,000\,$M_{\odot}$ stars belong to bound clusters. This mass is 36\,\% of the total stellar mass. 
Thus, about one-third of stars belong to bound clusters after the gas expulsion in our simulations.

\section{Summary}
We developed a new tree-direct hybrid code, \textsc{ASURA+BRIDGE} \citep{2021arXiv210105934F}, in which stars are realized as individual stars following a realistic mass function and are integrated using high-order integrators and sixth-order Hermite \citep{2006NewA...12..169N} and P$^3$T schemes \citep{PeTar2020MNRAS.497..536W}. 
In this code, we implemented stellar feedback schemes: thermal feedback due to radiation from massive stars and mechanical feedback due to the radiation and stellar wind. 
For the radiation, we adopted a model giving thermal feedback within a Str{\"o}mgren radius calculated from the gas density and radiation of massive stars. The ionization photon counts were estimated using OSTAR2002 \citep{2003ApJS..146..417L}. We also implemented the dusty feedback model of \citet{2016ApJ...819..137K}. 

Using this new code, we performed simulations of star cluster formation from turbulent molecular clouds. 
In these simulations, we adopted a probabilistic star formation scenario, in which star formation statistically happens. The star formation rate depends on the local density. The stellar mass was drawn from a given stellar mass function \citep{2020arXiv200512906H}. In this scheme, we were able to achieve a star-by-star treatment without using an extremely high resolution for the gas particle mass. Moreover, the integration of stellar orbits did not require softening. Thus, we were able to properly investigate the dynamical evolution of star clusters.

To compare our results with previous works, we performed simulations of models similar to those of \citet{2018ApJ...859...68K} and \citet{2020MNRAS.497.3830F} with/without gravitational softening. 
Without softening, some stars experienced close encounters with other stars in dense star-forming regions. These close encounters kicked out stars out of clusters. In particular, once tight binaries formed, even massive stars formed inside clusters were ejected. We found that such runaway massive stars escaped from the cluster center (dense region) to ionize the gas in the outskirts of the star-forming regions. With softening, in contrast, massive stars were kept inside clusters and ionization proceeded inside of the forming clusters. 

The difference with/without softening in the stellar mass evolution is not simple. When the system is super-virial, star formation is suppressed with softening for stars. In this case, ejected massive stars ionize the gas from the outskirts of the star-forming regions. This stops star formation at a later time. 
In sub-virial cases, in contrast, star formation is enhanced without softening for stars. In this case, one massive cluster forms in the cloud center, and it also ejects some massive stars, which ionize the gas as they reached outer, less dense regions. With this feedback, some gas is compressed. As a result, the star-formation rate increases. 
Our results are consistent with those of previous studies using radiation hydrodynamics simulations, in which the radiation was correctly treated but stars were treated as super-particles. With a softening length similar to that in \citet{2018AJ....156...84K}, we obtained a similar final stellar mass in the initially super-virial cloud model. In the case of sub-virial models, which were compared to those in \citet{2020MNRAS.497.3830F}, we obtained a higher star-formation efficiency. This may be due to the lack of high-mass stars in our simulations, which is also seen in observations of star clusters in the Milky Way. 

We further investigated the mass and radius relation of clusters formed in our simulations and compared them with observations and theoretical studies. Our results suggest that star clusters initially had densities much higher than those of currently observed open clusters. This result supports the analytical estimate reported in \citet{2012A&A...543A...8M}. Our results suggest that the densities of open clusters were $10^3$--$10^5 M_{\odot}$\,pc$^{-3}$ when they were embedded in their parental gas. However, the variation in the density is quite large. 

The star-formation history also has a large variation. The most massive cluster in our super-virial model contained stars with an age distribution in 5\,Myr. The distribution of this cluster showed three peaks caused by the hierarchical formation of star clusters. In contrast, another cluster showed a formation period of only $\sim 1$\,Myr. With our sub-virial initial condition one massive cluster was formed, and its mass and radius were comparable to those of young massive clusters in the Milky-Way galaxy. In both super- and sub-virial cases, about one-third of stars remain in the bound clusters after gas expulsion. We will further investigate the star-cluster-formation process using models with different initial conditions for gas in a forthcoming paper. 

Our model currently does not consider the mass loss from massive stars due to stellar wind, the supernova explosions and the formation of compact objects like black holes. These may affect the evolution of star clusters, especially in the cluster core. Supernova explosions may terminate the relatively slow star formation with a timescale longer than the present models. We will investigate these effects in the future work.

\begin{ack}
The authors thank Takashi Hosokawa for useful comments on the development of the code, Hajime Fukushima for fruitful discussion, and Steven Rieder for helping with AMUSE scripts.
The authors also thank the anonymous referee for their comments to improve the manuscript.
L.W. is grateful for the financial support from the JSPS International Research Fellowship (Graduate School of Science, The University of Tokyo).
Numerical computations were carried out on Cray
XC50 CPU-cluster at the Center for Computational Astrophysics (CfCA) of the National Astronomical Observatory of Japan and Oakbridge-CX system at the Information Technology Center, The University of Tokyo. 
This work was supported by JSPS KAKENHI Grant Number 19H01933, 20K14532, 21J00153, 21H04499, 21K03614, 21K03633 and MEXT as “Program for Promoting Researches on the Supercomputer Fugaku” (Toward a unified view of the universe: from large scale structures to planets, Revealing the formation history of the universe with large-scale simulations and astronomical big data), The University of Tokyo Excellent Young Researcher Program, the Initiative on Promotion of Supercomputing for Young or Women Researchers, Information Technology Center, The University of Tokyo, and the RIKEN Special Postdoctoral Researchers Program.
We would like to thank Editage (www.editage.com) for English language editing.
\end{ack}

\bibliography{reference}

\appendix 

\section{STARBENCH test}\label{sec:starbench}

To test our model for \HII region, we perform a series of simulations similar to STARBENCH runs \citep{2015MNRAS.453.1324B}. STARBENCH is a project to compare numerical simulations of expanding \HII region using different codes including both 1D and 3D grid-based and SPH codes.

\begin{table}
\begin{center}
  \tbl{Models for STARBENCH}{%
  \begin{tabular}{lcccc}
      \hline
      Name  & $m_{\rm g} (M_{\odot})$ & $\epsilon_{\rm g}$ (au) & Cooling & Other Feedback  \\ 
      \hline
      SB100K & 0.1 & 10000 & N & - \\
      SBCo & 0.1 & 10000 & Y & - \\
      SBCoRP & 0.1 & 10000 & Y & RP \\
      SBCoRPSW & 0.1 & 10000 & Y & RP, SW \\
      SB100K-H & 0.01 & 5000 & N & - \\
      SBCoRPSW-H & 0.01 & 5000 & Y & RP, SW \\
      \hline
    \end{tabular}}\label{tb:IC_SB}
\begin{tabnote}
`RP' and `SW' indicate radiation pressure (eq.~\ref{eq:kick_dusty}) and stellar wind (eq.~\ref{eq:kick_other}), respectively.
\end{tabnote}
\end{center}
\end{table}

We take an initial condition similar to that used in STARBENCH \citep{2015MNRAS.453.1324B} for SPH codes. Here, we describe our model, which is slightly different from STARBENCH initial condition. 
We assume a homogeneous sphere with a density of $5.21\times 10^{-21}$\,g\,cm$^{-3}$ with a radius of 5.82\,pc, which is twice as large as that adopted in STARBENCH. The resulting total gas mass is $6.4\times 10^4M_{\odot}$. We adopt a mass resolution of gas as $m_{\rm g}=0.1M_{\odot}$ and $0.01M_{\odot}$, which are the same as those we adopt for our cluster-formation simulations. The initial temperature of gas is set to 100\,K (model SB100K). We also perform a run with cooling (model SBCo). For models with cooling, we set the initial gas temperature to be 20\,K. 
We adopt a gravitational softening length of 10000\,au for the runs with a gas particle mass of $0.1 M_{\odot}$. For the runs with $0.01 M_{\odot}$, we set the softening length to $5,000$\,au. These models are summarized in table~\ref{tb:IC_SB}.

Instead of putting a source of $Q_0=10^{49}$\,photons per second, we put a star which emit photons with the same rate ($40 M_{\odot}$ in our model).
In addition, our simulations include self-gravity, and therefore the gas cloud collapses in its initial free-fall time.

\begin{figure}
 \begin{center}
  \includegraphics[width=7.8cm]{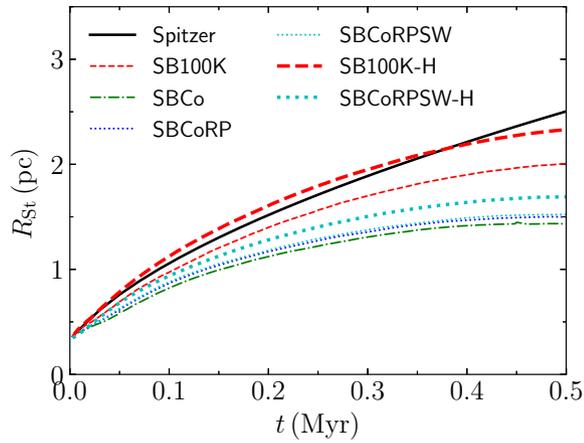}
 \end{center}
\caption{Evolution of Str{\"o}mgren radius for our models and Spitzer solution (eq.~\ref{eq:Spitzer}).}\label{fig:Stromgren_radius}
\end{figure}

\HII region expands because of the thermal pressure caused by the large temperature difference between the ionized region and the surrounding ISM.
This phase is called as Dense-type (D-type). \citet{1978ppim.book.....S} and \citet{1980pim..book.....D} obtained an analytic solution for the expansion as 
\begin{eqnarray}
R_{\mathrm{Sp}}(t)=R_{\mathrm{St}}\left(1+\frac{7}{4} \frac{c_{\mathrm{i}} t}{R_{\mathrm{St}}}\right)^{4 / 7},
\label{eq:Spitzer}
\end{eqnarray}
where $c_{\rm i}$ is the sound speed of ionized gas and $R_{\rm St}$ is the initial Str{\"o}mgren radius given as equation~\ref{eq:Stromgran_radius}.
This is the so-called Spitzer solution. 
Since we assume that $T_{\rm i}=10^4$\,K, $c_{\rm i}=12.85$\,km\,s$^{-1}$. 
The obtained analytic solution is shown in Fig.~\ref{fig:Stromgren_radius}.

In the same figure, we also show our results. Without cooling (models SB100K and SB100K-H), our results are consistent with the analytic formula. In STARBENCH simulations, the expansion stops at a radius, at which \HII region reaches in pressure equilibrium. The stagnation radius is analytically given as 
\begin{equation}
    R_{\mathrm{Stag}}=\left(\frac{c_{\mathrm{i}}}{c_{\mathrm{o}}}\right)^{4 / 3} R_{\mathrm{St}},
\end{equation}
where $c_{\rm o}$ is the sound speed of the outer gas and  $c_{\rm o}=0.91$\,km\,s$^{-1}$ for $T=100$\,K. In our model, $R_{\rm Stag}=10.7$\,pc. However, the Str{\"o}mgren radius stop to expand at $\sim 2.3$\,pc at 0.5\,Myr and then start to shrink. This is because our code include self-gravity. 
Compared with higher resolution model, we find that the \HII region is slightly larger if we improve the resolution. However, the difference is $\sim 10$\,\%.

We also test the case with cooling (model SBco), since our final goal is simulations with cooling. In addition, we test models with mechanical feedback due to radiation pressure (equation \ref{eq:kick_dusty}) (model SBcoRP) and also stellar wind (equation \ref{eq:kick_other}) (models SBCoRPSW). These results are also shown in Fig.~\ref{fig:Stromgren_radius}.
With cooling, the \HII region becomes smaller compared with the case without cooling. The radiation pressure and stellar wind slightly pushes the Str{\"o}mgren radius outer. 
We finally perform a run including everything with high resolution (model SBCoRPSW-H). Similar to the case without cooling, the Str{\"o}mgren radius is $\sim10$\,\% larger than the case with a lower resolution.

\section{Single-source test}\label{sec:single_source}

To compare our dusty \HII region model with previous work performed in \citet{2016ApJ...819..137K}, we perform a series of simulations similar to those in \citet{2016ApJ...819..137K}. They followed the description of the radiation pressure in dusty \HII region in \citet{2011ApJ...732..100D} and used \textsc{athena} to perform the radiation hydrodynamics (RHD) simulations of expanding dusty \HII region. Their result agrees with the analytic solution from \citet{2011ApJ...732..100D}.

We adopt molecular clouds with a uniform density ($n$) of $10^3$, $10^4$ and $10^5$\,cm$^{-3}$. 
For all models, the mass resolution ($m_{\rm g}$) is $0.1\,M_{\odot}$.
The softening length ($\epsilon_{\rm g}$) is set to be $10^4$\,au.

The initial gas temperature ($T_{\rm ini}$) is 20\,K and 100\,K for models with and without cooling, respectively. 
We put a star with a mass of $60\,M_{\odot}$ in the center of the gas distribution. 
The emission rates of ionization photons ($Q_0=\dot{\mathcal{N}}_{\rm LyC}$) are set to be $10^{49}$ and $10^{52}$\,s$^{-1}$ for models with $n=10^3$ and $10^4$\,cm$^{-3}$, respectively. We adopt a sufficiently large cloud size to remain most of the gas bound after the formation of \HII regions. This size depends on both the gas density and the emission rate. The masses and radii of gas distribution for our initial conditions are summarized in table~\ref{tb:IC_dusty}.
In all simulations, the feedback due to stellar wind is not included, but the self-gravity is included.

In order to test the effect of resolution, we perform a simulation with a higher resolution, named as DHQ50n1e4H. For this model, $Q_0=10^{52}$\,s$^{-1}$, $n=10^4$\,cm$^{-1}$, $m_{\rm g}=0.01 M_{\odot}$ and $\epsilon_{\rm g}=5000$\,au.
For this run, cooling is not included. 

\begin{table*}
\begin{center}
  \tbl{Models for dusty \HII region}{%
  \begin{tabular}{lcccccccc}
      \hline
      Name  & $m_{\rm g} (M_{\odot})$ & $\epsilon_{\rm g}$ (au) & n (cm$^{-3}$) & $T_{\rm ini} (K)$ & $M_{\rm g} (M_{\odot})$ & $R_{\rm g}$ (pc) & $Q_0$ (s$^{-1}$) & Cooling\\ 
      \hline
      DHQ49n1e3 & 0.1 & $10^4$ & $10^3$ & $100$ & $4\times 10^5$ & $14.0$ & $10^{49}$ & N\\
      DHQ49n1e3C & 0.1 & $10^4$ & $10^3$ & $20$ & $4\times 10^5$ & $14.0$ & $10^{49}$ & Y\\
      DHQ49n1e4 & 0.1 & $10^4$ & $10^4$ & $100$ & $2\times 10^5$ & $5.16$ & $10^{49}$ & N\\
      DHQ49n1e4C & 0.1 & $10^4$ & $10^4$ & $20$ & $2\times 10^5$ & $5.16$ & $10^{49}$ & Y\\
      DHQ49n1e5 & 0.1 & $10^4$ & $10^5$ & $100$ & $5\times 10^5$ & $3.25$ & $10^{49}$ & N\\
      DHQ49n1e5C & 0.1 & $10^4$ & $10^5$ & $20$ & $5\times 10^5$ & $3.25$ & $10^{49}$ & Y\\
      DHQ50n1e4 & 0.1 & $10^4$ & $10^4$ & $100$ & $2\times 10^5$ & $5.16$ & $10^{50}$ & N\\
      DHQ50n1e4H & 0.01 & $5\times10^3$ & $10^4$ & $100$ & $2\times 10^5$ & $5.16$ & $10^{50}$ & N\\
      DHQ50n1e4C & 0.1 & $10^4$ & $10^4$ & $20$ & $2\times 10^5$ & $5.16$ & $10^{50}$ & Y\\
      DHQ50n1e5 & 0.1 & $10^4$ & $10^5$ & $100$ & $5\times 10^5$ & $3.25$ & $10^{50}$ & N\\
      DHQ50n1e5C & 0.1 & $10^4$ & $10^5$ & $20$ & $5\times 10^5$ & $3.25$ & $10^{50}$ & Y\\
      DHQ52n1e4 & 0.1 & $10^4$ & $10^4$ & $100$ & $2\times 10^6$ & $11.1$ & $10^{52}$ & N\\
      DHQ52n1e4C & 0.1 & $10^4$ & $10^4$ & $20$ & $2\times 10^6$ & $11.1$ & $10^{52}$ & Y\\
      \hline
    \end{tabular}}\label{tb:IC_dusty}
\begin{tabnote}
From the left: model name, gas particle mass ($m_{\rm g}$), softening length for gas ($\epsilon_{\rm g}$), initial gas density ($n$), initial gas temperature ($T_{\rm ini}$), total gas mass ($M_{\rm g}$), initial gas radius ($R_{\rm g}$), ionizing photon emission rate $Q_0$, and cooling model. Model names include the value of $Q_0$, gas density ($n$), and the cooling.
\end{tabnote}
\end{center}
\end{table*}

\begin{figure*}
 \begin{center}
  \includegraphics[width=7.8cm]{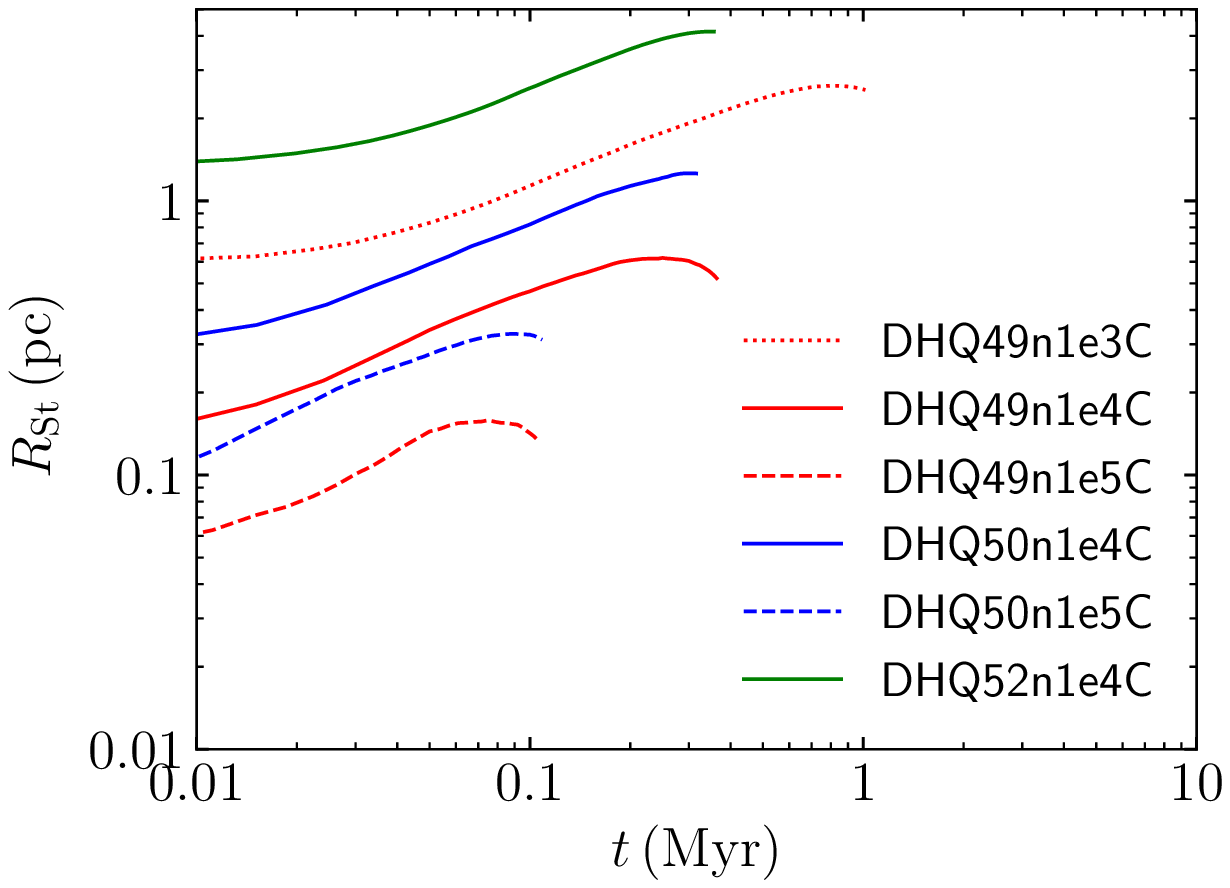}
  \includegraphics[width=7.8cm]{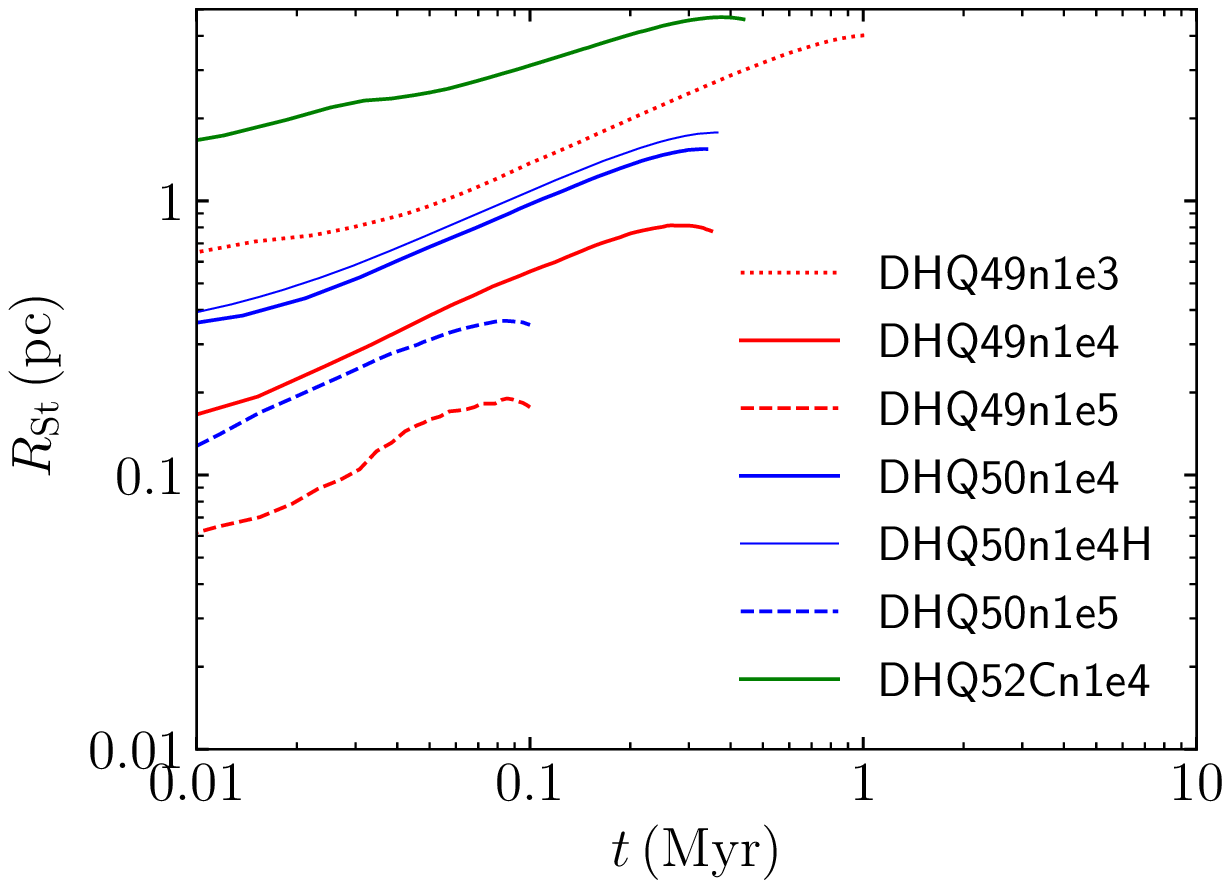}\\
  \includegraphics[width=7.8cm]{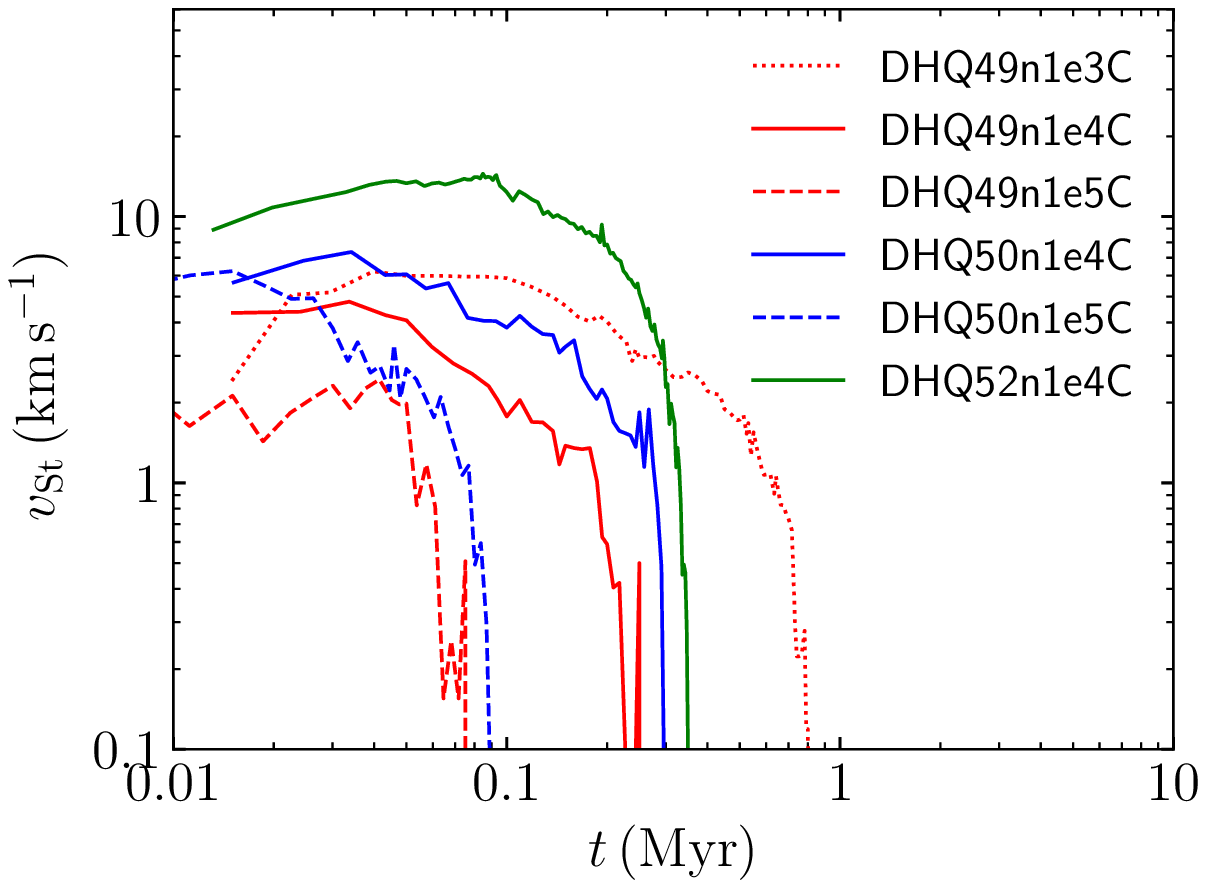}
  \includegraphics[width=7.8cm]{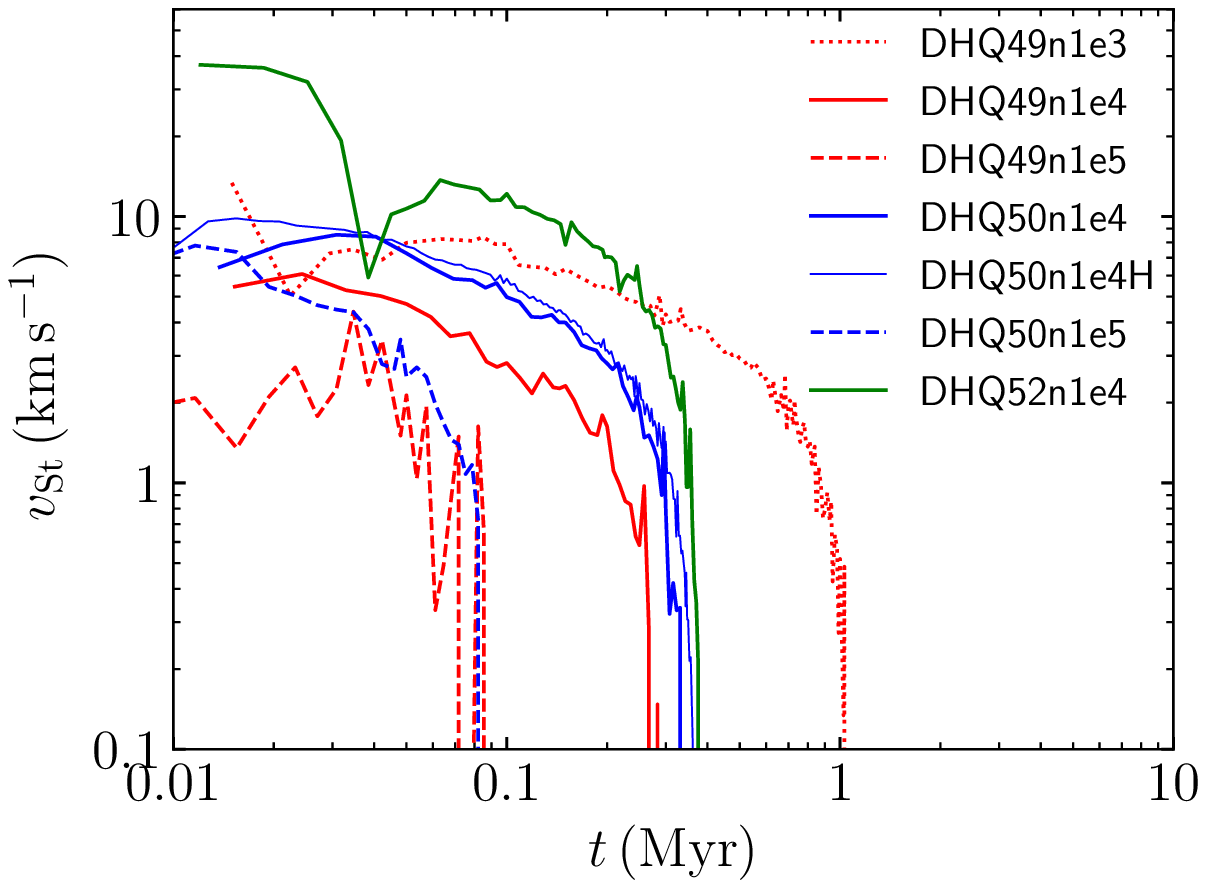}\\
 \end{center}
 \caption{Evolution of the Str{\"o}mgren radius and the expansion velocity of the shell for all models. Left and right panels are results with and without cooling.}
\label{fig:HII_dusty_all}
\end{figure*}

In Fig.~\ref{fig:HII_dusty_all}, we present the time evolution of the Str{\"o}mgren radius and its expansion velocity for all models. Since we include self-gravity, the expansion turns into contraction at one time. 
Once it happens, we stop the simulations.
The maximum radius and the time to reach there increases as the gas density decreases. 
As the ionizing photon emission rate increases, the maximum radius also increases, but the time to reach there does not.

With cooling, the maximum radius becomes slightly smaller compared with those without cooling. This difference becomes smaller as the Str{\"o}mgren radius becomes smaller. We also confirm that the resolution effect is sufficiently small by comparing models DHQ501e4 and DEQ501e4H.

We compare our results with those in \citet{2016ApJ...819..137K}. Their figure~4 shows the results of the models similar to DHQ50n1e4 and DHQ50n1e4C. In our simulations, the \HII region in the model DHQ50n1e4 expands up to $\sim 1.5$\,pc at $\sim 0.3$\,Myr, while that in \citet{2016ApJ...819..137K} expands to $\sim 2$\,pc at $\sim 0.8$\,Myr. Thus, our \HII region feedback model slightly underestimates the shell expansion. On the other hand, in a stronger emission case, i.e., models DHQ52n1e4 and DHQ52n1e4C, the \HII region in our models expands larger than that of \citet{2016ApJ...819..137K} ($\sim 3$\,pc). We note that such a large emission rate cannot be reached with ionizing photon emission from only one massive star (see Fig.~\ref{fig:Q_mass}).

\begin{figure}
 \begin{center}
  \includegraphics[width=7.8cm]{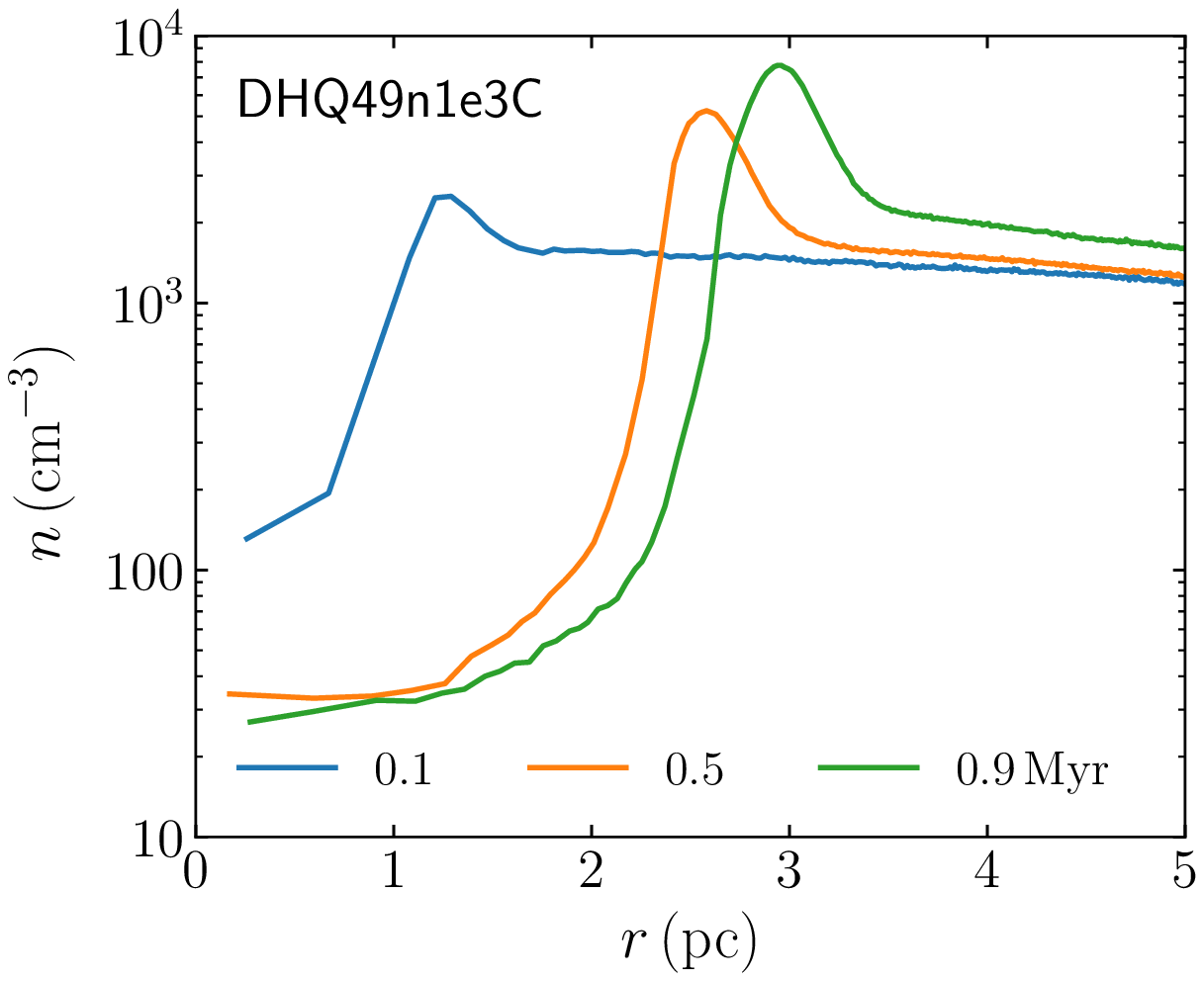}\\
  \includegraphics[width=7.8cm]{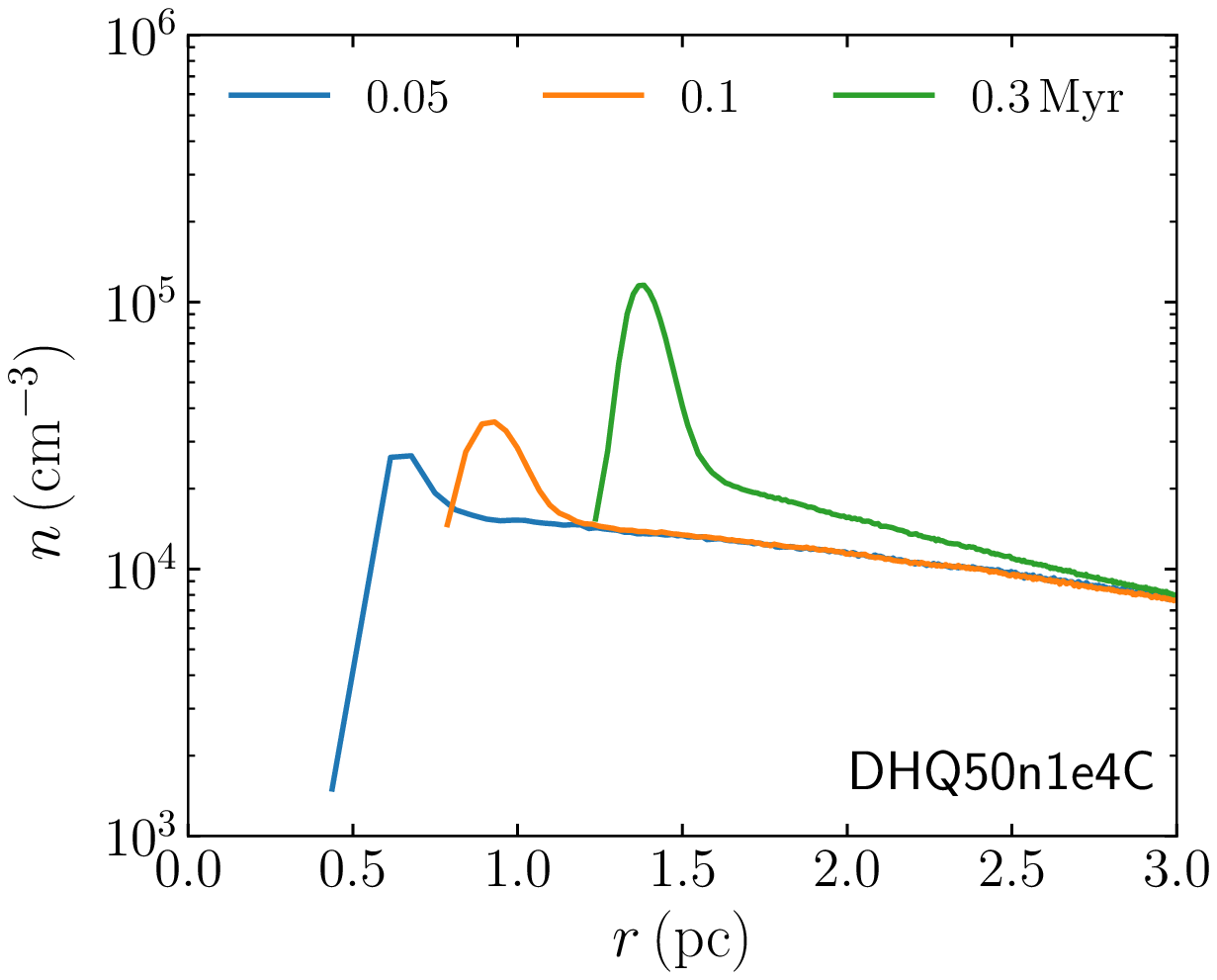}
  \includegraphics[width=7.8cm]{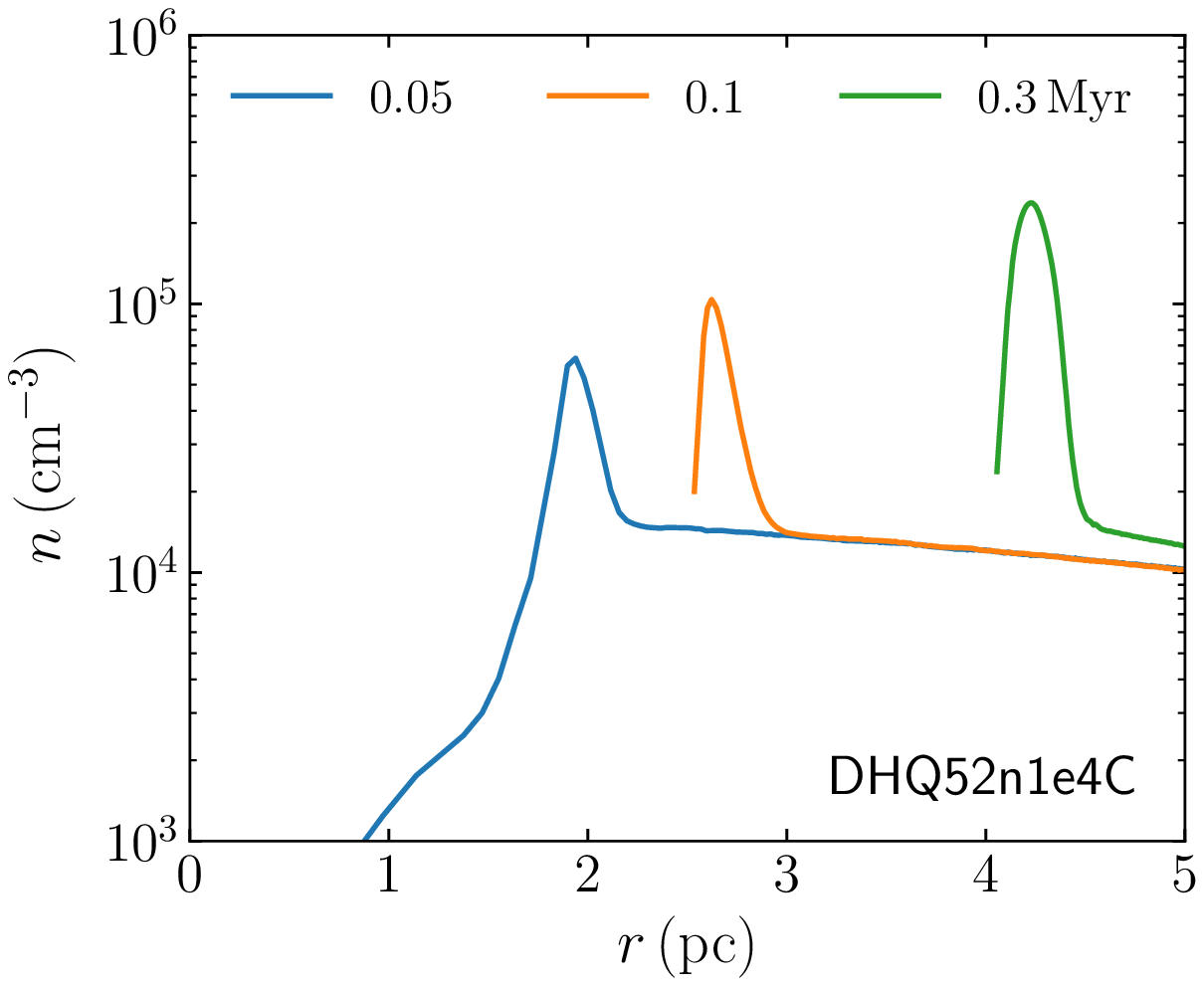}
 \end{center}
 \caption{Evolution of the radial density distributions of expanding \HII regions. From top to bottom, for models DHQ49n1e3C and DHQ50n1e4C, and DHQ52n1e4C.}
\label{fig:HII_dusty}
\end{figure}

In Fig.~\ref{fig:HII_dusty}, we present the time evolution of the radial density distributions of gas in our simulations. The models in the top (the model DHQ491e3C) and the bottom (the model DHQ521e4C) panels corresponds to the models A and C in \citet{2016ApJ...819..137K}. As is discussed in \citet{2016ApJ...819..137K}, this radiation pressure is expected to dominant to the thermal pressure when $nQ_{49} > 10^4$\,cm$^{-3}$, where $Q_{49}=Q_0/(10^{49}$\,s$^{-1}$). In such regime, a hole around the source is expected to form due the the radiation pressure. Indeed, our simulations also reproduce the hole with the same criterion (see the middle and bottom panels of Fig.~\ref{fig:HII_dusty}). 
The shell in our simulation is thicker than that of \citet{2016ApJ...819..137K}, but the density is roughly consistent. In \citet{2018ApJ...859...68K}, the shell density in \citet{2016ApJ...819..137K} decreases with time, because their simulations do not include self-gravity. In our simulations, on the other hand, the shell density increases with time since the cloud collapses due to the self-gravity. Thus, we confirm that our models sufficiently agree with those of \citet{2018ApJ...859...68K}.

\end{document}